\documentclass[twocolumn, twocolappendix]{aastex631}

\newcommand{\oiii}{[O\,\textsc{iii}]}
\newcommand{\oi}{[O\,\textsc{i}]}
\newcommand{\nii}{[N\,\textsc{ii}]}
\newcommand{\sii}{[S\,\textsc{ii}]}
\newcommand{\ha}{H$\alpha$}
\newcommand{\hb}{H$\beta$}
\newcommand{\hii}{H\,\textsc{ii}}

\newcommand{\targ}{GLIMPSE-16043}

\usepackage{hyperref}

\shorttitle{Exploring the low-mass BHs leveraging strong lensing}
\shortauthors{Fei et al.}
\graphicspath{{./}{figures/}}

\usepackage{amsmath}

\begin{document}

\title{
A GLIMPSE of Intermediate Mass Black holes in the epoch of reionization: \\
Witnessing the Descendants of Direct Collapse?
}
\author[0000-0001-7232-5355]{Qinyue Fei}
\affiliation{David A. Dunlap Department of Astronomy and Astrophysics, University of Toronto, 50 St. George Street, Toronto, Ontario, M5S 3H4, Canada}


\author[0000-0001-7201-5066]{Seiji Fujimoto}
\affiliation{David A. Dunlap Department of Astronomy and Astrophysics, University of Toronto, 50 St. George Street, Toronto, Ontario, M5S 3H4, Canada}
\affiliation{Dunlap Institute for Astronomy and Astrophysics, 50 St. George Street, Toronto, Ontario, M5S 3H4, Canada}

\author[0000-0003-3997-5705]{Rohan P.~Naidu}\altaffiliation{Hubble Fellow}
\affiliation{
MIT Kavli Institute for Astrophysics and Space Research, 70 Vassar Street, Cambridge, MA 02139, USA}

\author[0000-0002-0302-2577]{John Chisholm}
\affiliation{Department of Astronomy, The University of Texas at Austin, Austin, TX 78712, USA}
\affiliation{Cosmic Frontier Center, The University of Texas at Austin, Austin, TX 78712, USA}

\author[0000-0002-7570-0824]{Hakim Atek}
\affiliation{Institut d'Astrophysique de Paris, UMR 7095, CNRS, and Sorbonne Universit\'e, 98 bis boulevard Arago, 75014 Paris, France}

\author[0000-0003-2680-005X]{Gabriel Brammer}
\affiliation{Cosmic Dawn Center (DAWN), Niels Bohr Institute, University of Copenhagen, Jagtvej 128, K{\o}benhavn N, DK-2200, Denmark}


\author[0000-0003-3983-5438]{Yoshihisa Asada}
\affiliation{David A. Dunlap Department of Astronomy and Astrophysics, University of Toronto, 50 St. George Street, Toronto, Ontario, M5S 3H4, Canada}
\affiliation{Dunlap Institute for Astronomy and Astrophysics, 50 St. George Street, Toronto, Ontario, M5S 3H4, Canada}
\affiliation{Waseda Research Institute for Science and Engineering, Faculty of Science and Engineering, Waseda University, 3-4-1 Okubo, Shinjuku, Tokyo 169-8555, Japan}

\author[0000-0002-4153-053X]{Danielle A. Berg}
\affiliation{Department of Astronomy, The University of Texas at Austin, 2515 Speedway, Stop C1400, Austin, TX 78712, USA}
\affiliation{Cosmic Frontier Center, The University of Texas at Austin, Austin, TX 78712, USA} 

\author[0000-0003-0212-2979]{Volker Bromm}
\affiliation{Department of Astronomy, University of Texas, Austin, TX 78712, USA}
\affiliation{Weinberg Institute for Theoretical Physics, University of Texas, Austin, TX 78712, USA}
\affiliation{Cosmic Frontier Center, The University of Texas at Austin, Austin, TX 78712, USA}

\author[0000-0001-6278-032X]{Lukas J. Furtak}
\affiliation{Department of Physics, Ben-Gurion University of the Negev, P.O. Box 653, Be'er-Sheva 84105, Israel}

\author[0000-0002-5612-3427]{Jenny E. Greene}
\affiliation{Department of Astrophysical Sciences, Princeton University, 4 Ivy Lane, Princeton, NJ 08544, USA}

\author[0000-0003-4512-8705]{Tiger Yu-Yang Hsiao}
\affiliation{Department of Astronomy, The University of Texas at Austin, Austin, TX 78712, USA}
\affiliation{Cosmic Frontier Center, The University of Texas at Austin, Austin, TX 78712, USA}

\author[0000-0002-6038-5016]{Junehyoung Jeon}
\affiliation{Department of Astronomy, University of Texas, Austin, TX 78712, USA}
\affiliation{Cosmic Frontier Center, The University of Texas at Austin, Austin, TX 78712, USA}

\author[0000-0002-5588-9156]{Vasily Kokorev}
\affiliation{Department of Astronomy, University of Texas, Austin, TX 78712, USA}
\affiliation{Cosmic Frontier Center, The University of Texas at Austin, Austin, TX 78712, USA}

\author[0000-0003-2871-127X]{Jorryt Matthee} 
\affiliation{Institute of Science and Technology Austria (ISTA), Am Campus 1, 3400 Klosterneuburg, Austria}

\author[0000-0002-7809-0881]{Priyamvada Natarajan}
\affiliation{Department of Astronomy, Yale University, New Haven, CT 06511, USA}
\affiliation{Department of Physics, Yale University, New Haven, CT 06520, USA}
\affiliation{Black Hole Initiative, Harvard University, 20 Garden Street, Cambridge, MA 02138, USA}

\author[0000-0002-9651-5716]{Richard Pan}
\affiliation{Department of Physics \& Astronomy, Tufts University, MA 02155, USA}

\author[0000-0001-5492-1049]{Johan Richard}
\affiliation{Univ Lyon, Univ Lyon1, Ens de Lyon, CNRS, Centre de Recherche Astrophysique de Lyon UMR5574, 69230, Saint-Genis-Laval, France}

\author[0000-0001-8419-3062]{Alberto Saldana-Lopez}
\affiliation{Department of Astronomy, Oskar Klein Centre, Stockholm University, 106 91 Stockholm, Sweden}

\author[0000-0001-7144-7182]{Daniel Schaerer}
\affiliation{Observatoire de Gen\`eve, Universit\'e de Gen\`eve, Chemin Pegasi 51, 1290 Versoix, Switzerland}
\affiliation{CNRS, IRAP, 14 Avenue E. Belin, 31400 Toulouse, France}

\author[0000-0002-3216-1322]{Marta Volonteri}
\affiliation{Institut d’Astrophysique de Paris, UMR 7095, CNRS and Sorbonne Universit\'e, 98 bis Boulevard Arago, 75014 Paris, France}

\author[0000-0002-0350-4488]{Adi Zitrin}
\affiliation{Department of Physics, Ben-Gurion University of the Negev, P.O. Box 653, Be'er-Sheva 84105, Israel}



\begin{abstract}
JWST has revealed an abundance of supermassive black holes (BHs) in the early Universe, and yet the lowest mass seed black holes that gave rise to these populations remain elusive. 
Here we present a systematic search for broad-line Active Galactic Nuclei (AGNs) in some of the faintest high-$z$ galaxies surveyed yet by combining ultra-deep 
JWST/NIRSpec G395M spectroscopy with the strong lensing aid in Abell S1063.
By employing the profile of the \oiii$\lambda 5007$ emission lines as a template for narrow-line components and carefully cross-validating with mock observations, we identify a sample of ten broad-line AGNs at $4.5<z<7.0$ (eight secure, two tentative). 
The inferred BH masses from the broad \ha\ line explore the intermediate BH mass regime down to $\sim 10^{5.5}\,M_\odot$. 
The stellar mass ($M_*$) is estimated with a galaxy+AGN composite model, and we  
find the BH to stellar mass ratio spans down to 
$M_{\rm BH}/M_*\lesssim 0.1\%$, unveiling populations on the empirical $M_{\rm BH}-M*$ relation observed in the local universe. 
We also derive the black hole mass function and investigate its low-mass end at this epoch. 
While we confirm the agreement of our results with previous studies at $M_{\rm BH}\gtrsim10^{6.5}M_{\odot}$, we find the mass range of $\sim 10^{5.5}\,M_\odot$ features an enhanced abundance with respect to the extrapolated best-fit Schechter function.  
Comparison with theoretical models suggests that a possible origin for this enhanced abundance is the direct-collapse BH formation, supporting the scenario that the direct collapse of massive gas clouds is a significant pathway for the earliest supermassive BHs.
\end{abstract}


\section{Introduction}
\label{sec1: intro}

The discovery of luminous quasars at $z>6$, powered by supermassive black holes (SMBHs) already exceeding a billion solar masses ($M_\odot$), presented a fundamental challenge to our understanding of early galaxy formation \citep[e.g.,][]{Fan+01, Fan+03, Mortlock+11, Banados+18, Wang+21}. 
Occurring less than a billion years after the Big Bang, the existence of these compact massive objects requires near-continuous, highly efficient accretion onto lower mass BHs throughout the early stages of the universe (in this paper we refer to the $10^4-10^6\,M_\odot$ mass range as `lower mass', in contrast to stellar mass BHs that are approximately $10\,M_\odot$). 
Compounding this timing problem, these high-redshift quasars appear to host `overmassive' black holes \citep{Yue+24, LiR+25}, relative to the stellar mass of their host galaxies 
when compared to the well-established local $M_{\rm BH}-M_{\rm bulge}$ scaling relations \citep{Kormendy+13}. Even compared to the local $M_{\rm BH}-M_*$ relation for AGN host galaxies \citep[e.g., ][]{Reines+15}, these high-$z$ quasars seem to be systematically offset. 
This offset raises a key question: how did these SMBHs form so rapidly and seemingly outpace the growth of their host galaxies in the early universe \citep{2014ReviewSEEDS,Smith+19, Inayoshi+20}? 

If the SMBHs powering these high-$z$ quasars presumably originate through accretion onto lower mass BHs ($10^4\lesssim M_{\rm BH}\lesssim 10^6\,M_\odot$; intermediate mass black holes; IMBH), this implies that a population of lower mass BHs should exist at high-$z$. Therefore, a detailed study of the properties of IMBHs, including their mass, abundance, and host galaxy characteristics, can provide crucial insights into the initial BH seeding mechanisms and early growth pathways of SMBHs and quasars. 

The launch of the JWST has opened a new window into this field, unveiling a previously hidden population of fainter Active Galactic Nuclei (AGNs) at high redshifts \citep[e.g.,][]{Harikane+23, Maiolino+24, Larson+23, Kokorev+23, Kocevski+23, Maiolino+24b, Matthee+24, Stone+24, Lin+24, Taylor+25, Hviding+25}. These newly discovered AGNs are powered by significantly less massive BHs than luminous quasars, offering a glimpse into the more abundant AGN population. Yet these sources appear to be hosted in galactic nuclei where they are  `overmassive' compared to the local scaling relation. Many of these lower-luminosity AGNs reside in even lower-mass host galaxies ($\sim10^8\,M_\odot$), revealing black hole-to-stellar mass ratios ($M_{\rm BH}/M_*$) that deviate even more dramatically from the local relationship, with some objects exhibiting $M_{\rm BH}/M_*$ values 10 to 100 times higher than local relationship \citep[e.g., ][]{Maiolino+24}. This has deepened the question, suggesting that the `overmassive' nature of early BHs might be a widespread phenomenon. 

However, this apparent trend towards overmassive BHs at high redshift may be significantly influenced by selection bias \citep[e.g., ][]{LiJ+25b, Silverman+2025}. 
Even in the era of JWST, the AGN samples are often biased towards sources with higher AGN-galaxy luminosity ratio, depending on the NIRSpec observational strategy \citep[e.g., ][]{Greene+24}. 
This can create a selection effect that preferentially identifies systems with high $M_{\rm BH}/M_*$ ratios \citep{Lauer+07, Salviander+07}. The search for `undermassive' BHs residing in more massive galaxies, which can lie closer to or even below the local scaling relation, has been conducted to mitigate this offset issue \citep[e.g., ][]{Izumi+19, Ding+23, LiJ+25a}. 

While selection effects can alleviate the observed offset from local relations, they cannot fully explain the fundamental challenge of forming massive BHs in such low mass galaxies within such a short cosmic timescale. 
Two primary theoretical pathways have been proposed to explain their rapid formation. The first scenario is the `light seed', in which the BHs originate from the death of Population~III (Pop III) stars \citep[e.g.,][]{Madau2001, Smith+18, Haemmerle+20, Spinoso+23, Singh+23}. To grow into supermassive objects by $z>6$, these light seeds must subsequently undergo prolonged and efficient periods of accretion, likely at rates exceeding the Eddington limit. The second channel posits the formation of `heavy seeds' with much larger initial masses of $\sim 10^4-10^6\,M_\odot$ \citep[e.g.,][]{Lodato+06,Becerra_DCBH2018}. These objects are thought to form via the direct collapse of massive, primordial gas clouds\footnote{There is a second `flavor' of heavy seeds, invoking runaway collisions in dense stellar clusters \citep[e.g.,][]{Reinoso2023}. However, it is not clear whether such a merger-driven channel could operate at the required extremely high degree of efficiency.}, creating what are known as Direct Collapse Black Hole (DCBH) seeds \citep{Bromm+03, Begelman2006, Lodato+06, Lodato+07, Haemmerle+18, Haemmerle+20}. This pathway more easily explains the existence of the most luminous high-$z$ quasars, as it alleviates the need for continuous, highly efficient super-Eddington accretion \citep{Fan+23}, in effect providing a `head-start' in the SMBH assembly process. Furthermore, a DCBH origin requires environments of (at least initially) high densities and gas inflow rates, providing a natural explanation for the Compton-thick conditions inferred from the X-ray weakness of high-$z$ AGNs \citep[e.g.,][]{Pacucci_DCBH2015,Smith_DCBH2017, Cenci_DCBH2025, Yue+2024b, Ananna+2024}.

The situation mentioned above underscores the significance of exploring IMBHs at high-$z$. However, even with the sensitivity of JWST, this BH population has largely eluded detection in field surveys due to their intrinsic faintness \citep[for theoretical constraints, see][]{Jeon2023}. Based on the local calibrated virial estimators \citep[e.g., ][]{Greene+05, Reines+15}, previous studies, including those using stacking methods, have established a lower mass limit of $\sim 10^6\,M_\odot$ \citep{Maiolino+24, Geris+25}, with one notable candidate identified with a mass of $10^{5.65}\,M_\odot$ in a dual AGN system \citep{Maiolino+24}. 
The validity of these BH mass estimators at high-$z$ is being debated, as their physics may be different \citep[e.g., ][]{Chang+25, Liu+25}. 
Their real masses can be significantly changed by a factor of ten using a different mass estimator \citep[e.g., ][]{Rusakov+25, Greene+25}. 

A promising path to search for these IMBHs is the combination of deep NIRSpec spectra and the magnification provided by gravitational lensing. This strategy is expected to push the detection threshold down to BH masses of $\sim 10^5$ or even $10^4\,M_\odot$. Such observations not only enable the detection of some of the lowest-mass BHs in the early universe but also allow for an exploration of the `undermassive' region of the $M_{\rm BH}-M_*$ parameter space. 

In this paper, we present a systematic search for broad-line AGNs using deep JWST/NIRSpec G395M spectroscopic data. Our analysis is based on observations from the JWST Cycle 2 Large Program, GLIMPSE (PID~3293; PIs: H. Atek \& J. Chisholm) and Cycle 3 DDT program (PID~9223; PIs: S. Fujimoto \& R. Naidu), dubbed the GLIMPSE-\textit{D} survey. These observations focus on one of the best-studied massive galaxy cluster Abell S1063 (hereafter AS1063). By leveraging the strong gravitational lensing provided by the foreground cluster, these programs achieve unprecedented sensitivity to faint, high-redshift objects 
(down to an observed magnitude of 31, with delensed observed magnitudes of 32(34) with magnifications of 2(15), see more in the survey paper, Atek in prep). 
This unique combination of deep NIRSpec exposures and significant lensing magnification provides an ideal opportunity to detect and characterize AGNs powered by lower-mass BHs than those accessible in typical blank-field surveys.

The structure of this paper is as follows. In Section 2, we describe the data reduction process. In Section 3, we describe our spectral fitting methodology for identifying AGNs and performing the measurement of their properties. We present our main results, including the $M_{\rm BH}-M_*$ relation and BHMF, in Section 4. Finally, we summarize our conclusions in Section 5. Throughout this paper, we assume a flat universe with $\Omega_{\rm m}=0.3,\Omega_{\Lambda}=0.7,\sigma_8=0.8$, and $H_0=70\,\rm km\,s^{-1}\,Mpc^{-1}$. We use magnitudes in the AB system \citep{Oke+83}.

\section{Data}
\label{sec2: data}

\begin{figure*}
    \centering
    \includegraphics[width=\linewidth]{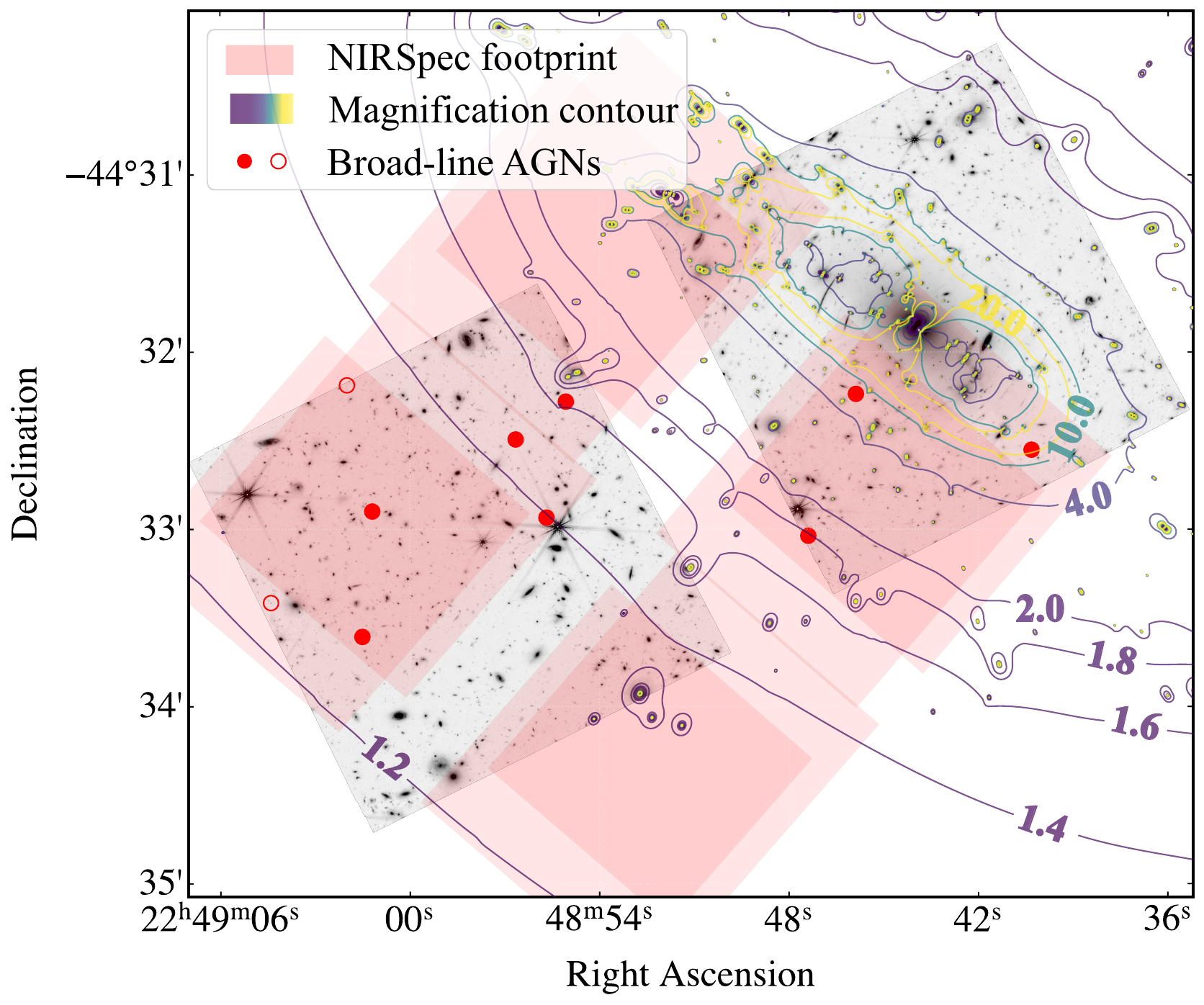}
    \caption{
    \textbf{Location of broad-line AGNs in the survey field.}
    The NIRSpec footprint is indicated by the red-shaded regions, which are overlaid on the NIRCam F444W image. The lensing magnification map for a source at $z_{\mathrm{s}}=6$ is denoted as the viridis contours. Red circles mark the positions of our identified broad-line AGNs, filled circles for robust objects, and open circles for tentative candidates.
    }
    \label{fig:footprint}
\end{figure*}

Detailed descriptions of the observations, data reduction, cluster light removal, and source extraction will be presented in the GLIMPSE survey overview paper (H. Atek et al. 2025, in preparation) and are briefly summarized below.

\subsection{GLIMPSE DDT}
\label{sec2.1: DDT}

The deep spectroscopic follow-up of the Pop~III galaxy candidate \targ\ \citep{Fujimoto+25} was obtained with \textit{JWST}/NIRSpec as part of the DDT program GLIMPSE-\textit{D} (PID~9223; PIs: S.~Fujimoto \& R.P.~Naidu), using the MOS mode with the G395M grating and F290LP filter ($R\sim1000$). 
Three separate MSA configurations were employed to deliver the full integration on the primary target while maximizing the number of filler sources observed with the deep G395M setup: \texttt{glimpse\_02} (42,016~s), \texttt{glimpse\_01} (33,613~s), and \texttt{glimpse\_01b} (33,613~s). 
The maximum on-source integration time (i.e., the sources included in all configurations) is 109,242~s ($\simeq30.4$~hr), while most filler targets were observed in only a single configuration. 
In cases where a source had multiple observations, each dataset was analyzed separately. 
For the final statistical analysis, we selected the observation with the highest data quality for each unique source to prevent its over-representation in the sample. 
The data were reduced with the \texttt{msaexp} pipeline \citep{Brammer2023} following standard procedures \citep[e.g.,][]{Heintz+25, deGraaff+25, Naidu+25, Valentino+25}. 
More details of the observational setup, data reduction, and calibration will be presented in S.~Fujimoto et al.\ (in prep.) and H.~Atek et al.\ (in prep.). 


\subsection{Filler Target Selection}
\label{sec2.2:selection}

In designing the NIRSpec MSA configurations, we followed a strategy similar to that adopted in the UNCOVER survey \citep{Bezanson+24, Price+25}, aiming both to secure the full integration on \targ\ and to maximize the multiplexing return of the deep G395M observations. 
To this end, we implemented a two–tiered prioritization scheme distinguishing between (1) extraordinary or otherwise scientifically unique individual sources, and (2) samples of galaxies of broader interest (e.g., statistical sets of $z>4$ galaxies). 
Within this framework, the ranking order was set such that high–priority individual sources were placed first, followed by interesting but less extraordinary individual galaxies, and finally sample-based sources with equal priority within a given catalog. 
This scheme was used in combination with our custom MSA optimization algorithm to maximize slit allocation efficiency. 

As a result, a total of 421 sources were allocated slits, corresponding to 262 unique objects across the three configurations. 
The largest sample consists of galaxies with $z_{\rm phot}=3$--9, targeted to detect strong rest-frame optical emission lines such as H$\alpha$ and [O\,\textsc{iii}]. 
Photometric redshifts were estimated using multiple codes of \textsc{EAZY} \citep{Brammer2008}, \texttt{BEAGLE} \citep{Chevallard+16}, \texttt{BAGPIPES} \citep{Carnall+18}, and \texttt{Prospector} \citep{Johnson+21}, and the final selections were determined by carefully cross-checking the outputs from each method. 

\section{Method}
\label{sec3: method}

We search for broad-line AGNs within a sample of 139 spectroscopically confirmed galaxies at redshifts $z \sim 4.5$–7, each of which passed ndependent quality control cross-checks of the 2D and 1D spectra by multiple team members. In this redshift range, the \oiii, \hb, and \ha\ emission lines all fall within the JWST/NIRSpec G395M coverage, making it an ideal window for identifying broad \ha\ emission. More than 75\% of the spectroscopic sample was originally selected from the blind $z_{\rm phot} = 3$–9 catalogs, providing a robust and largely unbiased basis for a blind search for broad-line features.
The remaining sources in this redshift range were preferentially targeted since they belonged to high priority classes -- e.g., high magnification MUSE LAEs, quiescent galaxy candidates (R.~Pan et al. in prep.), exceptionally blue Beta slope sources (Jecman et al. in prep.).
Our methodology involves fitting the continuum as well as rest-frame optical emission lines \hb, \oiii $\lambda\lambda4959,5007$, \ha, and \nii $\lambda\lambda6548,6584$. The continuum in each spectrum is modeled with a single power-law after masking prominent emission features. This fit is subsequently subtracted to isolate the emission line profiles.


\begin{figure*}
    \centering
    \includegraphics[width=0.95\linewidth]{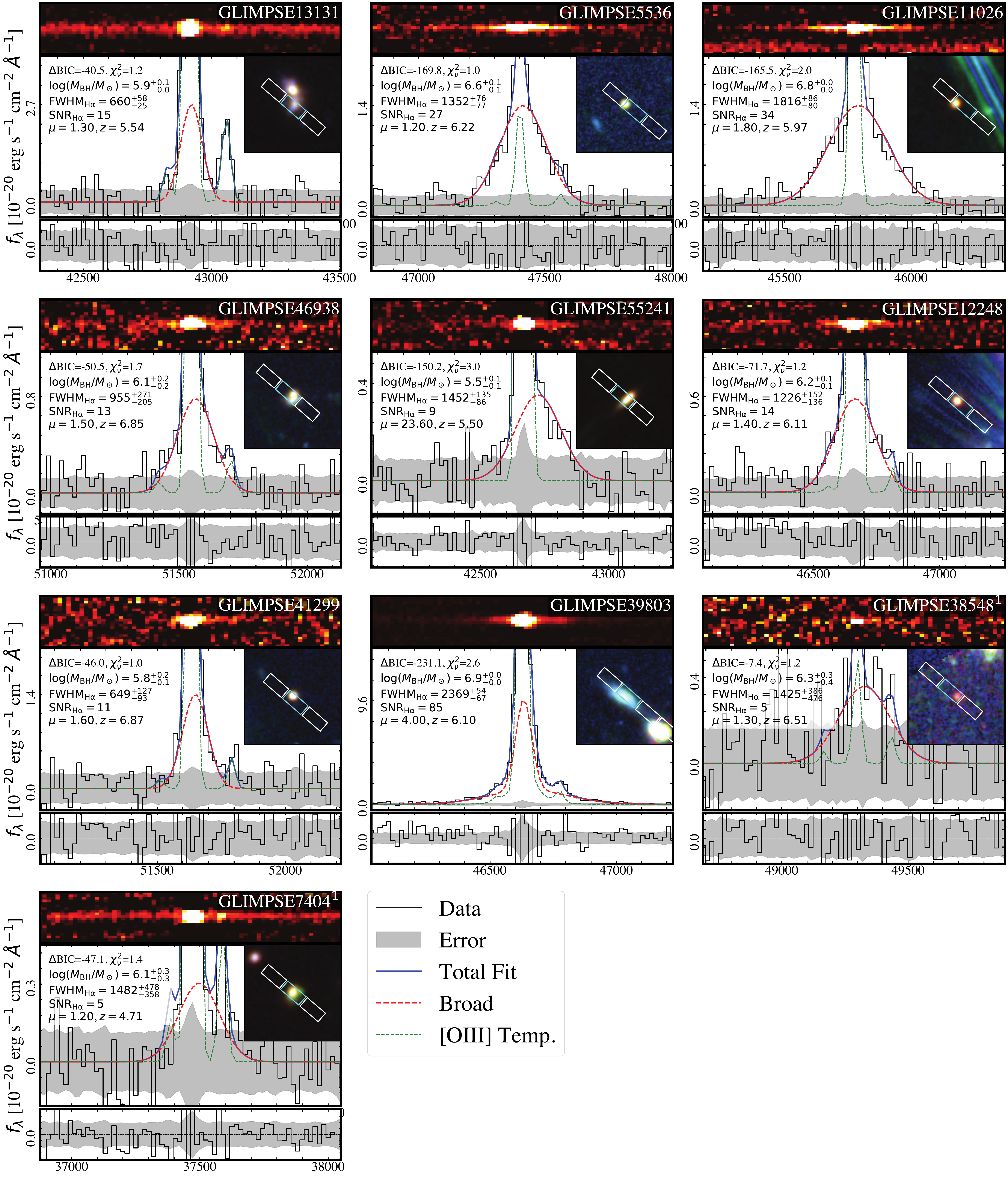}
    \caption{
    \textbf{The spectra of our broad line AGN and the best-fit result.} 
    For each source, the top panel displays the two-dimensional spectrum. The middle panel presents the corresponding one-dimensional spectrum, obtained through optimal extraction (black histogram). The uncertainty is represented by the gray shaded region. Overplotted on the 1D spectrum is the best-fit model (blue solid line), which is decomposed into narrow components from \oiii\ line template (green dotted line) and the broad \ha\ emission (red dashed line). The bottom panel presents the residual. The inset image presented in the middle panel shows the RGB image generated with JWST/NIRCam F150W, F277W, and F444W filters, and the corresponding NIRSpec/G395M shutters. The final two sub-panels are dedicated to presenting the spectra and corresponding model fits for two tentative detections, ID~38548 and ID~7404. 
    Basic fitting parameters and properties of sources are listed in the middle panel. $\Delta \rm BIC$ shows the difference between the BIC for the best-fit model and the narrow-only model, and $\chi_\nu^2$ is the reduced chi-squared value corresponding to the best-fit model. $\log (M_{\rm BH}/M_\odot)$ is the black hole mass, estimated from the properties of the broad \ha\ emission. $\rm FWHM_{\rm H\alpha}$ and $\rm SNR_{H\alpha}$ present the FWHM and the signal-to-noise ratio of the broad \ha\ emission. $\mu$ is the gravitational lensing magnification factor for the source, and $z$ is the redshift. 
    }
    \label{fig0: BLHA}
\end{figure*}

\subsection{Emission line fitting}
\label{sec3.1: line}

We perform the emission line fitting, in particular, for searching the broad line component following the method widely used in the local universe \citep[e.g.,][]{Greene+05} and high redshift \citep[e.g.,][]{Harikane+23}. The basic algorithm is to construct a model from the narrow forbidden lines, using this model as a template to generate narrow lines for \ha\ and \nii, and characterise the broad \ha\ line through separating the narrow component from the observed spectrum. 
This process allows for the identification of broad-line AGN through breaking the degeneracy between a broad-line component and emission from an outflow by comparing permitted and forbidden line profiles. Broad emission originating from the broad line region is seen only in permitted lines, as the critical densities for forbidden transitions are exceeded, suppressing their emission. Conversely, signatures of an outflow will be present in all emission lines arising from the lower-density outflowing medium, including both permitted and forbidden lines. 
In the local universe, \sii\ doublets usually serve as a narrow line template, while in our cases, \sii\ doublets have a relatively low signal-to-noise ratio (SNR) and are not appropriate for the template generation. Therefore, we use the \oiii\ doublet.

The primary line profile of \oiii\ doublets is modeled using a forward model \citep{Taylor+25}. The intrinsic line profile of \oiii\ doublets is generated with two Gaussian profiles, whose systemic velocity center and line width are tied to each other and the flux ratio is set to 2.98 ($f_{\rm \lambda 5007}/f_{\rm \lambda 4959}=2.98$) based on the theoretical calculation \citep{Storey+00}. The model is then convolved with the instrumental Line Spread Function (LSF) generated from \texttt{msafit} \citep{deGraaff+2024}, which accounts for the instrument's non-uniform resolving power that varies with wavelength, source morphology, and slit position. 

We then evaluate the potential for the galactic outflow by comparing a single-component Gaussian model (`core') against a two-component model (`core+outflow'). The latter incorporates the narrow core component plus a broader component to represent the outflow. The broader component of the \oiii\ emission line was initially modeled using one Gaussian function. While in some cases (in particular ID~39803), multiple Gaussians were required to characterize the broad wings of the profile. We subsequently determined that a single exponential function provides a superior fit to the broad \oiii\ component. Therefore, to ensure a consistent and uniform fitting across the entire sample, we adopted the exponential profile for modeling this component in all sources. 
The velocity and width of broad components are also kinematically tied across the \oiii\ doublet. 
Using the Bayesian Information Criterion ($\mathrm{BIC}\equiv \chi^2+k\ln n$, where $k$ is the number of free parameters and $n$ is the number of data points), we determine whether an outflow exists in the \oiii\ spectrum. The two-component model is adopted if it provided a better fit ($\Delta \rm BIC\equiv BIC_{core+outflow} - BIC_{core}<-10$; \citealt{Kass+95}).

To identify broad-line AGNs, we model the \ha+\nii\ emission-line complex using a two-step fitting procedure. In the first step, a narrow-line model is constructed by adopting the best-fit profile of the \oiii\ doublet as a kinematic template, i.e., the velocity and line width of the narrow \ha\ and \nii\ lines are fixed to the values derived from the \oiii\ best-fit result while their amplitudes are allowed to vary. The flux ratio of the \nii$\lambda 6584$ and \nii$\lambda 6548$ lines is fixed to its theoretical value of 3.02. Second, to test for the presence of an AGN, we fit the spectrum with a more complex model incorporating the narrow-line template plus a broad \ha\ component. We apply the following constraint to this component. First, its FWHM is required to be at least two times larger than that of the narrow component ($\rm FWHM_{broad}>2\times FWHM_{\rm narrow}$), while its central velocity is permitted to be offset from the systemic velocity defined by the \oiii\ template, which is common for high-redshift AGNs \citep{Harikane+23, Matthee+24}. Similar to the identification of \oiii\ outflow, the presence of broad-line AGN is confirmed with $\Delta \rm BIC \equiv BIC_{broad}-BIC_{narrow}<-10$. 

For a subset of sources, we found that this rigid template-fitting approach yielded a poor fit to the \ha+\nii\ complex. This difficulty is likely attributable to the physical conditions of the narrow-line region (NLR). The spatial distributions of \oiii, \ha, and \nii\ can be stratified by density, causing their intrinsic line profiles to differ. In these instances, the fitting strategy is modified. Instead of a fixed template, the \oiii\ profile is used as a prior, and the line widths of \ha\ and \nii\ lines are allowed to vary. This approach allowed for modest flexibility in the kinematic parameters of the narrow \ha\ component.

\subsection{Broad-line AGN identification}
\label{sec3.2: selection}

The spectral fitting procedure is conducted using the Python packages \texttt{lmfit} \citep{Newville+14}. We employ a two-stage process to ensure robust exploration of the parameter space and accurate uncertainty estimation.
First, a global optimization is performed using the differential evolution algorithm implemented within \texttt{lmfit} to identify the best-fit parameter set. This result is then used to initialize a comprehensive parameter estimation via Markov Chain Monte-Carlo (MCMC) using the \texttt{emcee} \citep{Foreman-Mackey+13}. The MCMC analysis involved two rounds of sampling to ensure robust convergence. We initiate a preliminary run with 64 walkers for 1000 steps, from which we identified the maximum a posteriori (MAP) values after a 500-step burn-in phase. These MAP values served as the starting point for a second MCMC run using the same configuration. For the final analysis, the first 500 steps of this second run are discarded as a burn-in, and the subsequent chains from all walkers are combined to construct the full posterior probability distribution (PDF) from 32,000 samples ($64\times500$). The median (50th percentile) of each marginalized posterior distribution is adopted as the best-fit value, with the $1\sigma$ uncertainties defined by the 16th and 84th percentiles. 

By applying the above-mentioned fitting algorithm to our parent sample of 139 sources in the redshift range $4.5<z<7.0$, we identified 10 broad-line AGN candidates. 
These sources are further categorized into eight secure detections and two tentative detections. 
The latter are characterized by a broad \ha\ component with a modest SNR ($\rm 3<SNR_{\rm H\alpha,\,broad}<5$). 
The final sample of these 10 AGNs and their properties are presented in Table~\ref{tab1:source}. Their spectra and the best-fit result are shown in Figure~\ref{fig0: BLHA}. The full spectrum and the best-fit result are shown in Appendix~\ref{app2: fitting}

\begin{deluxetable*}{cccccccc}
\tablenum{1}
\tablecaption{Properties of Our Broad Line AGNs \label{tab1:source}}
    \tablewidth{0pt}
    \tablehead{
    \colhead{Source ID} & \colhead{Redshift} & \colhead{R.A.} & \colhead{Dec.} & \colhead{$\mu$} & \colhead{$M_{\rm UV}$} & \colhead{$\rm SNR_{H\alpha,broad}$} & \colhead{$\Delta \rm BIC$} \\
    \colhead{} & \colhead{} & \colhead{(deg.)} & \colhead{(deg.)} & \colhead{} & \colhead{} & \colhead{} & \colhead{}
    }
    \decimalcolnumbers
    \startdata
    13131  & 5.538 & 342.254974 & -44.548328 & 1.3 & -19.04 & 17.22 & -40.53 \\
     5536  & 6.222 & 342.256287 & -44.560120 & 1.2 & -17.35 & 27.41 & -169.78 \\
    11026  & 5.972 & 342.197449 & -44.550621 & 1.8 & -18.38 & 33.97 & -165.45 \\
    46938  & 6.852 & 342.236053 & -44.541561 & 1.5 & -18.46 & 13.33 & -50.50 \\
    55241  & 5.500 & 342.167999 & -44.542530 & 23.6 & -16.52 & 9.27 & -150.17 \\
    12248  & 6.108 & 342.231995 & -44.548908 & 1.4 & -18.08 & 14.38 & -71.71 \\
    41299  & 6.865 & 342.229431 & -44.538002 & 1.6 & -18.70 & 11.06 & -46.05 \\
    39803  & 6.103 & 342.191162 & -44.537281 & 4.0 & -17.97 & 84.93 & -231.12 \\
    38548$^a$ & 6.508 & 342.258301 & -44.536457 & 1.3 & -16.05 & 5.49 & -7.37 \\
     7404$^a$ & 4.707 & 342.268341 & -44.556927 & 1.2 & -19.26 & 4.88 & -47.13 \\
     \enddata
     \tablecomments{(1) Source ID (also MSA ID); (2) Redshift from \oiii; (3) Right ascension; (4) Declination; (5) Magnification; (6) Absolute UV magnitude after lensing correction; (7) Signal-to-noise ratio (SNR) of the broad \ha\ emission; (8) Difference between BIC of the best-fit model and the pure-narrow line model.\\
    $^a$ Tentative sources in our sample due to lower SNR of broad \ha ($\rm SNR<5$) or small $\Delta \rm BIC$. }
\end{deluxetable*}

\begin{deluxetable*}{cccccccccc}
\tablenum{2}
\tablecaption{Derivation of the broad line AGNs \label{tab2: AGNs}}
    \tablewidth{0pt}
    \tablehead{
    \colhead{Source ID} & \colhead{$\rm FWHM_{H\alpha,broad}$} & \colhead{$f_{\rm H\alpha,broad}$} & \colhead{$\log L_{\rm H\alpha,broad}$} & \colhead{$\log L_{\lambda 5100}$} & \colhead{$\log L_{\rm bol}$} & \colhead{$\log M_{\rm BH}$} & \colhead{$\lambda_{\rm Edd}$} & \colhead{$\log M_{\rm *, SED}$} & \colhead{$\log M_{\rm *,UV}$} \\
    \colhead{} & \colhead{($\rm km\,s^{-1}$)} & \colhead{($10^{-20}\rm erg\,s^{-1}\,cm^{-2}$)} & \colhead{($\rm erg\,s^{-1}$)} & \colhead{($\rm erg\,s^{-1}$)} & \colhead{($\rm erg\,s^{-1}$)} & \colhead{($M_\odot$)} & \colhead{} & \colhead{($M_\odot$)} & \colhead{($M_\odot$)}
    }
    \decimalcolnumbers
    \startdata
    13131 & $659.6_{-25.0}^{+58.3}$ & $221.3_{-26.0}^{+26.9}$ & $41.88_{-0.05}^{+0.05}$ & $43.28_{-0.05}^{+0.05}$ & $44.29_{-0.05}^{+0.05}$ & $5.86_{-0.04}^{+0.07}$ & $1.80_{-0.21}^{+0.22}$ & $8.77_{-0.90}^{+0.27}$ & 8.16\\
     5536$^a$& $1352.3_{-77.2}^{+75.9}$ & $264.3_{-12.8}^{+11.6}$ & $42.08_{-0.02}^{+0.02}$ & $43.45_{-0.02}^{+0.02}$ & $44.46_{-0.02}^{+0.02}$ & $6.60_{-0.06}^{+0.05}$ & $0.48_{-0.02}^{+0.02}$ & $9.74_{-0.03}^{+0.03}$ & 7.20\\
    11026$^a$& $1815.5_{-79.8}^{+86.0}$ & $228.6_{-7.5}^{+7.4}$ & $41.97_{-0.01}^{+0.01}$ & $43.36_{-0.01}^{+0.01}$ & $44.37_{-0.01}^{+0.01}$ & $6.81_{-0.04}^{+0.04}$ & $0.24_{-0.01}^{+0.01}$ & $7.81_{-0.23}^{+0.15}$ & 7.79\\
    46938 & $954.7_{-205.4}^{+271.0}$ & $93.5_{-9.2}^{+11.0}$ & $41.72_{-0.05}^{+0.05}$ & $43.14_{-0.05}^{+0.05}$ & $44.15_{-0.05}^{+0.05}$ & $6.09_{-0.23}^{+0.25}$ & $0.77_{-0.08}^{+0.09}$ & $7.45_{-1.30}^{+0.29}$ & 7.84\\
    55241 & $1452.0_{-86.0}^{+135.0}$ & $3.3_{-0.3}^{+0.3}$ & $40.05_{-0.05}^{+0.04}$ & $41.69_{-0.05}^{+0.04}$ & $42.71_{-0.05}^{+0.04}$ & $5.55_{-0.06}^{+0.09}$ & $0.10_{-0.01}^{+0.01}$ & $6.60_{-0.10}^{+0.08}$ & 6.73\\
    12248$^a$ & $1226.1_{-135.7}^{+152.3}$ & $85.7_{-7.5}^{+7.8}$ & $41.57_{-0.04}^{+0.04}$ & $43.01_{-0.04}^{+0.04}$ & $44.02_{-0.04}^{+0.04}$ & $6.24_{-0.11}^{+0.11}$ & $0.40_{-0.04}^{+0.04}$ & $7.93_{-0.33}^{+0.19}$ & 7.62\\
    41299 & $648.9_{-92.8}^{+127.4}$ & $109.3_{-13.1}^{+13.4}$ & $41.79_{-0.06}^{+0.05}$ & $43.20_{-0.06}^{+0.05}$ & $44.21_{-0.06}^{+0.05}$ & $5.79_{-0.15}^{+0.17}$ & $1.77_{-0.21}^{+0.22}$ & $7.61_{-0.26}^{+0.16}$ &7.97 \\
    39803 & $2369.4_{-66.8}^{+53.9}$ & $128.4_{-5.6}^{+5.3}$ & $41.75_{-0.02}^{+0.02}$ & $43.16_{-0.02}^{+0.02}$ & $44.17_{-0.02}^{+0.02}$ & $6.92_{-0.02}^{+0.02}$ & $0.12_{-0.01}^{+0.00}$ & $7.71_{-0.66}^{+0.25}$ & 7.56\\
    38548$^a$ & $1425.0_{-476.2}^{+385.8}$ & $62.8_{-15.9}^{+15.9}$ & $41.50_{-0.13}^{+0.10}$ & $42.95_{-0.13}^{+0.10}$ & $43.96_{-0.13}^{+0.10}$ & $6.32_{-0.41}^{+0.26}$ & $0.29_{-0.07}^{+0.07}$ & $7.66_{-0.57}^{+0.57}$ & 8.29\\
     7404 & $1482.1_{-357.6}^{+478.3}$ & $51.7_{-11.6}^{+12.0}$ & $41.09_{-0.11}^{+0.09}$ & $42.59_{-0.11}^{+0.09}$ & $43.60_{-0.11}^{+0.09}$ & $6.13_{-0.25}^{+0.27}$ & $0.20_{-0.04}^{+0.05}$ & $8.24_{-0.34}^{+0.19}$ & 6.46\\
     \enddata
     \tablecomments{(1) Source ID; (2) FWHM of the broad \ha\ emission line; (3) Flux of the broad \ha\ emission line with lensing effect correction; (4) The \ha\ luminosity calculated from the \ha\ flux; (5) The 5100\AA\ monochromatic luminosity estimated from \ha\ luminosity; (6) The bolometric luminosity; (7) The BH mass (without considering the systematic uncertainties introduced by the BH mass estimator); (8) The Eddington ratio; (9) The stellar mass derived from SED fitting; (10) The stellar mass derived from the UV-magnitude listed in Table~\ref{tab1:source} with the $M_*-M_{\rm UV}$ relationship \citep{Stefanon+21}, uncertainties are not shown in this column.\\ 
     $^a$ Point-like sources.\\
     }
\end{deluxetable*}

\begin{deluxetable*}{ccccccc}
    \tablenum{3}
    \tablecaption{Narrow line properties of our targets \label{tab3: NLR}}
    \tablewidth{0pt}
    \tablehead{
    \colhead{Source ID} & \colhead{$f_{\rm H\alpha}$} & \colhead{$f_{\rm H\beta}$} & \colhead{$f_{\rm [OIII]}$} & \colhead{$f_{\rm [NII]}$} & \colhead{$f_{\rm [SII]}$} & \colhead{$f_{\rm [OI]}$} \\
    \colhead{} & \colhead{($10^{-20}\,\rm erg\,s^{-1}\,cm^{-2}$)} & \colhead{($10^{-20}\,\rm erg\,s^{-1}\,cm^{-2}$)} & \colhead{($10^{-20}\,\rm erg\,s^{-1}\,cm^{-2}$)} & \colhead{($10^{-20}\,\rm erg\,s^{-1}\,cm^{-2}$)} & \colhead{($10^{-20}\,\rm erg\,s^{-1}\,cm^{-2}$)} & \colhead{($10^{-20}\,\rm erg\,s^{-1}\,cm^{-2}$)}
    }
    \decimalcolnumbers
    \startdata
    13131 & $ 1054.86_{-27.05}^{+27.63}$ & $ 332.20_{-9.97}^{+9.24}$ & $ 2158.22_{-13.41}^{+9.56}$ & $ 96.92_{-10.50}^{+9.92}$ & $ 215.22_{-23.19}^{+10.12}$ & $ 56.77_{-4.53}^{+4.94}$\\
    5536 & $ 57.99_{-3.95}^{+4.92}$ & $ 18.84_{-4.83}^{+5.02}$ & $ 100.30_{-12.54}^{+7.46}$ & $<12.51$ & $<4.82$ & $<11.46$\\
    11026 & $ 258.44_{-5.45}^{+5.62}$ & $ 86.72_{-5.45}^{+5.62}$ & $ 295.27_{-8.85}^{+15.69}$ & $<4.68$ & $<5.75$ & $<2.43$\\
    46938 & $ 291.18_{-9.43}^{+8.72}$ & $ 119.23_{-4.14}^{+4.39}$ & $ 624.41_{-29.63}^{+33.76}$ & $<17.61$ & $<11.94$ & $<6.55$\\
    55241 & $ 1750.26_{-164.03}^{+121.80}$ & $ 374.55_{-5.31}^{+5.17}$ & $ 1663.40_{-8.48}^{+68.58}$ & $<10.92$ & $<8.13$ & $<5.37$\\
    12248 & $ 380.41_{-46.86}^{+60.59}$ & $ 105.23_{-2.86}^{+2.75}$ & $ 485.06_{-3.48}^{+0.13}$ & $<7.65$ & $<10.87$ & $<57.45$\\
    41299 & $ 415.90_{-16.57}^{+17.16}$ & $ 139.57_{-16.57}^{+17.16}$ & $ 759.39_{-50.44}^{+75.32}$ & $<21.04$ & $<63.04$ & $<8.37$\\
    39803 & $ 2815.90_{-65.14}^{+57.39}$ & $ 972.22_{-18.49}^{+17.06}$ & $ 6255.46_{-133.10}^{+98.89}$ & $ 47.86_{-3.79}^{+3.67}$ & $ 32.59_{-4.10}^{+2.39}$ & $ 18.75_{-1.75}^{+2.40}$\\
    38548 & $ 18.81_{-4.01}^{+5.32}$ & $ 9.33_{-5.10}^{+4.19}$ & $<13.99$ & $<14.99$ & $<17.21$ & $<9.60$\\
    7404 & $ 524.22_{-10.57}^{+10.33}$ & $ 120.29_{-21.09}^{+25.38}$ & $ 1098.54_{-41.99}^{+50.12}$ & $ 19.47_{-4.44}^{+4.84}$ & $ 61.02_{-4.33}^{+6.87}$ & $ 14.64_{-3.00}^{+3.55}$\\
    \enddata
    \tablecomments{(1) Source ID; (2) Flux of the narrow component of \ha\ emission line; (3) Flux of the \hb\ emission line; (4) Flux of the \oiii$\lambda5007$ emission line; (5) Flux of the \nii$\lambda6584$ emission line; (6) Flux of the \sii\ doublet lines; (7) Flux of the \oi\ emission line. Both measurements and uncertainties are given by \texttt{lmfit} fitting result. }
\end{deluxetable*}

\subsection{BH mass estimate}
\label{sec3.3: BH mass}

For each source with broad \ha\ line detection, we estimate its black hole mass ($M_{\rm BH}$) using a single-epoch virial mass estimator based on the properties of the broad \ha\ emission line. We employ the calibration from \cite{Greene+05}, which is established for AGNs in the local universe ($z\sim 0$):

\begin{equation}
\begin{split}
    M_{\rm BH} &= 2.0\times 10^6\,M_\odot \\ 
    & \times \left(\frac{L_{\rm H\alpha,broad}}{10^{42}\rm\,erg\,s^{-1}}\right)^{0.55} \left(\frac{\rm FWHM_{H\alpha,broad}}{10^3\rm\, km\,s^{-1}}\right)^{2.06},
\end{split}
\end{equation}
where $L_{\rm H\alpha,broad}$ is the \ha\ luminosity which should contain both broad line and narrow line in principle. However, regarding the origin of the narrow \ha\ line in our sources is ambiguous, here we adopt the luminosity of the broad \ha\ emission following previous studies \citep[e.g., ][]{Harikane+23, Kocevski+23, Lin+24}. The lensing effect is also considered when calculating the \ha\ luminosities. 
The \ha-based BH mass estimator has an intrinsic scatter of $\sim$0.3\,dex \citep{Shen+24}, and $M_{\rm BH}$ uncertainties in this work are computed as the quadratic sum of the intrinsic scatter and the statistical uncertainty propagated from the line width and flux measurements in spectral fitting. 
The lensing uncertainties are not included in the \ha\ luminosity and BH mass estimates. 


Since de-blending the continuum from the AGN and the host galaxy is challenging, especially faint AGN systems potentially with low $M_{\rm BH}/M_{\star}$, the derivation of AGN bolometric luminosity ($L_{\rm bol}$) from wide-band photometry is not straightforward for our sources. Therefore we estimate $L_{\rm bol}$ and the monochromatic luminosity at $5100\,\AA$ ($L_{\rm \lambda 5100}$) with the magnification-corrected $L_{\rm H\alpha}$ from our fitting result using the empirical relationship in the local universe \citep{Greene+05, Richards+06}:
\begin{equation}\label{eq2}
\begin{split}
    L_{\lambda 5100}&=10^{44}\left(\frac{L_{\rm H\alpha}}{5.25\times 10^{42}\,\rm erg\,s^{-1}}\right)^{\frac{1}{1.157}}\,\rm erg\,s^{-1}, \\
    L_{\rm bol} &= 10.33\times L_{\lambda 5100}.
\end{split}
\end{equation}
where $L_{\rm H\alpha}$ is the luminosity of broad \ha\ emission line since whether the narrow \ha\ line originates from the narrow line region or \hii\ region is unknown. 
Following the commonly used method in recent JWST studies \citep[e.g., ][]{Harikane+23, Maiolino+24, Lin+24}, we estimate the Eddington ratio ($\lambda_{\rm Edd}$) using $L_{\rm bol}/L_{\rm Edd}$, where $L_{\rm Edd}$ is the theoretical maximum luminosity achievable when radiation pressure and gravity are balanced in a spherical geometry, which can be calculated following \cite{Trakhtenbrot+11}:
\begin{equation}
    L_{\rm Edd} = 1.5\times 10^{38}\, M_{\rm BH}/M_\odot.
\end{equation}

The derived physical properties of our AGNs are listed in Table~\ref{tab2: AGNs}. The $\lambda_{\rm Edd}$ of our sources are around the range of $0.10\sim 0.50$, comparable to those of $z\gtrsim 5$ quasars \citep[e.g., ][]{Shen+19, Farina+22} and JWST discovered faint AGNs \citep[e.g., ][]{Harikane+23, Maiolino+24, Lin+24}, except for two sources with $M_{\rm BH}\lesssim 10^{6}\,M_\odot$ (ID~13131 and ID~41299), whose Eddington ratios are 2.82 and 1.77, respectively. However, we need to be cautious that the BH masses and bolometric luminosities of our targets are derived from the best-fit \ha\ parameters using empirical relationships, therefore, the Eddington ratio derived from this approach reflects the relation between the \ha\ luminosity and \ha\ line width and might have systematic errors. 

We acknowledge that the BH mass estimates for recently discovered JWST AGN, in particular, little red dots (LRDs), are subject to significant uncertainties. These uncertainties arise because the accretion mode in these high-$z$ sources may differ from the AGN in the local universe that anchor the virial mass calibrations \citep[e.g.,][]{Naidu+25, Rusakov+25, Chang+25}.
Indeed, some studies suggest that applying local relationships to these populations could lead to BH mass overestimations by as much as an order of magnitude \citep[e.g.,][]{Greene+25}.
To ensure a fair comparison with other recent JWST AGN studies \citep[e.g., ][]{Harikane+23}, which is a primary scope of this work, we adopt BH masses derived from local relationships in this work. 
The final parameters, including BH mass, are listed in Table~\ref{tab2: AGNs}.

\subsection{Galaxy mass estimate}
\label{sec3.4: galaxy mass}

\begin{figure}
    \centering
    \includegraphics[width=\linewidth]{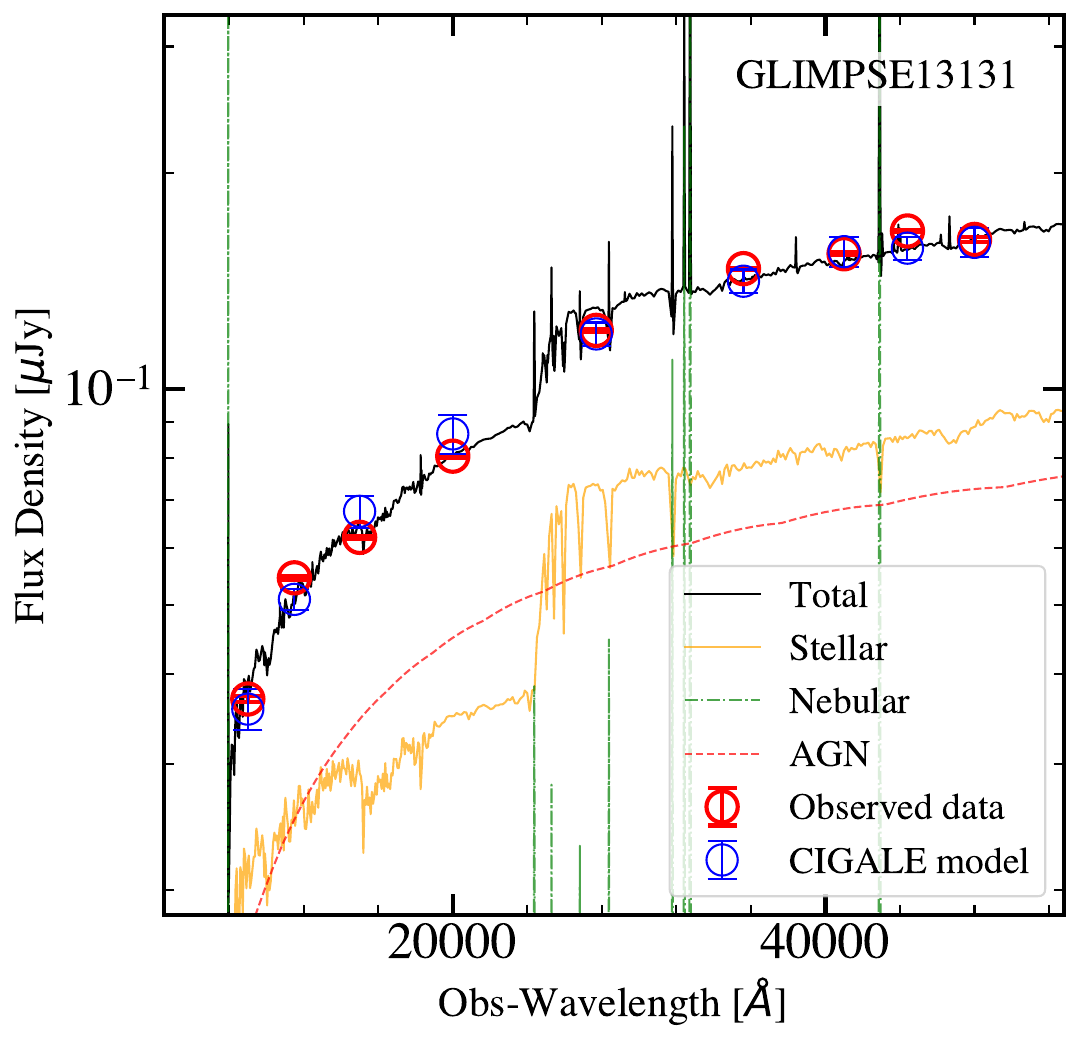}
    \caption{
    \textbf{Example of the best-fit galaxy and AGN composite SED of our sample, ID13131}.  
    The red data points present the flux measured from JWST/NIRCam images, and the blue data points present the model value. The black solid line presents the best-fit spectrum from \texttt{CIGALE}. The yellow solid line, green dash-dotted line, and red dashed line present the stellar emission, the nebular emission, and the AGN contribution, respectively.
    }
    \label{fig2: sed}
\end{figure}

To derive the physical properties of the host galaxies, particularly their stellar masses ($M_*$), we perform spectral energy distribution (SED) fitting on the available photometric data using the Code Investigating GALaxy Emission (\texttt{CIGALE})
\citep[e.g., ][]{Boquien+19, Yang+22}. The photometry is measured within a 0\farcs3 diameter aperture, which is determined to be sufficiently encompass the total emission from each host galaxy. Our sources' redshifts are determined from the best-fit \oiii\ result.

For the stellar population, we adopt the single stellar population models of \cite{Bruzual+03} (\texttt{bc03} module), assuming a \cite{Chabrier+03} initial mass function (IMF). The SFH is modeled using a delayed-$\tau$ function. We also include a burst component in the model whose SFH can be modeled as an exponential profile. The mass fraction of the late starburst population is allowed to vary between 0 and 0.5 in step of 0.1. Following the established procedures of \cite{Zhuang+23} and \cite{Tanaka+25}, we set an upper limit on the age of the stellar population, requiring it to be less than 95\% of the age of the universe at the galaxy's redshift ($t_{\rm age}\leq 0.95\,t_{\rm H}$). We include contributions from ionized gas using the \texttt{nebular} module. To account for the potentially extreme interstellar medium (ISM) conditions in high-redshift galaxies, we allow the ionization parameter ($\log U$) to vary between -3 and -1 and the gas-phase metallicity to range from 10\% to 100\% of the solar value. The dust attenuation is also included in the fitting, following a modified \cite{Calzetti+03} attenuation law (\texttt{dustatt\_modified\_starburst} module). Given the confirmed presence of broad lines in our sources, we include an AGN component using the default \texttt{skirtor} module in \texttt{CIGALE} \citep{Stalevski+16, Boquien+19}. We fix the viewing angle to a face-on orientation ($i=0^\circ$) in the fitting and allow the AGN fraction to vary uniformly from 0.1 to 0.9. We notice that the AGN template utilized in our analysis may possess a redder spectral slope than some observed quasar populations \citep{Young+25}, this uncertainty does not significantly impact the stellar mass estimation. The stellar mass is primarily determined by the rest-frame optical portion of the SED, where the host galaxy's stellar emission is the dominant component.

An example of our SED fitting results is shown in Figure~\ref{fig2: sed}. The full SED fitting result is shown in Figure~\ref{figapp3: seds}. The final derived stellar masses for our sample are listed in Table~\ref{tab2: AGNs}. 

\section{Results \& Discussion}
\label{sec4: results}

\subsection{Broad line AGNs}
\label{sec4.1: AGN sample}

\begin{figure}
    \centering
    \includegraphics[width=\linewidth]{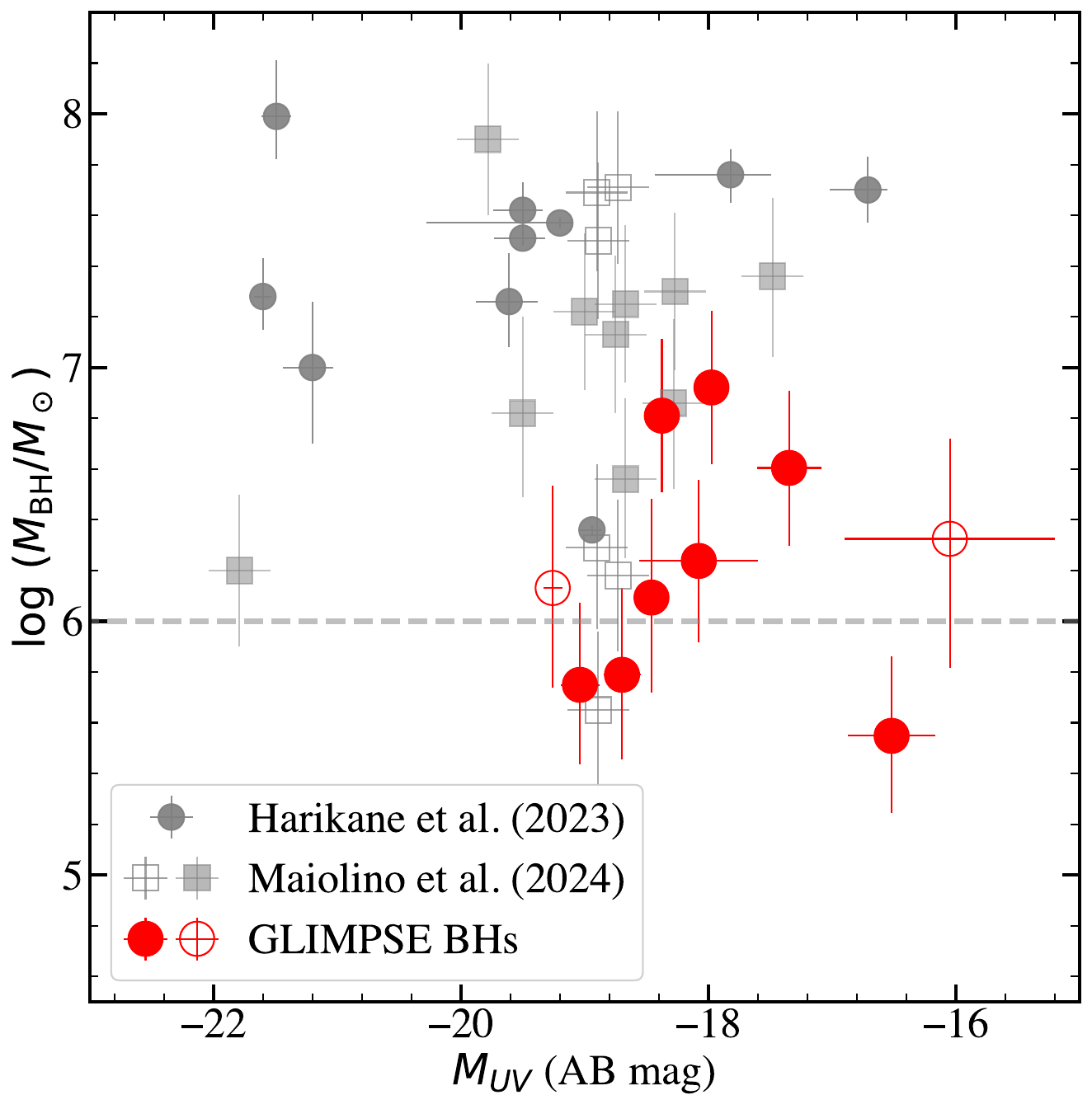}
    \caption{BH mass and UV magnitude of our selected broad-line AGNs (red circles), where filled circles present a confident sample and open circles present the sample of tentative sources. The gray circles and orange squares denote other AGNs identified in other studies \citep{Harikane+23, Maiolino+24}. The open squares present the dual AGNs classified in \cite{Maiolino+24}.}
    \label{fig2: pspace}
\end{figure}

Figure~\ref{fig2: pspace} shows the UV magnitudes and BH mass of our selected AGNs. 
The absolute magnitudes of these sources are $-20\lesssim M_{\rm UV}\lesssim -16$, which is consistent with with faintest end of those AGNs discovered by JWST \citep[e.g., ][]{Harikane+23, Kocevski+23, Kokorev+23, Furtak+24, Greene+24, Maiolino+24, Matthee+24}, while they are at least two orders of magnitude fainter than traditional quasars found through ground-based telescopes \citep[$-28\lesssim M_{\rm UV} \lesssim -22$; e.g., ][]{Wang+16, Wang+19, Wang+21, Yang+19, Yang+20, Matsuoka+16, Matsuoka+18a, Matsuoka+18b, Onoue+17, Onoue+19}. 

Among the ten sources analyzed, we identified three hosting BHs with masses below $10^6\,M_\odot$, in contrast to the previous successful detection down to $10^{6.2}\,M_{\odot}$ in  single AGN systems\footnote{One is reported at a similar redshift in a dual AGN system \citep{Maiolino+24}}. Our work, leveraged by the deep G395M observations and the strong lensing, therefore demonstrates the power of exploring such a low-mass BH regime in the early universe. The typical magnifications of our sources are 1.2$\sim$1.8. With a magnification level of 1.8, to achieve the SNR reported here for an unlensed source would require an equivalent integration time of 100 hours. 
Considering the FWHMs of broad \ha\ lines in these sources are around $1000\,\rm km\,s^{-1}$, there are several possible additional scenario to explain their origin. We further discuss these possibilities in Section~\ref{sec4.4: Caveats}.

Based on the JWST/NIRCam images, we find that four sources present point-like morphologies even using the shortest wavelength filters (ID~5536, ID~11026, ID~12248, and ID~38548), while the remaining sources present clear extended emission. By applying the `V-shape' color selection criteria for LRDs as defined in \cite{Greene+24}, we identify three sources (ID~5536, ID~12248, and ID~38548) in our sample consistent with the LRD classification \citep{Labbe+23, Matthee+24}. An additional source, ID~11026, also meets the criteria if the $\rm F277W-F356W>0.7$ threshold is relaxed.
While a detailed study of LRDs is beyond the scope of this work, their unique nature suggests that $M_*$ derived from standard SED fitting may have significant uncertainties, particularly under the `black-hole star' scenario \citep{Naidu+25}, which suggests that the optical continuum is dominated by BH accretion instead of stellar light. To evaluate the stellar mass estimates, we therefore compared them against the established $M_{\rm UV}-M_*$ relation at this redshift \citep{Duncan+14, Song+16, Stefanon+21}.
We find that the vast majority of our sample (8 out of 10 sources) are consistent with this relation (Figure~\ref{figapp3:mm-comparison}). However,  the LRD ID~5536 (and a tentative source, ID~7404), deviates significantly from the 1:1 relation. For these objects, the $M_*$ inferred from their UV luminosity is approximately two orders of magnitude (2~dex) lower than the value derived from our SED fitting. 
For these sources, we use this UV-based $M_*$ in the following comparison. But we acknowledge that the $M_*-M_{\rm UV}$ relation has a significant systematic uncertainty, which can be approximately 1\,dex \citep[e.g., ][]{Stefanon+21}.

\subsection{BPT diagram}
\label{sec4.2: BPT}

\begin{figure*}
    \centering
    \includegraphics[width=\linewidth]{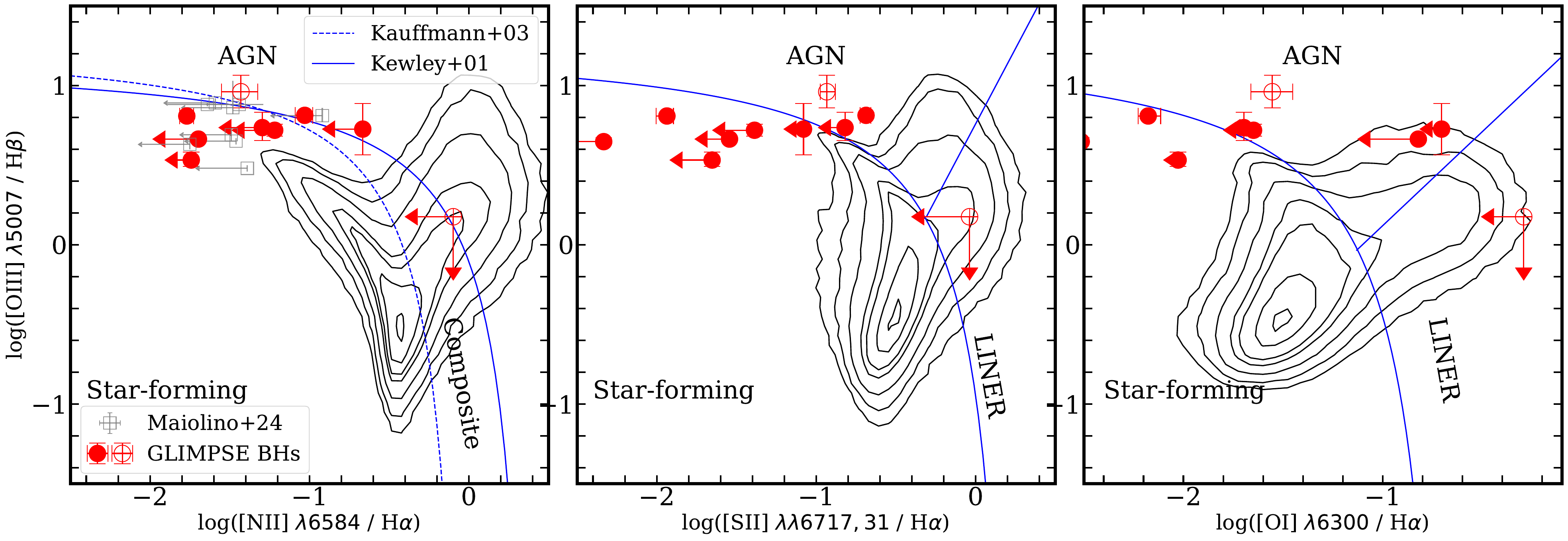}
    \caption{
    \textbf{BPT diagrams of our targets.}
    The distribution of local galaxies from the Sloan Digital Sky Survey (SDSS) DR7 is shown as background contours, illustrating the star-forming sequence and AGN branch. Demarcation lines from \cite{Kewley+01} (solid blue) and \cite{Kauffmann+03} (dashed blue) separate star-forming galaxies from AGNs in the left panel. In the right panels, these lines distinguish between star-forming galaxies, AGNs, and LINERs. Our sources are plotted as red circles, plotted with other JWST-discovered faint AGNs from \cite{Maiolino+24} (gray squares).
    }
    \label{fig: bpt}
\end{figure*}

The emission line fitting is also performed for the forbidden \sii\ and \oi\ emission lines for emission line diagnostics. Following the same methodology used for the narrow Balmer lines, the best-fit profile of \oiii\ is used as a template to model these emission lines. In instances where this template resulted in significant residuals, the line parameters are fit independently to achieve a more accurate description of these lines.

The resulting emission line flux ratios are used to construct the Baldwin, Phillips \& Terlevich (BPT) diagnostic diagram \citep{Baldwin+81}, which is presented in Figure~\ref{fig: bpt}. For a subset of our sample, robust measurements of all lines are not possible. Specifically, the narrow forbidden lines (e.g., \nii, \sii, and \oi) are often not detected due to their intrinsic faintness (e.g., ID~12248, ID~41299), while for some objects, the \hb\ emission lines are not covered. In such cases, upper limits are placed on the relevant diagnostic ratios. For non-coverage of \hb, its flux is estimated from the measured \ha\ narrow line flux, assuming a theoretical Case B recombination ratio of $f_{\rm H\alpha}/f_{\rm H\beta}=2.86$.

Based on the classification threshold from the local universe \citep{Kewley+01, Kauffmann+03, Kewley+06}, we find our sources are located on the boundary between AGN and star-forming galaxies defined from the local universe in the \sii\ and \nii\ based diagram, which is consistent with previous faint AGNs discovered by JWST \citep{Kocevski+23, Ubler+23, Harikane+23, Maiolino+24, Chisholm+24}. This result suggests the local classification criterion is not appropriate in discriminating AGNs from star-forming galaxies at high-$z$. While through the \oi-based diagram, three of our sources () are located within the AGN-regime. 

For those two tentative sources identified through our broad line AGN selection threshold (ID~7404 and ID~38948), one source (ID~7404) is also identified as an AGN with these diagnostics, which independently supports its identification as an AGN. Despite marginal detection of the broad component, this independently supports their identification as the broad-line AGNs.
While for ID~38948, a relatively low \oiii/\hb\ ratio caused by the marginal detection ($2\sigma$) of \oiii\ lines suggests its consistent with the SF locus. 

\subsection{$M_{\rm BH}-M_*$ relation}
\label{sec4.2: MM relation}

\begin{figure*}
    \centering
    \includegraphics[width=\linewidth]{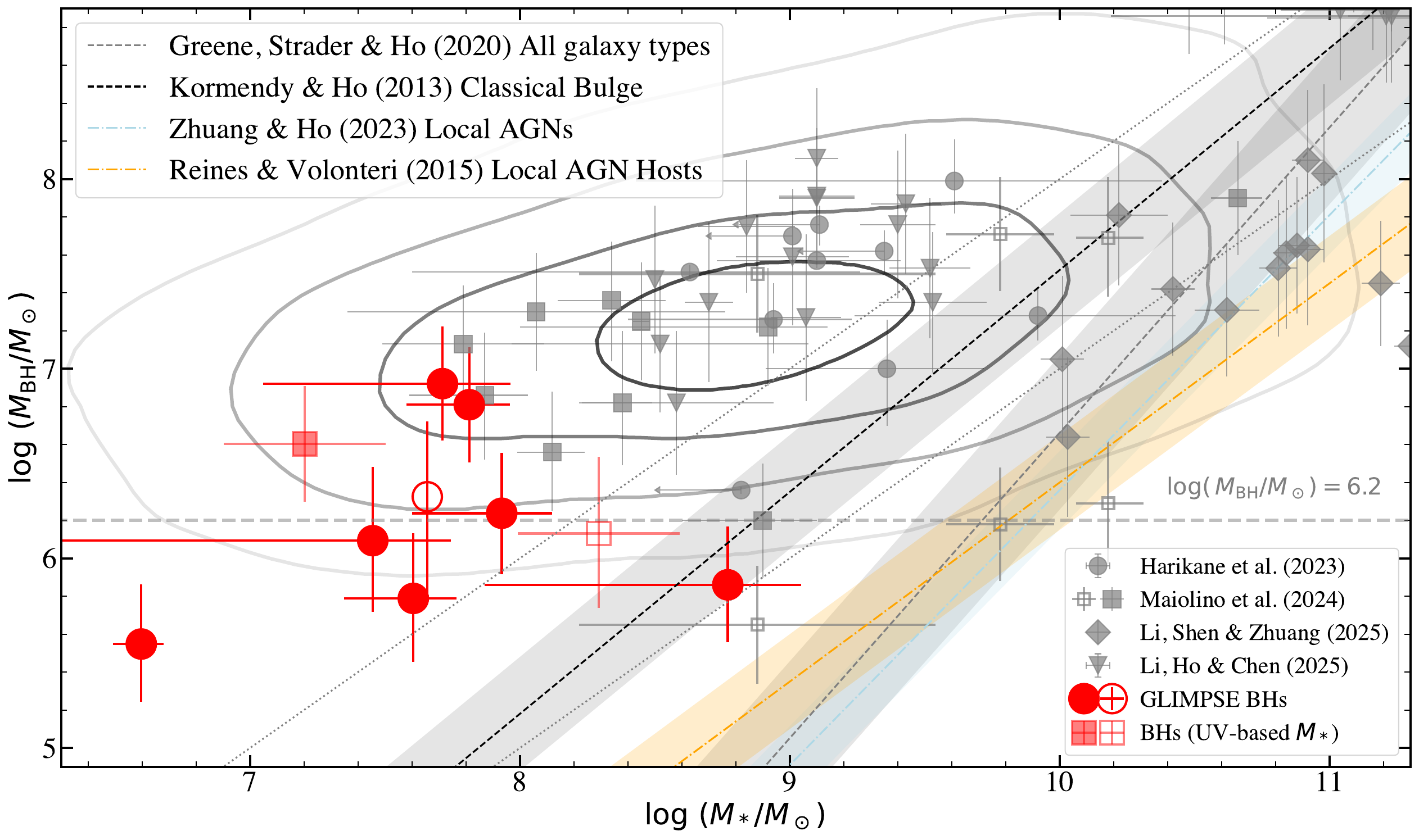}
    \caption{Relation between BH mass and galaxies' stellar mass for our sources. Our sources are shown as red circles and squares, where circles denote sources whose $M_*$ were derived from SED fitting, and squares denote those $M_*$ estimated based on UV magnitude. Filled circles indicate objects with robust broad H$\alpha$ line detections, whereas open circles denote sources with more marginal detections. The gray transparent circles, squares, diamonds and triangles depict the stellar mass and black hole from the literature for high-$z$ AGNs \citep{Harikane+23, Maiolino+24, LiJ+25a, LiR+25}. The gray and black dashed lines present the $M_{\rm BH}-M_*$ relationship in the local universe for inactive galaxies \citep{Kormendy+13, Greene+20}, the orange and lightblue dash-dotted lines present the relationship for local AGN host galaxies \citep{Reines+15, Zhuang+23}, respectively. The black dotted lines denote constant $M_{\rm BH}/M_*$ of 0.1\% and 1\%. The gray contour represents the selection bias inherent to the $M_{\rm BH}/M_*$ relation, adapted from the result from \cite{LiJ+25b}. }
    \label{fig3: relation}
\end{figure*}

The scaling relations observed in the local universe between BH masses and bulge properties, such as the $M_{\rm BH}-\sigma$ and $M_{\rm BH}-M_{\rm bulge}$ relations \citep{Magorrian+98, Ferrarese+00, Gebhardt+00, Kormendy+13}, strongly suggest a coupled evolutionary history \citep{Madau+14}. A similar correlation, though characterized by greater scatter, is observed between $M_{\rm BH}$ and the total host galaxy stellar mass, $M_*$ \citep[e.g., ][]{Reines+15}. This co-evolution is presumed to be particularly critical during the early stages of galaxy and BH growth. In contrast to these well-established local benchmarks, AGNs observed at high redshifts ($4.5<z<7$) systematically lie above the local relations and exhibit significant scatter on the $M_{\rm BH}-M_*$ plane \citep[e.g., ][]{Harikane+23, Maiolino+24}. 

A prevailing explanation for this discrepancy has been observational selection bias, where flux-limited surveys are inherently biased toward detecting only the most luminous and massive BHs whose emission dominates that of their host galaxy \citep[e.g., ][]{Shen+15, LiJ+21, LiJ+25b, Zhang+23, Tanaka+25}.  However, recent deep JWST surveys focusing on low-luminosity AGNs at high redshift have argued that this selection effect is insufficient to fully account for the large observed offset, implying a genuine physical difference in the galaxy-BH connection at high redshift \citep[e.g., ][]{Pacucci+23, Maiolino+24, Stone+24}.

Figure~\ref{fig3: relation} presents the relationship between $M_{\rm BH}$ and stellar mass ($M_*$) for our sample. We also present comparison samples of high-redshift luminous quasars and other JWST-identified faint AGNs from recent literature \citep{Harikane+23, Maiolino+24, LiJ+25a, LiR+25}, as well as empirical relationships from the local universe \citep{Reines+15, Zhuang+23, Kormendy+13, Greene+20}.

One source has a modest $M_{\rm BH}/M_*\sim 0.1\%$. 
This value is considerably lower and aligns more closely with the relation observed in the local universe \citep[$\sim 0.5\%$; ][]{Kormendy+13}. This result corroborates recent discoveries of low-luminosity AGNs in massive galaxies that also follow the local trend \citep{LiJ+25a}.
Four sources (ID~11026, ID~46938, ID~12248, and ID~39803) have a significantly higher BH-to-stellar mass ratio of $M_{\rm BH}/M_* \geq 2\%$,  which are consistent with faint AGNs identified in previous studies \citep[e.g., ][]{Harikane+23, Maiolino+24}. 

Furthermore, we assess our sample in the context of the $M_{\rm BH}-M_*$ relation that is purely induced by observational bias, as modeled by \cite{LiJ+25b}. With the exception of two objects, our sources are consistently located along the selection boundary predicted by their model. This result strongly indicates that our sample probes a population of AGNs that falls below the detection limits of typical surveys, thereby underscoring the critical role of selection effects in shaping observed scaling relations. 

Taken together, these results support the hypothesis that selection bias plays a crucial role in the frequent observation of claimed overmassive BHs in earlier studies. By sampling a fainter population of AGNs, our work suggests that some galaxies at high-$z$ may host BHs that are not unusually massive. The BH-to-stellar mass ratio for these BHs appears to be consistent with local scaling relations.

\subsection{BH mass function}
\label{sec4.3: BHMF}

\begin{deluxetable}{ccc}
    \tablenum{3}
    \tablecaption{Black Hole Mass Function}
    \tablewidth{0pt}
    \tablehead{
    \colhead{$\log M_{\rm BH}$} & \colhead{$\Phi$} & \colhead{$N_{\rm eff}$} \\
    \colhead{$(M_\odot)$} & \colhead{($10^{-3}\rm Mpc^{-3}\,dex^{-1}$)} & \colhead{} 
    }
    \decimalcolnumbers
    \startdata
    5.50 & $12.70 \pm 6.73$ ($4.45 \pm 2.36$) & 12.86 (4.50) \\
    6.25 & $0.64 \pm 0.29$ ($0.22 \pm 0.10$) & 7.74 (2.71) \\
    6.75 & $0.50 \pm 0.25$ ($0.17 \pm 0.09$) & 6.01 (2.10) \\
    7.25 & $0.18 \pm 0.11$ ($0.06 \pm 0.04$) & 2.13 (0.74) \\
    \enddata
    \tablecomments{(1) The logarithm of the central value of each BH mass bin; (2) The number density of AGNs per BH mass bin. The values in parentheses presents the measurement without considering the observational incompleteness; (3) The effective number of broad-line AGNs within the corresponding BH mass bin (with the original accounting number without considering the observational incompleteness in the parentheses).}
\end{deluxetable}
\begin{figure*}
    \centering
    \includegraphics[width=\linewidth]{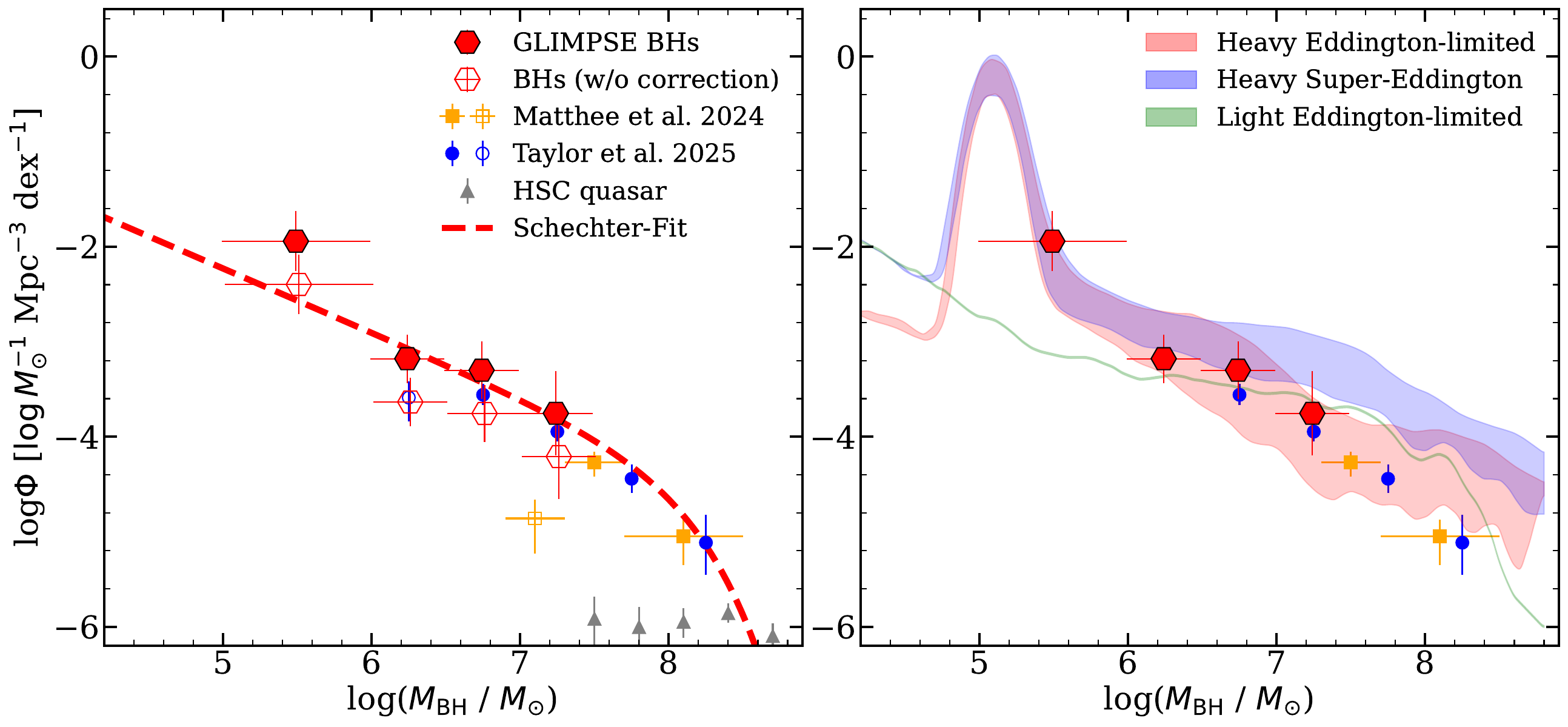}
    \caption{
    \textbf{Black hole mass function derived from this study. }
    \textit{Left:} The measured BHMF and the empirical relation. 
    Our sources are shown as red hexagons, where filled hexagons denote the analysis with applying the completeness correction and open hexagons not. Previous studies are shown as blue circles and yellow squares \citep{Matthee+24, Taylor+25}, where open symbols denote the data points with high incompleteness in their analysis. 
    The gray triangles present the BHs number density calculated from quasars at $z\sim 4$ from HSC observations \citep{He+24}.
    Our fitting using the Schechter function is shown as the red dashed line. 
    \textit{Right:} Comparison between our measurement and a recent theoretical prediction including heavy seeds. 
    The blue and red shaded regions denote the BHMF from SAM predictions for the heavy seeds model, while the green shaded region represents the BHMF from the light seeds model \citep{Jeon+25}.}
    \label{fig4: bhmf}
\end{figure*}


Though the broad \ha\ can have different origins, in particular for those sources with narrower FWHM (e.g., ID~13131), here we assume that these broad lines originate from the AGN, and we use the BH masses derived from broad \ha\ properties to estimate the BH abundance at these redshifts. We begin by comparing the high-mass end of our derived BHMF with existing results from the literature, a regime that corresponds to the lowest mass range in those studies. Furthermore, the detection of three BHs with masses of $M_{\rm BH}\lesssim 10^6\,M_\odot$ enables us to place initial constraints on the abundance of BHs in this low-mass regime, providing critical insights for understanding their formation pathways.

To accurately calculate the BHMF, we must first quantify the selection completeness of our AGN sample. This completeness is primarily affected by two factors: observational completeness and line-detection completeness. 
%
Line-detection incompleteness arises when the broad line is either too faint to be detected above the noise or is rejected by our specific selection criteria (see Section~\ref{sec3.2: selection}). To correct for this incompleteness, we adopt a methodology similar to previous studies \citep[e.g., ][]{Taylor+25}, with slight modifications for our dataset. The core of this method involves generating a large suite of mock spectra that thoroughly sample the relevant AGN parameter space. We then subject these mock spectra to our full fitting and selection threshold to precisely determine the recovery fraction as a function of the AGN properties. A comprehensive description of our completeness simulation and its results is provided in Appendix~\ref{app1: incompleteness}.

The observational incompleteness arises from sources that lie within the JWST/NIRSpec field of view but are not allocated a slitlet for spectroscopy due to instrumental limitations in the MSA target configuration. Consequently, a fraction of the total galaxy population within our redshift range is not observed. 
To account for these untargeted galaxies, we make the standard assumption that the intrinsic AGN fraction is identical for both the observed and unobserved populations. 
This correction is calculated as follows:
\begin{eqnarray}
    n_{\rm AGN,cor}=\frac{n_{\rm AGN}}{n_{\rm gal}}\times n_{\rm gal, tot},
\end{eqnarray}
where $n_{\rm AGN}$ represents the number of spectroscopically confirmed AGNs, while $n_{\rm gal}$ denotes the total number of galaxies within the redshift range of $4.5<z<7$ that have been observed with JWST/NIRSpec. The ratio ($f_{\rm AGN} = n_{\rm AGN}/n_{\rm gal}$) provides the observed AGN fraction within our spectroscopic sample. 
$n_{\rm gal, tot}$ is the total number of all known galaxies in the same redshift range that lie within the NIRSpec footprint, including both targeted and untargeted sources. Following this method and the same photometric redshift catalog used in the MSA target selection, we identified 397 sources whose photometric redshift and LW observed magnitudes are aligned with our AGN sample. 

We define bins of $\log_{10}M_{\rm BH}$ from 5 to 7.5 with first binwidth of 1\,dex and remaining 0.5\,dex. 
To construct the BHMF, we adopt a weighting scheme for each broad-line AGN. The best-fit BH mass from Table~\ref{tab2: AGNs} is taken as the central value for its probability distribution. We then used the inverse square of its uncertainty as the weight. This method allows each AGN to contribute fractionally to multiple mass bins rather than being assigned exclusively to a single bin. The contribution of a given AGN to any specific mass bin is determined by integrating its probability density function, which is defined by its measured $M_{\rm BH}$ and uncertainty, over the logarithmic range of that bin. The total `effective' number of AGNs in each bin, $N_{\rm eff}$, is the sum of these fractional contributions from all AGNs in the sample. Finally, the comoving number density, $\Phi$, for each mass bin is calculated by dividing the effective number count by the bin width ($\Delta \log M_{\rm BH}$) and the total comoving volume ($V_c$) of the survey, as shown in the equation:
\begin{eqnarray}
    \Phi = \frac{N_{\rm eff}}{V_c\cdot \Delta \log M_{\rm BH}}.
\end{eqnarray}
We further implement the Poisson uncertainty \citep{Gehrels+86} and the uncertainty of cosmic variance \citep{Trenti+08}. 
To perform a fair comparison with previous studies of high-$z$ AGNs, we do not apply a duty cycle correction to the BHMF estimation. It is important to remember that since the black hole accretion is episodic, the BHMF derived from the AGN can be systematically different from the intrinsic BHMF of the total black hole population. We will discuss this effect later. 

Our final BHMF for the redshift range $4.5<z<7.0$ is presented in Figure~\ref{fig4: bhmf}. The final, fully corrected data points, generated from eight confident sources, are shown as red-filled hexagons. We also plot the results derived without the observational incompleteness correction as open red hexagons. These results are compared with previous JWST-based BHMF determinations from \cite{Matthee+24} (yellow squares) and \cite{Taylor+25} (blue circles), which are based on observations of unlensed fields \citep{Kashino+23, Oesch+23, Finkelstein+25, deGraaff+25}. At our high-mass end (low-mass end for their studies), our derived BHMF is consistent with these previous studies within the measurement uncertainties.

We fit a Schechter function to the BHMF, using the data points both from our measurement (except the lowest mass bin) and literature down to $10^{6}\,M_{\odot}$ with high completeness (i.e., three highest-mass bins from our sample and two from \citealt{Taylor+25}). 
Interestingly, we find an enhancement of BHMF in the low-mass end bin ($10^5<M_{\rm BH}<10^6\,M_\odot$): the BH number density we measured in this bin is $\sim 0.3\,$dex higher than the fitting result, which corresponds to about twice the measurement uncertainty at this mass bin. 
We confirmed that the fit result is not significantly changed against the choice of data. Including additional complete data from the literature \citep{Matthee+24, Taylor+25} did not significantly change the result, consistently showing an excess in our lowest-mass bin. It should be noted that data from ground-based luminous quasar surveys \citep[e.g., ][]{Fan+03, Matsuoka+16, Wang+16, Shen+19, He+24} were excluded from our analysis due to significant differences in sample selection and BH mass estimation methods. But we further confirmed that including one high-mass data point ($M_{\rm BH}>10^{9}\,M_{\odot}$) from the HSC discovered quasar \cite{He+24} does not change our best-fit result, and the excess in the lowest-mass bin still remains.

Another consideration of our BHMF calculation is the duty cycle of high-$z$ AGNs. If we simply assume a mass-independent duty cycle, the derived BHMF would experience a vertical shift while preserving its intrinsic shape. 
However, we need to consider the possibility of mass-dependent duty cycle \citep{Tanaka+09, Inayoshi+20}. Specifically, the existence of massive quasars at $z>6$ necessitates a prolonged and efficient accretion history, implying their duty cycles must approach unity, whereas this requirement is relaxed for lower-mass black holes. 
In such a scenario, the incompleteness correction would be most significant in the low-mass bins. Therefore, the true number density at the low-mass end would be even higher than observed, strengthening the deviation from empirical relations extrapolated from the high-mass regime.

The observed deviation at the low-mass end may be attributable to a population of black holes formed through the direct collapse channel (heavy seed model). As demonstrated by the semi-analytic models (SAMs) of \cite{Jeon+25}, the collapse of primordial gas might create an enhanced abundance of black holes in the mass range of $10^5-10^6\,M_\odot$ \citep{Lodato+07, Volonteri+10, Ricarte+18, Spinoso+23}. This would naturally lead to a feature in the BHMF that deviates from a standard Schechter form, a result consistent with the `bump' seen in their SAM models that include DCBHs (Figure~1 in \citealt{Jeon+25}), which cannot be produced through a light seed-only model (high-$z$ SMBHs originate through accretion onto BHs with $M_{\rm BH}\sim 10^2-10^3\,M_\odot$). 
While previous studies focusing on higher-mass BHs lacked the leverage to distinguish between these formation scenarios, our new data point at the low-mass end provides a critical test.
Our measurement reveals a 4-$\sigma$ deviation from the predictions of the light-seed-only model, while remaining fully consistent with the DCBH model presented in \citet{Jeon+25}. Although this single data point cannot resolve the full shape of the characteristic peak, its significant deviation from the light seed-only model provides strong supportive evidence that the direct collapse channel is a possible formation pathway for high-$z$ AGNs.

It is also important to notice the metallicity of the host galaxies of our lowest-mass BHs with $M_{\rm BH}<10^{6}\,M_{\odot}$. 
The canonical formation channel for DCBHs requires pristine, extremely low-metallicity gas clouds, considering their similar origin with Pop~III stars \citep{Bromm+03, Lodato+06, Omukai+08}. However, the initial near-zero metallicity is predicted to rapidly increase during the subsequent build-up of a stellar component in the DCBH host system \citep[e.g.,][]{2017ApJ...838..117N,Jeon2025_lrd}. The highly unusual Abell2744-QSO1 system, recently captured by JWST \citep{Maiolino+25}, exhibits such pristine conditions, but it is also extremely overmassive, even with respect to the host dynamical (virial) mass. At even earlier cosmic epochs, the JWST detection, measurements and inferred properties of the $z>10$ accreting sources UHZ1 and GHZ9 support the existence of heavy seeds that form from direct collapse and are hence also overmassive as predicted and expected \cite{2017ApJ...838..117N}.  An alternate origin for such systems has thus been suggested, implicating massive primordial black hole (PBH) seeds \citep[e.g.,][]{Dayal_PBH2024, Zhang_PBH2025}. More generally, PBHs that may not contribute significantly to the dark matter budget but may produce the observed excess in the correlated cosmic X-ray-IR backgrounds have been postulated to provide seeds for SMBHs detected at $z<8$ \citep{2022ApJ...926..205C, 2025arXiv250613858Q}. 
In contrast, our lowest-mass BH systems show \oiii/H$\beta$ $>1$ in the narrow line components (see Figure \ref{fig: bpt}), indicating a moderately metal-enriched nature of the host galaxies. However, 
our lowest-mass BH objects are not necessarily captured immediately after their births,  
and recent theoretical models confirm the presence of evolved DCBHs in moderately metal-enriched host galaxies, but still with $M_{\rm BH}< 10^{6}\,M_{\odot}$ \citep{Jeon2025_lrd}. 
Therefore, the enhanced abundance of our lowest-mass BHs is not in contradiction with their DCBH origin. 

\subsection{Caveats}
\label{sec4.4: Caveats}

The FWHM of the broad H$\alpha$ line of these objects are around $1000\,\rm km\,s^{-1}$ (in particular, $\rm FWHM_{H\alpha,broad}\approx 660\rm km\,s^{-1}$ for ID~13131 and $\rm FWHM_{H\alpha,broad}\approx 650\rm km\,s^{-1}$ for ID~41299). This value is significantly lower than the broad line widths in previously discovered high-$z$ AGNs, which are often selected based on criteria requiring $\rm FWHM > 2000\,\rm km\,s^{-1}$ \citep[e.g.,][]{Greene+24}. 
This is a natural consequence since we are exploring the lowest mass BH regime, which should have a narrower broad line width in principle. With the careful completeness modeling, we find that our observational setup has the capability of exploring BHs in this mass regime through identifying `narrow' broad lines ($\rm FWHM_{H\alpha,broad}\sim 500-700\,km\,s^{-1}$; more details in Sec.~\ref{app1: incompleteness}). While for BHs with lower masses ($M_{\rm BH}\lesssim 10^5\,M_\odot$), the much narrower broad line is hard to separate from the narrow line. Definitely confirming the nature of this narrow broad-line component will require deep, high-resolution follow-up observations. We have tested the robustness of our results against the potential misclassification of this source. Excluding it from our sample induces only a minor change of 0.1 dex to the BHMF in our lowest mass bin. Consequently, the observed deviation at the low-mass end is not dominated by this single object.
Here, we discuss several other possibilities that can produce emission lines with this line width. 

Intense star formation and galactic mergers are also known to produce broad emission lines. However, these mechanisms are unlikely to explain our observations. Galactic mergers typically broaden both allowed and forbidden lines, and these emission lines are often resolved into multiple narrow kinematic components ($\rm FWHM\lesssim 500\,km\,s^{-1}$) \citep[e.g.,][]{Freeman+19, Hashimoto+19, Romano+21, Xu+22}. In contrast, our spectra reveal a single broad component solely in the \ha\ line. A highly tuned low-metallicity outflow could theoretically reproduce this signature, but such an event remains observationally unconfirmed even in the low-metallicity galaxies \citep{Xu+22}.

A significant caveat, however, is the possibility of collisional de-excitation. If the outflowing gas possesses a density similar to that of an AGN broad-line region, the emission from forbidden lines would be naturally suppressed. Although this scenario has not been observed in the local universe, we cannot fully rule out its possibility in the high-$z$ universe where gas densities can be much higher than local. Supernovae can also provide broad Balmer lines \citep[e.g., ][]{Baldassare+16}, which cannot be fully ruled out in the current stage. The time-domain follow-up for our sources is necessary to solidly distinguish the AGN and supernova scenario, as the brightness of the broad line driven by supernova should monochromatically decrease.  

\section{Conclusion}
\label{sec5: conclusion}

In this study, we present the results of a systematic search for broad-line Active Galactic Nuclei (AGNs) at high redshift ($z>4.5$) by leveraging deep JWST/NIRSpec spectroscopy of the gravitationally lensed field Abell S1063. This strategy, combining the power of JWST with the magnification from a massive foreground galaxy cluster, allows us to probe an unprecedentedly faint population of AGNs powered by lower-mass black holes than are typically accessible in blank-field surveys. Our main findings are summarized as follows:
\begin{enumerate}
    \item Our systematic search has identified ten broad-line AGNs, consisting of eight secure and two tentative candidate detections. Of particular significance is the fact that this sample includes three candidate Intermediate-Mass Black Holes (IMBHs), with masses estimated to be between $10^{5.5}-10^6\,M_\odot$. These are the first candidate IMBHs identified in single-host systems at these early cosmic epochs.
    \item We investigate the location of these sources on the BPT diagram, and find that consistent with other high-$z$ AGNs found by JWST, these are located on the boundary between the star-forming galaxies and AGN regions as defined by local galaxies. The two tentative sources are located within the AGN region, which further supports their classification of broad-line AGNs.
    \item By comparing the determined BH masses to their host galaxy stellar masses, we find that in addition to six sources with large $M_{\rm BH}/M_*$; two sources lie on or below the local $M_{\rm BH}-M_*$ relationship. This finding of the range of loci for these sources provides strong evidence that selection bias has significantly skewed our initial understanding of the high-$z$ AGN population that were reported to be overmassive compared to the local scaling relation.
    \item We construct the first empirically determined black hole mass function (BHMF) at $4.5<z<7$ that extends down to the IMBH mass regime of $10^5-10^6\,M_\odot$. Interestingly, the measured number density of IMBHs is higher than the predictions of the light seeds model. This estimated abundance is, meanwhile, in good agreement with the predictions of the heavy seed scenario, which posits the formation of a large population of black holes in this specific mass range \cite{Lodato+07}. Here we interpret these observations as suggesting that direct collapse of massive gas clouds may represent a possible fairly efficient channel for the formation of  massive BHs in the early universe. And therefore, multiple pathways exist for the formation of black hole seeds.
\end{enumerate}

Our investigation reveals a glimpse of the elusive IMBH population at $4.5<z<7.0$ and provides a critical new insight into the formation of the first supermassive black holes in the universe.


\section*{Acknowledgements}
We thank Kohei Inayoshi for helpful discussions. 
This work is based on observations made with the NASA/ESA/CSA James Webb Space Telescope. The data were obtained from the Mikulski Archive for Space Telescopes at the Space Telescope Science Institute, which is operated by the Association of Universities for Research in Astronomy, Inc., under NASA contract NAS 5-03127 for JWST. 
These observations are associated with programs \# 3293 and \# 9223.
SF acknowledges support from the Dunlap Institute that is funded through an endowment established by the David Dunlap family and the University of Toronto.
ASL acknowledges support from the Knut and Alice Wallenberg Foundation.
AZ acknowledges support by Grant No. 2020750 from the United States-Israel Binational Science Foundation (BSF) and Grant No. 2109066 from the United States National Science Foundation (NSF); and by the Israel Science Foundation Grant No. 864/23.
\textit{Facility:} JWST

\textit{Software:} \texttt{LMFIT} \citep{Newville+14} \texttt{msafit} \citep{deGraaff+2024}


\bibliography{sample631}{}

\begin{thebibliography}{}
\expandafter\ifx\csname natexlab\endcsname\relax\def\natexlab#1{#1}\fi
\providecommand{\url}[1]{\href{#1}{#1}}
\providecommand{\dodoi}[1]{doi:~\href{http://doi.org/#1}{\nolinkurl{#1}}}
\providecommand{\doeprint}[1]{\href{http://ascl.net/#1}{\nolinkurl{http://ascl.net/#1}}}
\providecommand{\doarXiv}[1]{\href{https://arxiv.org/abs/#1}{\nolinkurl{https://arxiv.org/abs/#1}}}

\bibitem[{{Ananna} {et~al.}(2024){Ananna}, {Bogd{\'a}n}, {Kov{\'a}cs},
  {Natarajan}, \& {Hickox}}]{Ananna+2024}
{Ananna}, T.~T., {Bogd{\'a}n}, {\'A}., {Kov{\'a}cs}, O.~E., {Natarajan}, P., \&
  {Hickox}, R.~C. 2024, \apjl, 969, L18, \dodoi{10.3847/2041-8213/ad5669}

\bibitem[{{Ba{\~n}ados} {et~al.}(2018){Ba{\~n}ados}, {Venemans},
  {Mazzucchelli}, {Farina}, {Walter}, {Wang}, {Decarli}, {Stern}, {Fan},
  {Davies}, {Hennawi}, {Simcoe}, {Turner}, {Rix}, {Yang}, {Kelson}, {Rudie}, \&
  {Winters}}]{Banados+18}
{Ba{\~n}ados}, E., {Venemans}, B.~P., {Mazzucchelli}, C., {et~al.} 2018, \nat,
  553, 473, \dodoi{10.1038/nature25180}

\bibitem[{{Baldassare} {et~al.}(2016){Baldassare}, {Reines}, {Gallo}, {Greene},
  {Graur}, {Geha}, {Hainline}, {Carroll}, \& {Hickox}}]{Baldassare+16}
{Baldassare}, V.~F., {Reines}, A.~E., {Gallo}, E., {et~al.} 2016, \apj, 829,
  57, \dodoi{10.3847/0004-637X/829/1/57}

\bibitem[{{Baldwin} {et~al.}(1981){Baldwin}, {Phillips}, \&
  {Terlevich}}]{Baldwin+81}
{Baldwin}, J.~A., {Phillips}, M.~M., \& {Terlevich}, R. 1981, \pasp, 93, 5,
  \dodoi{10.1086/130766}

\bibitem[{{Becerra} {et~al.}(2018){Becerra}, {Marinacci}, {Bromm}, \&
  {Hernquist}}]{Becerra_DCBH2018}
{Becerra}, F., {Marinacci}, F., {Bromm}, V., \& {Hernquist}, L.~E. 2018,
  \mnras, 480, 5029, \dodoi{10.1093/mnras/sty2210}

\bibitem[{{Begelman} {et~al.}(2006){Begelman}, {Volonteri}, \&
  {Rees}}]{Begelman2006}
{Begelman}, M.~C., {Volonteri}, M., \& {Rees}, M.~J. 2006, \mnras, 370, 289,
  \dodoi{10.1111/j.1365-2966.2006.10467.x}

\bibitem[{{Bezanson} {et~al.}(2024){Bezanson}, {Labbe}, {Whitaker}, {Leja},
  {Price}, {Franx}, {Brammer}, {Marchesini}, {Zitrin}, {Wang}, {Weaver},
  {Furtak}, {Atek}, {Coe}, {Cutler}, {Dayal}, {van Dokkum}, {Feldmann},
  {F{\"o}rster Schreiber}, {Fujimoto}, {Geha}, {Glazebrook}, {de Graaff},
  {Greene}, {Juneau}, {Kassin}, {Kriek}, {Khullar}, {Maseda}, {Mowla},
  {Muzzin}, {Nanayakkara}, {Nelson}, {Oesch}, {Pacifici}, {Pan}, {Papovich},
  {Setton}, {Shapley}, {Smit}, {Stefanon}, {Taylor}, \&
  {Williams}}]{Bezanson+24}
{Bezanson}, R., {Labbe}, I., {Whitaker}, K.~E., {et~al.} 2024, \apj, 974, 92,
  \dodoi{10.3847/1538-4357/ad66cf}

\bibitem[{{Boquien} {et~al.}(2019){Boquien}, {Burgarella}, {Roehlly}, {Buat},
  {Ciesla}, {Corre}, {Inoue}, \& {Salas}}]{Boquien+19}
{Boquien}, M., {Burgarella}, D., {Roehlly}, Y., {et~al.} 2019, \aap, 622, A103,
  \dodoi{10.1051/0004-6361/201834156}

\bibitem[{{Bouwens} {et~al.}(2014){Bouwens}, {Illingworth}, {Oesch},
  {Labb{\'e}}, {van Dokkum}, {Trenti}, {Franx}, {Smit}, {Gonzalez}, \&
  {Magee}}]{Bouwens+14}
{Bouwens}, R.~J., {Illingworth}, G.~D., {Oesch}, P.~A., {et~al.} 2014, \apj,
  793, 115, \dodoi{10.1088/0004-637X/793/2/115}

\bibitem[{{Brammer}(2023)}]{Brammer2023}
{Brammer}, G. 2023, {msaexp: NIRSpec analyis tools}, 0.6.17,  Zenodo,
  \dodoi{10.5281/zenodo.8319596}

\bibitem[{{Brammer} {et~al.}(2008){Brammer}, {van Dokkum}, \&
  {Coppi}}]{Brammer2008}
{Brammer}, G.~B., {van Dokkum}, P.~G., \& {Coppi}, P. 2008, \apj, 686, 1503,
  \dodoi{10.1086/591786}

\bibitem[{{Bromm} \& {Loeb}(2003)}]{Bromm+03}
{Bromm}, V., \& {Loeb}, A. 2003, \apj, 596, 34, \dodoi{10.1086/377529}

\bibitem[{{Bruzual} \& {Charlot}(2003)}]{Bruzual+03}
{Bruzual}, G., \& {Charlot}, S. 2003, \mnras, 344, 1000,
  \dodoi{10.1046/j.1365-8711.2003.06897.x}

\bibitem[{{Calzetti} {et~al.}(2000){Calzetti}, {Armus}, {Bohlin}, {Kinney},
  {Koornneef}, \& {Storchi-Bergmann}}]{Calzetti+03}
{Calzetti}, D., {Armus}, L., {Bohlin}, R.~C., {et~al.} 2000, \apj, 533, 682,
  \dodoi{10.1086/308692}

\bibitem[{{Cappelluti} {et~al.}(2022){Cappelluti}, {Hasinger}, \&
  {Natarajan}}]{2022ApJ...926..205C}
{Cappelluti}, N., {Hasinger}, G., \& {Natarajan}, P. 2022, \apj, 926, 205,
  \dodoi{10.3847/1538-4357/ac332d}

\bibitem[{{Carnall} {et~al.}(2018){Carnall}, {McLure}, {Dunlop}, \&
  {Dav{\'e}}}]{Carnall+18}
{Carnall}, A.~C., {McLure}, R.~J., {Dunlop}, J.~S., \& {Dav{\'e}}, R. 2018,
  \mnras, 480, 4379, \dodoi{10.1093/mnras/sty2169}

\bibitem[{{Cenci} \& {Habouzit}(2025)}]{Cenci_DCBH2025}
{Cenci}, E., \& {Habouzit}, M. 2025, arXiv e-prints, arXiv:2508.14897.
\newblock \doarXiv{2508.14897}

\bibitem[{{Chabrier}(2003)}]{Chabrier+03}
{Chabrier}, G. 2003, \pasp, 115, 763, \dodoi{10.1086/376392}

\bibitem[{{Chang} {et~al.}(2025){Chang}, {Gronke}, {Matthee}, \&
  {Mason}}]{Chang+25}
{Chang}, S.-J., {Gronke}, M., {Matthee}, J., \& {Mason}, C. 2025, arXiv
  e-prints, arXiv:2508.08768, \dodoi{10.48550/arXiv.2508.08768}

\bibitem[{{Chevallard} \& {Charlot}(2016)}]{Chevallard+16}
{Chevallard}, J., \& {Charlot}, S. 2016, \mnras, 462, 1415,
  \dodoi{10.1093/mnras/stw1756}

\bibitem[{{Chisholm} {et~al.}(2024){Chisholm}, {Berg}, {Endsley}, {Gazagnes},
  {Richardson}, {Lambrides}, {Greene}, {Finkelstein}, {Flury}, {Guseva},
  {Henry}, {Hutchison}, {Izotov}, {Marques-Chaves}, {Oesch}, {Papovich},
  {Saldana-Lopez}, {Schaerer}, \& {Stephenson}}]{Chisholm+24}
{Chisholm}, J., {Berg}, D.~A., {Endsley}, R., {et~al.} 2024, \mnras, 534, 2633,
  \dodoi{10.1093/mnras/stae2199}

\bibitem[{{Dayal}(2024)}]{Dayal_PBH2024}
{Dayal}, P. 2024, \aap, 690, A182, \dodoi{10.1051/0004-6361/202451481}

\bibitem[{{de Graaff} {et~al.}(2024){de Graaff}, {Rix}, {Carniani}, {Suess},
  {Charlot}, {Curtis-Lake}, {Arribas}, {Baker}, {Boyett}, {Bunker}, {Cameron},
  {Chevallard}, {Curti}, {Eisenstein}, {Franx}, {Hainline}, {Hausen}, {Ji},
  {Johnson}, {Jones}, {Maiolino}, {Maseda}, {Nelson}, {Parlanti}, {Rawle},
  {Robertson}, {Tacchella}, {{\"U}bler}, {Williams}, {Willmer}, \&
  {Willott}}]{deGraaff+2024}
{de Graaff}, A., {Rix}, H.-W., {Carniani}, S., {et~al.} 2024, \aap, 684, A87,
  \dodoi{10.1051/0004-6361/202347755}

\bibitem[{{de Graaff} {et~al.}(2025){de Graaff}, {Brammer}, {Weibel}, {Lewis},
  {Maseda}, {Oesch}, {Bezanson}, {Boogaard}, {Cleri}, {Cooper}, {Gottumukkala},
  {Greene}, {Hirschmann}, {Hviding}, {Katz}, {Labb{\'e}}, {Leja}, {Matthee},
  {McConachie}, {Miller}, {Naidu}, {Price}, {Rix}, {Setton}, {Suess}, {Wang},
  {Whitaker}, \& {Williams}}]{deGraaff+25}
{de Graaff}, A., {Brammer}, G., {Weibel}, A., {et~al.} 2025, \aap, 697, A189,
  \dodoi{10.1051/0004-6361/202452186}

\bibitem[{{Ding} {et~al.}(2023){Ding}, {Onoue}, {Silverman}, {Matsuoka},
  {Izumi}, {Strauss}, {Jahnke}, {Phillips}, {Li}, {Volonteri}, {Haiman},
  {Andika}, {Aoki}, {Baba}, {Bieri}, {Bosman}, {Bottrell}, {Eilers},
  {Fujimoto}, {Habouzit}, {Imanishi}, {Inayoshi}, {Iwasawa}, {Kashikawa},
  {Kawaguchi}, {Kohno}, {Lee}, {Lupi}, {Lyu}, {Nagao}, {Overzier}, {Schindler},
  {Schramm}, {Shimasaku}, {Toba}, {Trakhtenbrot}, {Trebitsch}, {Treu},
  {Umehata}, {Venemans}, {Vestergaard}, {Walter}, {Wang}, \& {Yang}}]{Ding+23}
{Ding}, X., {Onoue}, M., {Silverman}, J.~D., {et~al.} 2023, \nat, 621, 51,
  \dodoi{10.1038/s41586-023-06345-5}

\bibitem[{{Duncan} {et~al.}(2014){Duncan}, {Conselice}, {Mortlock}, {Hartley},
  {Guo}, {Ferguson}, {Dav{\'e}}, {Lu}, {Ownsworth}, {Ashby}, {Dekel},
  {Dickinson}, {Faber}, {Giavalisco}, {Grogin}, {Kocevski}, {Koekemoer},
  {Somerville}, \& {White}}]{Duncan+14}
{Duncan}, K., {Conselice}, C.~J., {Mortlock}, A., {et~al.} 2014, \mnras, 444,
  2960, \dodoi{10.1093/mnras/stu1622}

\bibitem[{{Dunlop} {et~al.}(2012){Dunlop}, {McLure}, {Robertson}, {Ellis},
  {Stark}, {Cirasuolo}, \& {de Ravel}}]{Dunlop+12}
{Dunlop}, J.~S., {McLure}, R.~J., {Robertson}, B.~E., {et~al.} 2012, \mnras,
  420, 901, \dodoi{10.1111/j.1365-2966.2011.20102.x}

\bibitem[{{Fan} {et~al.}(2023){Fan}, {Ba{\~n}ados}, \& {Simcoe}}]{Fan+23}
{Fan}, X., {Ba{\~n}ados}, E., \& {Simcoe}, R.~A. 2023, \araa, 61, 373,
  \dodoi{10.1146/annurev-astro-052920-102455}

\bibitem[{{Fan} {et~al.}(2001){Fan}, {Narayanan}, {Lupton}, {Strauss}, {Knapp},
  {Becker}, {White}, {Pentericci}, {Leggett}, {Haiman}, {Gunn}, {Ivezi{\'c}},
  {Schneider}, {Anderson}, {Brinkmann}, {Bahcall}, {Connolly}, {Csabai}, {Doi},
  {Fukugita}, {Geballe}, {Grebel}, {Harbeck}, {Hennessy}, {Lamb}, {Miknaitis},
  {Munn}, {Nichol}, {Okamura}, {Pier}, {Prada}, {Richards}, {Szalay}, \&
  {York}}]{Fan+01}
{Fan}, X., {Narayanan}, V.~K., {Lupton}, R.~H., {et~al.} 2001, \aj, 122, 2833,
  \dodoi{10.1086/324111}

\bibitem[{{Fan} {et~al.}(2003){Fan}, {Strauss}, {Schneider}, {Becker}, {White},
  {Haiman}, {Gregg}, {Pentericci}, {Grebel}, {Narayanan}, {Loh}, {Richards},
  {Gunn}, {Lupton}, {Knapp}, {Ivezi{\'c}}, {Brandt}, {Collinge}, {Hao},
  {Harbeck}, {Prada}, {Schaye}, {Strateva}, {Zakamska}, {Anderson},
  {Brinkmann}, {Bahcall}, {Lamb}, {Okamura}, {Szalay}, \& {York}}]{Fan+03}
{Fan}, X., {Strauss}, M.~A., {Schneider}, D.~P., {et~al.} 2003, \aj, 125, 1649,
  \dodoi{10.1086/368246}

\bibitem[{{Farina} {et~al.}(2022){Farina}, {Schindler}, {Walter},
  {Ba{\~n}ados}, {Davies}, {Decarli}, {Eilers}, {Fan}, {Hennawi},
  {Mazzucchelli}, {Meyer}, {Trakhtenbrot}, {Volonteri}, {Wang}, {Worseck},
  {Yang}, {Gutcke}, {Venemans}, {Bosman}, {Costa}, {De Rosa}, {Drake}, \&
  {Onoue}}]{Farina+22}
{Farina}, E.~P., {Schindler}, J.-T., {Walter}, F., {et~al.} 2022, \apj, 941,
  106, \dodoi{10.3847/1538-4357/ac9626}

\bibitem[{{Ferrarese} \& {Merritt}(2000)}]{Ferrarese+00}
{Ferrarese}, L., \& {Merritt}, D. 2000, \apjl, 539, L9, \dodoi{10.1086/312838}

\bibitem[{{Finkelstein} {et~al.}(2025){Finkelstein}, {Bagley}, {Arrabal Haro},
  {Dickinson}, {Ferguson}, {Kartaltepe}, {Kocevski}, {Koekemoer}, {Lotz},
  {Papovich}, {P{\'e}rez-Gonz{\'a}lez}, {Pirzkal}, {Somerville}, {Trump},
  {Yang}, {Yung}, {Fontana}, {Grazian}, {Grogin}, {Kewley}, {Kirkpatrick},
  {Larson}, {Pentericci}, {Ravindranath}, {Wilkins}, {Almaini}, {Amor{\'\i}n},
  {Barro}, {Bhatawdekar}, {Bisigello}, {Brooks}, {Buat}, {Buitrago},
  {Burgarella}, {Calabr{\`o}}, {Castellano}, {Cheng}, {Cleri}, {Cole},
  {Cooper}, {Cooper}, {Costantin}, {Cox}, {Croton}, {Daddi}, {Davis}, {Dekel},
  {Elbaz}, {Fern{\'a}ndez}, {Fujimoto}, {Gandolfi}, {Gardner}, {Gawiser},
  {Giavalisco}, {G{\'o}mez-Guijarro}, {Guo}, {Gupta}, {Hathi}, {Harish},
  {Henry}, {Hirschmann}, {Hu}, {Hutchison}, {Iyer}, {Jaskot}, {Jha}, {Jung},
  {Kassin}, {Kokorev}, {Kurczynski}, {Leung}, {Llerena}, {Long}, {Lucas}, {Lu},
  {McGrath}, {McIntosh}, {Merlin}, {Mobasher}, {Morales}, {Napolitano},
  {Pacucci}, {Pandya}, {Rafelski}, {Rodighiero}, {Rose}, {Santini},
  {Seill{\'e}}, {Simons}, {Shen}, {Straughn}, {Tacchella}, {Taylor},
  {Vanderhoof}, {Vega-Ferrero}, {Weiner}, {Willmer}, {Zhu}, {Bell}, {Wuyts},
  {Holwerda}, {Wang}, {Wang}, {Zavala}, \& {CEERS
  Collaboration}}]{Finkelstein+25}
{Finkelstein}, S.~L., {Bagley}, M.~B., {Arrabal Haro}, P., {et~al.} 2025,
  \apjl, 983, L4, \dodoi{10.3847/2041-8213/adbbd3}

\bibitem[{{Foreman-Mackey} {et~al.}(2013){Foreman-Mackey}, {Hogg}, {Lang}, \&
  {Goodman}}]{Foreman-Mackey+13}
{Foreman-Mackey}, D., {Hogg}, D.~W., {Lang}, D., \& {Goodman}, J. 2013, \pasp,
  125, 306, \dodoi{10.1086/670067}

\bibitem[{{Freeman} {et~al.}(2019){Freeman}, {Siana}, {Kriek}, {Shapley},
  {Reddy}, {Coil}, {Mobasher}, {Muratov}, {Azadi}, {Leung}, {Sanders},
  {Shivaei}, {Price}, {DeGroot}, \& {Kere{\v{s}}}}]{Freeman+19}
{Freeman}, W.~R., {Siana}, B., {Kriek}, M., {et~al.} 2019, \apj, 873, 102,
  \dodoi{10.3847/1538-4357/ab0655}

\bibitem[{{Fujimoto} {et~al.}(2025){Fujimoto}, {Naidu}, {Chisholm}, {Atek},
  {Endsley}, {Kokorev}, {Furtak}, {Pan}, {Liu}, {Bromm}, {Venditti}, {Visbal},
  {Sarmento}, {Weibel}, {Oesch}, {Brammer}, {Schaerer}, {Adamo}, {Berg},
  {Bezanson}, {Bouwens}, {Chemerynska}, {Claeyssens}, {Dessauges-Zavadsky},
  {Frebel}, {Korber}, {Labbe}, {Marques-Chaves}, {Matthee}, {McQuinn},
  {Mu{\~n}oz}, {Natarajan}, {Saldana-Lopez}, {Suess}, {Volonteri}, \&
  {Zitrin}}]{Fujimoto+25}
{Fujimoto}, S., {Naidu}, R.~P., {Chisholm}, J., {et~al.} 2025, \apj, 989, 46,
  \dodoi{10.3847/1538-4357/ade9a1}

\bibitem[{{Furtak} {et~al.}(2024){Furtak}, {Labb{\'e}}, {Zitrin}, {Greene},
  {Dayal}, {Chemerynska}, {Kokorev}, {Miller}, {Goulding}, {de Graaff},
  {Bezanson}, {Brammer}, {Cutler}, {Leja}, {Pan}, {Price}, {Wang}, {Weaver},
  {Whitaker}, {Atek}, {Bogd{\'a}n}, {Charlot}, {Curtis-Lake}, {van Dokkum},
  {Endsley}, {Feldmann}, {Fudamoto}, {Fujimoto}, {Glazebrook}, {Juneau},
  {Marchesini}, {Maseda}, {Nelson}, {Oesch}, {Plat}, {Setton}, {Stark}, \&
  {Williams}}]{Furtak+24}
{Furtak}, L.~J., {Labb{\'e}}, I., {Zitrin}, A., {et~al.} 2024, \nat, 628, 57,
  \dodoi{10.1038/s41586-024-07184-8}

\bibitem[{{Gebhardt} {et~al.}(2000){Gebhardt}, {Bender}, {Bower}, {Dressler},
  {Faber}, {Filippenko}, {Green}, {Grillmair}, {Ho}, {Kormendy}, {Lauer},
  {Magorrian}, {Pinkney}, {Richstone}, \& {Tremaine}}]{Gebhardt+00}
{Gebhardt}, K., {Bender}, R., {Bower}, G., {et~al.} 2000, \apjl, 539, L13,
  \dodoi{10.1086/312840}

\bibitem[{{Gehrels}(1986)}]{Gehrels+86}
{Gehrels}, N. 1986, \apj, 303, 336, \dodoi{10.1086/164079}

\bibitem[{{Geris} {et~al.}(2025){Geris}, {Maiolino}, {Isobe}, {Scholtz},
  {D'Eugenio}, {Ji}, {Juodzbalis}, {Simmonds}, {Dayal}, {Trinca}, {Schneider},
  {Arribas}, {Bhatawdekar}, {Bunker}, {Carniani}, {Charlot}, {Chevallard},
  {Curtis-Lake}, {Johnson}, {Parlanti}, {Rinaldi}, {Robertson}, {Tacchella},
  {Uebler}, {Venturi}, {Williams}, \& {Witstok}}]{Geris+25}
{Geris}, S., {Maiolino}, R., {Isobe}, Y., {et~al.} 2025, arXiv e-prints,
  arXiv:2506.22147, \dodoi{10.48550/arXiv.2506.22147}

\bibitem[{{Greene} \& {Ho}(2005)}]{Greene+05}
{Greene}, J.~E., \& {Ho}, L.~C. 2005, \apj, 630, 122, \dodoi{10.1086/431897}

\bibitem[{{Greene} {et~al.}(2020){Greene}, {Strader}, \& {Ho}}]{Greene+20}
{Greene}, J.~E., {Strader}, J., \& {Ho}, L.~C. 2020, \araa, 58, 257,
  \dodoi{10.1146/annurev-astro-032620-021835}

\bibitem[{{Greene} {et~al.}(2024){Greene}, {Labbe}, {Goulding}, {Furtak},
  {Chemerynska}, {Kokorev}, {Dayal}, {Volonteri}, {Williams}, {Wang}, {Setton},
  {Burgasser}, {Bezanson}, {Atek}, {Brammer}, {Cutler}, {Feldmann}, {Fujimoto},
  {Glazebrook}, {de Graaff}, {Khullar}, {Leja}, {Marchesini}, {Maseda},
  {Matthee}, {Miller}, {Naidu}, {Nanayakkara}, {Oesch}, {Pan}, {Papovich},
  {Price}, {van Dokkum}, {Weaver}, {Whitaker}, \& {Zitrin}}]{Greene+24}
{Greene}, J.~E., {Labbe}, I., {Goulding}, A.~D., {et~al.} 2024, \apj, 964, 39,
  \dodoi{10.3847/1538-4357/ad1e5f}

\bibitem[{{Greene} {et~al.}(2025){Greene}, {Setton}, {Furtak}, {Naidu},
  {Volonteri}, {Dayal}, {Labbe}, {van Dokkum}, {Bezanson}, {Brammer}, {Cutler},
  {Glazebrook}, {de Graaff}, {Hirschmann}, {Hviding}, {Kokorev}, {Leja}, {Liu},
  {Ma}, {Matthee}, {Nanayakkara}, {Oesch}, {Pan}, {Price}, {Spilker}, {Wang},
  {Weaver}, {Whitaker}, {Williams}, \& {Zitrin}}]{Greene+25}
{Greene}, J.~E., {Setton}, D.~J., {Furtak}, L.~J., {et~al.} 2025, arXiv
  e-prints, arXiv:2509.05434, \dodoi{10.48550/arXiv.2509.05434}

\bibitem[{{Haemmerl{\'e}} {et~al.}(2020){Haemmerl{\'e}}, {Mayer}, {Klessen},
  {Hosokawa}, {Madau}, \& {Bromm}}]{Haemmerle+20}
{Haemmerl{\'e}}, L., {Mayer}, L., {Klessen}, R.~S., {et~al.} 2020, \ssr, 216,
  48, \dodoi{10.1007/s11214-020-00673-y}

\bibitem[{{Haemmerl{\'e}} {et~al.}(2018){Haemmerl{\'e}}, {Woods}, {Klessen},
  {Heger}, \& {Whalen}}]{Haemmerle+18}
{Haemmerl{\'e}}, L., {Woods}, T.~E., {Klessen}, R.~S., {Heger}, A., \&
  {Whalen}, D.~J. 2018, \mnras, 474, 2757, \dodoi{10.1093/mnras/stx2919}

\bibitem[{{Harikane} {et~al.}(2023){Harikane}, {Zhang}, {Nakajima}, {Ouchi},
  {Isobe}, {Ono}, {Hatano}, {Xu}, \& {Umeda}}]{Harikane+23}
{Harikane}, Y., {Zhang}, Y., {Nakajima}, K., {et~al.} 2023, \apj, 959, 39,
  \dodoi{10.3847/1538-4357/ad029e}

\bibitem[{{Hashimoto} {et~al.}(2019){Hashimoto}, {Inoue}, {Mawatari}, {Tamura},
  {Matsuo}, {Furusawa}, {Harikane}, {Shibuya}, {Knudsen}, {Kohno}, {Ono},
  {Zackrisson}, {Okamoto}, {Kashikawa}, {Oesch}, {Ouchi}, {Ota}, {Shimizu},
  {Taniguchi}, {Umehata}, \& {Watson}}]{Hashimoto+19}
{Hashimoto}, T., {Inoue}, A.~K., {Mawatari}, K., {et~al.} 2019, \pasj, 71, 71,
  \dodoi{10.1093/pasj/psz049}

\bibitem[{{He} {et~al.}(2024){He}, {Akiyama}, {Enoki}, {Ichikawa}, {Inayoshi},
  {Kashikawa}, {Kawaguchi}, {Matsuoka}, {Nagao}, {Onoue}, {Oogi}, {Schulze},
  {Toba}, \& {Ueda}}]{He+24}
{He}, W., {Akiyama}, M., {Enoki}, M., {et~al.} 2024, \apj, 962, 152,
  \dodoi{10.3847/1538-4357/ad1518}

\bibitem[{{Heintz} {et~al.}(2025){Heintz}, {Brammer}, {Watson}, {Oesch},
  {Keating}, {Hayes}, {Abdurro'uf}, {Arellano-C{\'o}rdova}, {Carnall},
  {Christiansen}, {Cullen}, {Dav{\'e}}, {Dayal}, {Ferrara}, {Finlator},
  {Fynbo}, {Flury}, {Gelli}, {Gillman}, {Gottumukkala}, {Gould}, {Greve},
  {Hardin}, {Hsiao}, {Hutter}, {Jakobsson}, {Killi}, {Khosravaninezhad},
  {Laursen}, {Lee}, {Magdis}, {Matthee}, {Naidu}, {Narayanan}, {Pollock},
  {Prescott}, {Rusakov}, {Shuntov}, {Sneppen}, {Smit}, {Tanvir}, {Terp},
  {Toft}, {Valentino}, {Vijayan}, {Weaver}, {Wise}, \& {Witstok}}]{Heintz+25}
{Heintz}, K.~E., {Brammer}, G.~B., {Watson}, D., {et~al.} 2025, \aap, 693, A60,
  \dodoi{10.1051/0004-6361/202450243}

\bibitem[{{Hviding} {et~al.}(2025){Hviding}, {de Graaff}, {Miller}, {Setton},
  {Greene}, {Labb{\'e}}, {Brammer}, {Bezanson}, {Boogaard}, {Cleri}, {Leja},
  {Maseda}, {McConachie}, {Matthee}, {Naidu}, {Oesch}, {Wang}, {Whitaker}, \&
  {Williams}}]{Hviding+25}
{Hviding}, R.~E., {de Graaff}, A., {Miller}, T.~B., {et~al.} 2025, arXiv
  e-prints, arXiv:2506.05459, \dodoi{10.48550/arXiv.2506.05459}

\bibitem[{{Inayoshi} {et~al.}(2020){Inayoshi}, {Visbal}, \&
  {Haiman}}]{Inayoshi+20}
{Inayoshi}, K., {Visbal}, E., \& {Haiman}, Z. 2020, \araa, 58, 27,
  \dodoi{10.1146/annurev-astro-120419-014455}

\bibitem[{{Izumi} {et~al.}(2019){Izumi}, {Onoue}, {Matsuoka}, {Nagao},
  {Strauss}, {Imanishi}, {Kashikawa}, {Fujimoto}, {Kohno}, {Toba}, {Umehata},
  {Goto}, {Ueda}, {Shirakata}, {Silverman}, {Greene}, {Harikane}, {Hashimoto},
  {Ikarashi}, {Iono}, {Iwasawa}, {Lee}, {Minezaki}, {Nakanishi}, {Tamura},
  {Tang}, \& {Taniguchi}}]{Izumi+19}
{Izumi}, T., {Onoue}, M., {Matsuoka}, Y., {et~al.} 2019, \pasj, 71, 111,
  \dodoi{10.1093/pasj/psz096}

\bibitem[{{Jeon} {et~al.}(2023){Jeon}, {Liu}, {Bromm}, \&
  {Finkelstein}}]{Jeon2023}
{Jeon}, J., {Liu}, B., {Bromm}, V., \& {Finkelstein}, S.~L. 2023, \mnras, 524,
  176, \dodoi{10.1093/mnras/stad1877}

\bibitem[{{Jeon} {et~al.}(2025{\natexlab{a}}){Jeon}, {Liu}, {Taylor},
  {Kokorev}, {Chisholm}, {Kocevski}, {Finkelstein}, \& {Bromm}}]{Jeon+25}
{Jeon}, J., {Liu}, B., {Taylor}, A.~J., {et~al.} 2025{\natexlab{a}}, \apj, 988,
  110, \dodoi{10.3847/1538-4357/ade2e1}

\bibitem[{{Jeon} {et~al.}(2025{\natexlab{b}}){Jeon}, {Liu}, {Bromm},
  {Fujimoto}, {Taylor}, {Kokorev}, {Larson}, {Chisholm}, {Finkelstein}, \&
  {Kocevski}}]{Jeon2025_lrd}
{Jeon}, J., {Liu}, B., {Bromm}, V., {et~al.} 2025{\natexlab{b}}, arXiv
  e-prints, arXiv:2508.14155.
\newblock \doarXiv{2508.14155}

\bibitem[{{Johnson} {et~al.}(2021){Johnson}, {Leja}, {Conroy}, \&
  {Speagle}}]{Johnson+21}
{Johnson}, B.~D., {Leja}, J., {Conroy}, C., \& {Speagle}, J.~S. 2021, \apjs,
  254, 22, \dodoi{10.3847/1538-4365/abef67}

\bibitem[{{Kashino} {et~al.}(2023){Kashino}, {Lilly}, {Matthee}, {Eilers},
  {Mackenzie}, {Bordoloi}, \& {Simcoe}}]{Kashino+23}
{Kashino}, D., {Lilly}, S.~J., {Matthee}, J., {et~al.} 2023, \apj, 950, 66,
  \dodoi{10.3847/1538-4357/acc588}

\bibitem[{Kass \& Raftery(1995)}]{Kass+95}
Kass, R.~E., \& Raftery, A.~E. 1995, Journal of the American Statistical
  Association, 90, 773, \dodoi{10.1080/01621459.1995.10476572}

\bibitem[{{Kauffmann} {et~al.}(2003){Kauffmann}, {Heckman}, {Tremonti},
  {Brinchmann}, {Charlot}, {White}, {Ridgway}, {Brinkmann}, {Fukugita}, {Hall},
  {Ivezi{\'c}}, {Richards}, \& {Schneider}}]{Kauffmann+03}
{Kauffmann}, G., {Heckman}, T.~M., {Tremonti}, C., {et~al.} 2003, \mnras, 346,
  1055, \dodoi{10.1111/j.1365-2966.2003.07154.x}

\bibitem[{{Kewley} {et~al.}(2006){Kewley}, {Groves}, {Kauffmann}, \&
  {Heckman}}]{Kewley+06}
{Kewley}, L.~J., {Groves}, B., {Kauffmann}, G., \& {Heckman}, T. 2006, \mnras,
  372, 961, \dodoi{10.1111/j.1365-2966.2006.10859.x}

\bibitem[{{Kewley} {et~al.}(2001){Kewley}, {Heisler}, {Dopita}, \&
  {Lumsden}}]{Kewley+01}
{Kewley}, L.~J., {Heisler}, C.~A., {Dopita}, M.~A., \& {Lumsden}, S. 2001,
  \apjs, 132, 37, \dodoi{10.1086/318944}

\bibitem[{{Kocevski} {et~al.}(2023){Kocevski}, {Onoue}, {Inayoshi}, {Trump},
  {Arrabal Haro}, {Grazian}, {Dickinson}, {Finkelstein}, {Kartaltepe},
  {Hirschmann}, {Aird}, {Holwerda}, {Fujimoto}, {Juneau}, {Amor{\'\i}n},
  {Backhaus}, {Bagley}, {Barro}, {Bell}, {Bisigello}, {Calabr{\`o}}, {Cleri},
  {Cooper}, {Ding}, {Grogin}, {Ho}, {Hutchison}, {Inoue}, {Jiang}, {Jones},
  {Koekemoer}, {Li}, {Li}, {McGrath}, {Molina}, {Papovich},
  {P{\'e}rez-Gonz{\'a}lez}, {Pirzkal}, {Wilkins}, {Yang}, \&
  {Yung}}]{Kocevski+23}
{Kocevski}, D.~D., {Onoue}, M., {Inayoshi}, K., {et~al.} 2023, \apjl, 954, L4,
  \dodoi{10.3847/2041-8213/ace5a0}

\bibitem[{{Kokorev} {et~al.}(2023){Kokorev}, {Fujimoto}, {Labbe}, {Greene},
  {Bezanson}, {Dayal}, {Nelson}, {Atek}, {Brammer}, {Caputi}, {Chemerynska},
  {Cutler}, {Feldmann}, {Fudamoto}, {Furtak}, {Goulding}, {de Graaff}, {Leja},
  {Marchesini}, {Miller}, {Nanayakkara}, {Oesch}, {Pan}, {Price}, {Setton},
  {Smit}, {Stefanon}, {Wang}, {Weaver}, {Whitaker}, {Williams}, \&
  {Zitrin}}]{Kokorev+23}
{Kokorev}, V., {Fujimoto}, S., {Labbe}, I., {et~al.} 2023, \apjl, 957, L7,
  \dodoi{10.3847/2041-8213/ad037a}

\bibitem[{{Kormendy} \& {Ho}(2013)}]{Kormendy+13}
{Kormendy}, J., \& {Ho}, L.~C. 2013, \araa, 51, 511,
  \dodoi{10.1146/annurev-astro-082708-101811}

\bibitem[{{Labb{\'e}} {et~al.}(2023){Labb{\'e}}, {van Dokkum}, {Nelson},
  {Bezanson}, {Suess}, {Leja}, {Brammer}, {Whitaker}, {Mathews}, {Stefanon}, \&
  {Wang}}]{Labbe+23}
{Labb{\'e}}, I., {van Dokkum}, P., {Nelson}, E., {et~al.} 2023, \nat, 616, 266,
  \dodoi{10.1038/s41586-023-05786-2}

\bibitem[{{Larson} {et~al.}(2023){Larson}, {Finkelstein}, {Kocevski},
  {Hutchison}, {Trump}, {Arrabal Haro}, {Bromm}, {Cleri}, {Dickinson},
  {Fujimoto}, {Kartaltepe}, {Koekemoer}, {Papovich}, {Pirzkal}, {Tacchella},
  {Zavala}, {Bagley}, {Behroozi}, {Champagne}, {Cole}, {Jung}, {Morales},
  {Yang}, {Zhang}, {Zitrin}, {Amor{\'\i}n}, {Burgarella}, {Casey}, {Ch{\'a}vez
  Ortiz}, {Cox}, {Chworowsky}, {Fontana}, {Gawiser}, {Grazian}, {Grogin},
  {Harish}, {Hathi}, {Hirschmann}, {Holwerda}, {Juneau}, {Leung}, {Lucas},
  {McGrath}, {P{\'e}rez-Gonz{\'a}lez}, {Rigby}, {Seill{\'e}}, {Simons}, {de La
  Vega}, {Weiner}, {Wilkins}, {Yung}, \& {Ceers Team}}]{Larson+23}
{Larson}, R.~L., {Finkelstein}, S.~L., {Kocevski}, D.~D., {et~al.} 2023, \apjl,
  953, L29, \dodoi{10.3847/2041-8213/ace619}

\bibitem[{{Lauer} {et~al.}(2007){Lauer}, {Tremaine}, {Richstone}, \&
  {Faber}}]{Lauer+07}
{Lauer}, T.~R., {Tremaine}, S., {Richstone}, D., \& {Faber}, S.~M. 2007, \apj,
  670, 249, \dodoi{10.1086/522083}

\bibitem[{{Li} {et~al.}(2025{\natexlab{a}}){Li}, {Shen}, \& {Zhuang}}]{LiJ+25a}
{Li}, J., {Shen}, Y., \& {Zhuang}, M.-Y. 2025{\natexlab{a}}, arXiv e-prints,
  arXiv:2502.05048, \dodoi{10.48550/arXiv.2502.05048}

\bibitem[{{Li} {et~al.}(2021){Li}, {Silverman}, {Ding}, {Strauss}, {Goulding},
  {Schramm}, {Yesuf}, {Sun}, {Xue}, {Birrer}, {Shi}, {Toba}, {Nagao}, \&
  {Imanishi}}]{LiJ+21}
{Li}, J., {Silverman}, J.~D., {Ding}, X., {et~al.} 2021, \apj, 922, 142,
  \dodoi{10.3847/1538-4357/ac2301}

\bibitem[{{Li} {et~al.}(2025{\natexlab{b}}){Li}, {Silverman}, {Shen},
  {Volonteri}, {Jahnke}, {Zhuang}, {Scoggins}, {Ding}, {Harikane}, {Onoue}, \&
  {Tanaka}}]{LiJ+25b}
{Li}, J., {Silverman}, J.~D., {Shen}, Y., {et~al.} 2025{\natexlab{b}}, \apj,
  981, 19, \dodoi{10.3847/1538-4357/ada603}

\bibitem[{{Li} {et~al.}(2025{\natexlab{c}}){Li}, {Ho}, \& {Chen}}]{LiR+25}
{Li}, R., {Ho}, L.~C., \& {Chen}, C.-H. 2025{\natexlab{c}}, arXiv e-prints,
  arXiv:2505.12867, \dodoi{10.48550/arXiv.2505.12867}

\bibitem[{{Lin} {et~al.}(2024){Lin}, {Wang}, {Fan}, {Cai}, {Champagne}, {Sun},
  {Volonteri}, {Yang}, {Hennawi}, {Ba{\~n}ados}, {Barth}, {Eilers}, {Farina},
  {Liu}, {Jin}, {Jun}, {Lupi}, {Kakiichi}, {Mazzucchelli}, {Onoue}, {Pan},
  {Pizzati}, {Rojas-Ruiz}, {Schindler}, {Trakhtenbrot}, {Shen}, {Trebitsch},
  {Zhuang}, {Endsley}, {Meyer}, {Li}, {Li}, {Pudoka}, {Tee}, {Wu}, \&
  {Zhang}}]{Lin+24}
{Lin}, X., {Wang}, F., {Fan}, X., {et~al.} 2024, \apj, 974, 147,
  \dodoi{10.3847/1538-4357/ad6565}

\bibitem[{{Liu} {et~al.}(2025){Liu}, {Jiang}, {Quataert}, {Greene}, \&
  {Ma}}]{Liu+25}
{Liu}, H., {Jiang}, Y.-F., {Quataert}, E., {Greene}, J.~E., \& {Ma}, Y. 2025,
  arXiv e-prints, arXiv:2507.07190, \dodoi{10.48550/arXiv.2507.07190}

\bibitem[{{Lodato} \& {Natarajan}(2006)}]{Lodato+06}
{Lodato}, G., \& {Natarajan}, P. 2006, \mnras, 371, 1813,
  \dodoi{10.1111/j.1365-2966.2006.10801.x}

\bibitem[{{Lodato} \& {Natarajan}(2007)}]{Lodato+07}
---. 2007, \mnras, 377, L64, \dodoi{10.1111/j.1745-3933.2007.00304.x}

\bibitem[{{Madau} \& {Dickinson}(2014)}]{Madau+14}
{Madau}, P., \& {Dickinson}, M. 2014, \araa, 52, 415,
  \dodoi{10.1146/annurev-astro-081811-125615}

\bibitem[{{Madau} \& {Rees}(2001)}]{Madau2001}
{Madau}, P., \& {Rees}, M.~J. 2001, \apjl, 551, L27, \dodoi{10.1086/319848}

\bibitem[{{Magorrian} {et~al.}(1998){Magorrian}, {Tremaine}, {Richstone},
  {Bender}, {Bower}, {Dressler}, {Faber}, {Gebhardt}, {Green}, {Grillmair},
  {Kormendy}, \& {Lauer}}]{Magorrian+98}
{Magorrian}, J., {Tremaine}, S., {Richstone}, D., {et~al.} 1998, \aj, 115,
  2285, \dodoi{10.1086/300353}

\bibitem[{{Maiolino} {et~al.}(2024{\natexlab{a}}){Maiolino}, {Scholtz},
  {Curtis-Lake}, {Carniani}, {Baker}, {de Graaff}, {Tacchella}, {{\"U}bler},
  {D'Eugenio}, {Witstok}, {Curti}, {Arribas}, {Bunker}, {Charlot},
  {Chevallard}, {Eisenstein}, {Egami}, {Ji}, {Jones}, {Lyu}, {Rawle},
  {Robertson}, {Rujopakarn}, {Perna}, {Sun}, {Venturi}, {Williams}, \&
  {Willott}}]{Maiolino+24}
{Maiolino}, R., {Scholtz}, J., {Curtis-Lake}, E., {et~al.} 2024{\natexlab{a}},
  \aap, 691, A145, \dodoi{10.1051/0004-6361/202347640}

\bibitem[{{Maiolino} {et~al.}(2024{\natexlab{b}}){Maiolino}, {Scholtz},
  {Witstok}, {Carniani}, {D'Eugenio}, {de Graaff}, {{\"U}bler}, {Tacchella},
  {Curtis-Lake}, {Arribas}, {Bunker}, {Charlot}, {Chevallard}, {Curti},
  {Looser}, {Maseda}, {Rawle}, {Rodr{\'\i}guez del Pino}, {Willott}, {Egami},
  {Eisenstein}, {Hainline}, {Robertson}, {Williams}, {Willmer}, {Baker},
  {Boyett}, {DeCoursey}, {Fabian}, {Helton}, {Ji}, {Jones}, {Kumari},
  {Laporte}, {Nelson}, {Perna}, {Sandles}, {Shivaei}, \& {Sun}}]{Maiolino+24b}
{Maiolino}, R., {Scholtz}, J., {Witstok}, J., {et~al.} 2024{\natexlab{b}},
  \nat, 627, 59, \dodoi{10.1038/s41586-024-07052-5}

\bibitem[{{Maiolino} {et~al.}(2025){Maiolino}, {Uebler}, {D'Eugenio},
  {Scholtz}, {Juodzbalis}, {Ji}, {Perna}, {Bromm}, {Dayal}, {Koudmani}, {Liu},
  {Schneider}, {Sijacki}, {Valiante}, {Trinca}, {Zhang}, {Volonteri},
  {Inayoshi}, {Carniani}, {Nakajima}, {Isobe}, {Witstok}, {Jones}, {Tacchella},
  {Arribas}, {Bunker}, {Cataldi}, {Charlot}, {Cresci}, {Curti}, {Fabian},
  {Katz}, {Kumari}, {Laporte}, {Mazzolari}, {Robertson}, {Sun}, {Rodriguez Del
  Pino}, \& {Venturi}}]{Maiolino+25}
{Maiolino}, R., {Uebler}, H., {D'Eugenio}, F., {et~al.} 2025, arXiv e-prints,
  arXiv:2505.22567, \dodoi{10.48550/arXiv.2505.22567}

\bibitem[{{Matsuoka} {et~al.}(2016){Matsuoka}, {Onoue}, {Kashikawa}, {Iwasawa},
  {Strauss}, {Nagao}, {Imanishi}, {Niida}, {Toba}, {Akiyama}, {Asami}, {Bosch},
  {Foucaud}, {Furusawa}, {Goto}, {Gunn}, {Harikane}, {Ikeda}, {Kawaguchi},
  {Kikuta}, {Komiyama}, {Lupton}, {Minezaki}, {Miyazaki}, {Morokuma},
  {Murayama}, {Nishizawa}, {Ono}, {Ouchi}, {Price}, {Sameshima}, {Silverman},
  {Sugiyama}, {Tait}, {Takada}, {Takata}, {Tanaka}, {Tang}, \&
  {Utsumi}}]{Matsuoka+16}
{Matsuoka}, Y., {Onoue}, M., {Kashikawa}, N., {et~al.} 2016, \apj, 828, 26,
  \dodoi{10.3847/0004-637X/828/1/26}

\bibitem[{{Matsuoka} {et~al.}(2018{\natexlab{a}}){Matsuoka}, {Onoue},
  {Kashikawa}, {Iwasawa}, {Strauss}, {Nagao}, {Imanishi}, {Lee}, {Akiyama},
  {Asami}, {Bosch}, {Foucaud}, {Furusawa}, {Goto}, {Gunn}, {Harikane}, {Ikeda},
  {Izumi}, {Kawaguchi}, {Kikuta}, {Kohno}, {Komiyama}, {Lupton}, {Minezaki},
  {Miyazaki}, {Morokuma}, {Murayama}, {Niida}, {Nishizawa}, {Oguri}, {Ono},
  {Ouchi}, {Price}, {Sameshima}, {Schulze}, {Shirakata}, {Silverman},
  {Sugiyama}, {Tait}, {Takada}, {Takata}, {Tanaka}, {Tang}, {Toba}, {Utsumi},
  \& {Wang}}]{Matsuoka+18a}
---. 2018{\natexlab{a}}, \pasj, 70, S35, \dodoi{10.1093/pasj/psx046}

\bibitem[{{Matsuoka} {et~al.}(2018{\natexlab{b}}){Matsuoka}, {Strauss},
  {Kashikawa}, {Onoue}, {Iwasawa}, {Tang}, {Lee}, {Imanishi}, {Nagao},
  {Akiyama}, {Asami}, {Bosch}, {Furusawa}, {Goto}, {Gunn}, {Harikane}, {Ikeda},
  {Izumi}, {Kawaguchi}, {Kato}, {Kikuta}, {Kohno}, {Komiyama}, {Lupton},
  {Minezaki}, {Miyazaki}, {Murayama}, {Niida}, {Nishizawa}, {Noboriguchi},
  {Oguri}, {Ono}, {Ouchi}, {Price}, {Sameshima}, {Schulze}, {Shirakata},
  {Silverman}, {Sugiyama}, {Tait}, {Takada}, {Takata}, {Tanaka}, {Toba},
  {Utsumi}, {Wang}, \& {Yamashita}}]{Matsuoka+18b}
{Matsuoka}, Y., {Strauss}, M.~A., {Kashikawa}, N., {et~al.} 2018{\natexlab{b}},
  \apj, 869, 150, \dodoi{10.3847/1538-4357/aaee7a}

\bibitem[{{Matthee} {et~al.}(2023){Matthee}, {Mackenzie}, {Simcoe}, {Kashino},
  {Lilly}, {Bordoloi}, \& {Eilers}}]{Matthee+23}
{Matthee}, J., {Mackenzie}, R., {Simcoe}, R.~A., {et~al.} 2023, \apj, 950, 67,
  \dodoi{10.3847/1538-4357/acc846}

\bibitem[{{Matthee} {et~al.}(2024){Matthee}, {Naidu}, {Brammer}, {Chisholm},
  {Eilers}, {Goulding}, {Greene}, {Kashino}, {Labbe}, {Lilly}, {Mackenzie},
  {Oesch}, {Weibel}, {Wuyts}, {Xiao}, {Bordoloi}, {Bouwens}, {van Dokkum},
  {Illingworth}, {Kramarenko}, {Maseda}, {Mason}, {Meyer}, {Nelson}, {Reddy},
  {Shivaei}, {Simcoe}, \& {Yue}}]{Matthee+24}
{Matthee}, J., {Naidu}, R.~P., {Brammer}, G., {et~al.} 2024, \apj, 963, 129,
  \dodoi{10.3847/1538-4357/ad2345}

\bibitem[{{Meyer} {et~al.}(2024){Meyer}, {Oesch}, {Giovinazzo}, {Weibel},
  {Brammer}, {Matthee}, {Naidu}, {Bouwens}, {Chisholm}, {Covelo-Paz},
  {Fudamoto}, {Maseda}, {Nelson}, {Shivaei}, {Xiao}, {Herard-Demanche},
  {Illingworth}, {Kerutt}, {Kramarenko}, {Labbe}, {Leonova}, {Magee},
  {Matharu}, {Prieto Lyon}, {Reddy}, {Schaerer}, {Shapley}, {Stefanon},
  {Wozniak}, \& {Wuyts}}]{Meyer+24}
{Meyer}, R.~A., {Oesch}, P.~A., {Giovinazzo}, E., {et~al.} 2024, \mnras, 535,
  1067, \dodoi{10.1093/mnras/stae2353}

\bibitem[{{Mortlock} {et~al.}(2011){Mortlock}, {Warren}, {Venemans}, {Patel},
  {Hewett}, {McMahon}, {Simpson}, {Theuns}, {Gonz{\'a}les-Solares}, {Adamson},
  {Dye}, {Hambly}, {Hirst}, {Irwin}, {Kuiper}, {Lawrence}, \&
  {R{\"o}ttgering}}]{Mortlock+11}
{Mortlock}, D.~J., {Warren}, S.~J., {Venemans}, B.~P., {et~al.} 2011, \nat,
  474, 616, \dodoi{10.1038/nature10159}

\bibitem[{{Naidu} {et~al.}(2025){Naidu}, {Matthee}, {Katz}, {de Graaff},
  {Oesch}, {Smith}, {Greene}, {Brammer}, {Weibel}, {Hviding}, {Chisholm},
  {Labb\textbackslash'e}, {Simcoe}, {Witten}, {Atek}, {Baggen}, {Belli},
  {Bezanson}, {Boogaard}, {Bose}, {Covelo-Paz}, {Dayal}, {Fudamoto}, {Furtak},
  {Giovinazzo}, {Goulding}, {Gronke}, {Heintz}, {Hirschmann}, {Illingworth},
  {Inoue}, {Johnson}, {Leja}, {Leonova}, {McConachie}, {Maseda}, {Natarajan},
  {Nelson}, {Setton}, {Shivaei}, {Sobral}, {Stefanon}, {Tacchella}, {Toft},
  {Torralba}, {van Dokkum}, {van der Wel}, {Volonteri}, {Walter}, {Wang}, \&
  {Watson}}]{Naidu+25}
{Naidu}, R.~P., {Matthee}, J., {Katz}, H., {et~al.} 2025, arXiv e-prints,
  arXiv:2503.16596, \dodoi{10.48550/arXiv.2503.16596}

\bibitem[{{Natarajan}(2014)}]{2014ReviewSEEDS}
{Natarajan}, P. 2014, General Relativity and Gravitation, 46, 1702,
  \dodoi{10.1007/s10714-014-1702-6}

\bibitem[{{Natarajan} {et~al.}(2017){Natarajan}, {Pacucci}, {Ferrara},
  {Agarwal}, {Ricarte}, {Zackrisson}, \& {Cappelluti}}]{2017ApJ...838..117N}
{Natarajan}, P., {Pacucci}, F., {Ferrara}, A., {et~al.} 2017, \apj, 838, 117,
  \dodoi{10.3847/1538-4357/aa6330}

\bibitem[{{Newville} {et~al.}(2014){Newville}, {Stensitzki}, {Allen}, \&
  {Ingargiola}}]{Newville+14}
{Newville}, M., {Stensitzki}, T., {Allen}, D.~B., \& {Ingargiola}, A. 2014,
  {LMFIT: Non-Linear Least-Square Minimization and Curve-Fitting for Python},
  0.8.0,  Zenodo, \dodoi{10.5281/zenodo.11813}

\bibitem[{{Oesch} {et~al.}(2023){Oesch}, {Brammer}, {Naidu}, {Bouwens},
  {Chisholm}, {Illingworth}, {Matthee}, {Nelson}, {Qin}, {Reddy}, {Shapley},
  {Shivaei}, {van Dokkum}, {Weibel}, {Whitaker}, {Wuyts}, {Covelo-Paz},
  {Endsley}, {Fudamoto}, {Giovinazzo}, {Herard-Demanche}, {Kerutt},
  {Kramarenko}, {Labbe}, {Leonova}, {Lin}, {Magee}, {Marchesini}, {Maseda},
  {Mason}, {Matharu}, {Meyer}, {Neufeld}, {Prieto Lyon}, {Schaerer}, {Sharma},
  {Shuntov}, {Smit}, {Stefanon}, {Wyithe}, \& {Xiao}}]{Oesch+23}
{Oesch}, P.~A., {Brammer}, G., {Naidu}, R.~P., {et~al.} 2023, \mnras, 525,
  2864, \dodoi{10.1093/mnras/stad2411}

\bibitem[{{Oke} \& {Gunn}(1983)}]{Oke+83}
{Oke}, J.~B., \& {Gunn}, J.~E. 1983, \apj, 266, 713, \dodoi{10.1086/160817}

\bibitem[{{Omukai} {et~al.}(2008){Omukai}, {Schneider}, \&
  {Haiman}}]{Omukai+08}
{Omukai}, K., {Schneider}, R., \& {Haiman}, Z. 2008, \apj, 686, 801,
  \dodoi{10.1086/591636}

\bibitem[{{Onoue} {et~al.}(2017){Onoue}, {Kashikawa}, {Willott}, {Hibon}, {Im},
  {Furusawa}, {Harikane}, {Imanishi}, {Ishikawa}, {Kikuta}, {Matsuoka},
  {Nagao}, {Niino}, {Ono}, {Ouchi}, {Tanaka}, {Tang}, {Toshikawa}, \&
  {Uchiyama}}]{Onoue+17}
{Onoue}, M., {Kashikawa}, N., {Willott}, C.~J., {et~al.} 2017, \apjl, 847, L15,
  \dodoi{10.3847/2041-8213/aa8cc6}

\bibitem[{{Onoue} {et~al.}(2019){Onoue}, {Kashikawa}, {Matsuoka}, {Kato},
  {Izumi}, {Nagao}, {Strauss}, {Harikane}, {Imanishi}, {Ito}, {Iwasawa},
  {Kawaguchi}, {Lee}, {Noboriguchi}, {Suh}, {Tanaka}, \& {Toba}}]{Onoue+19}
{Onoue}, M., {Kashikawa}, N., {Matsuoka}, Y., {et~al.} 2019, \apj, 880, 77,
  \dodoi{10.3847/1538-4357/ab29e9}

\bibitem[{{Pacucci} {et~al.}(2015){Pacucci}, {Ferrara}, {Volonteri}, \&
  {Dubus}}]{Pacucci_DCBH2015}
{Pacucci}, F., {Ferrara}, A., {Volonteri}, M., \& {Dubus}, G. 2015, \mnras,
  454, 3771, \dodoi{10.1093/mnras/stv2196}

\bibitem[{{Pacucci} {et~al.}(2023){Pacucci}, {Nguyen}, {Carniani}, {Maiolino},
  \& {Fan}}]{Pacucci+23}
{Pacucci}, F., {Nguyen}, B., {Carniani}, S., {Maiolino}, R., \& {Fan}, X. 2023,
  \apjl, 957, L3, \dodoi{10.3847/2041-8213/ad0158}

\bibitem[{{Price} {et~al.}(2025){Price}, {Bezanson}, {Labbe}, {Furtak}, {de
  Graaff}, {Greene}, {Kokorev}, {Setton}, {Suess}, {Brammer}, {Cutler}, {Leja},
  {Pan}, {Wang}, {Weaver}, {Whitaker}, {Atek}, {Burgasser}, {Chemerynska},
  {Dayal}, {Feldmann}, {F{\"o}rster Schreiber}, {Fudamoto}, {Fujimoto},
  {Glazebrook}, {Goulding}, {Khullar}, {Kriek}, {Marchesini}, {Maseda},
  {Miller}, {Muzzin}, {Nanayakkara}, {Nelson}, {Oesch}, {Shipley}, {Smit},
  {Taylor}, {Dokkum}, {Williams}, \& {Zitrin}}]{Price+25}
{Price}, S.~H., {Bezanson}, R., {Labbe}, I., {et~al.} 2025, \apj, 982, 51,
  \dodoi{10.3847/1538-4357/adaec1}

\bibitem[{{Qin} {et~al.}(2025){Qin}, {Kumar}, {Natarajan}, \&
  {Weiner}}]{2025arXiv250613858Q}
{Qin}, W., {Kumar}, S., {Natarajan}, P., \& {Weiner}, N. 2025, arXiv e-prints,
  arXiv:2506.13858, \dodoi{10.48550/arXiv.2506.13858}

\bibitem[{{Reines} \& {Volonteri}(2015)}]{Reines+15}
{Reines}, A.~E., \& {Volonteri}, M. 2015, \apj, 813, 82,
  \dodoi{10.1088/0004-637X/813/2/82}

\bibitem[{{Reinoso} {et~al.}(2023){Reinoso}, {Klessen}, {Schleicher}, {Glover},
  \& {Solar}}]{Reinoso2023}
{Reinoso}, B., {Klessen}, R.~S., {Schleicher}, D., {Glover}, S. C.~O., \&
  {Solar}, P. 2023, \mnras, 521, 3553, \dodoi{10.1093/mnras/stad790}

\bibitem[{{Ricarte} \& {Natarajan}(2018)}]{Ricarte+18}
{Ricarte}, A., \& {Natarajan}, P. 2018, \mnras, 481, 3278,
  \dodoi{10.1093/mnras/sty2448}

\bibitem[{{Richards} {et~al.}(2006){Richards}, {Lacy}, {Storrie-Lombardi},
  {Hall}, {Gallagher}, {Hines}, {Fan}, {Papovich}, {Vanden Berk}, {Trammell},
  {Schneider}, {Vestergaard}, {York}, {Jester}, {Anderson}, {Budav{\'a}ri}, \&
  {Szalay}}]{Richards+06}
{Richards}, G.~T., {Lacy}, M., {Storrie-Lombardi}, L.~J., {et~al.} 2006, \apjs,
  166, 470, \dodoi{10.1086/506525}

\bibitem[{{Romano} {et~al.}(2021){Romano}, {Cassata}, {Morselli}, {Jones},
  {Ginolfi}, {Zanella}, {B{\'e}thermin}, {Capak}, {Faisst}, {Le F{\`e}vre},
  {Schaerer}, {Silverman}, {Yan}, {Bardelli}, {Boquien}, {Cimatti},
  {Dessauges-Zavadsky}, {Enia}, {Fujimoto}, {Gruppioni}, {Hathi}, {Ibar},
  {Koekemoer}, {Lemaux}, {Rodighiero}, {Vergani}, {Zamorani}, \&
  {Zucca}}]{Romano+21}
{Romano}, M., {Cassata}, P., {Morselli}, L., {et~al.} 2021, \aap, 653, A111,
  \dodoi{10.1051/0004-6361/202141306}

\bibitem[{{Rusakov} {et~al.}(2025){Rusakov}, {Watson}, {Nikopoulos}, {Brammer},
  {Gottumukkala}, {Harvey}, {Heintz}, {Nielsen}, {Sim}, {Sneppen}, {Vijayan},
  {Adams}, {Austin}, {Conselice}, {Goolsby}, {Toft}, \& {Witstok}}]{Rusakov+25}
{Rusakov}, V., {Watson}, D., {Nikopoulos}, G.~P., {et~al.} 2025, arXiv
  e-prints, arXiv:2503.16595, \dodoi{10.48550/arXiv.2503.16595}

\bibitem[{{Salviander} {et~al.}(2007){Salviander}, {Shields}, {Gebhardt}, \&
  {Bonning}}]{Salviander+07}
{Salviander}, S., {Shields}, G.~A., {Gebhardt}, K., \& {Bonning}, E.~W. 2007,
  \apj, 662, 131, \dodoi{10.1086/513086}

\bibitem[{{Shen} {et~al.}(2015){Shen}, {Greene}, {Ho}, {Brandt}, {Denney},
  {Horne}, {Jiang}, {Kochanek}, {McGreer}, {Merloni}, {Peterson}, {Petitjean},
  {Schneider}, {Schulze}, {Strauss}, {Tao}, {Trump}, {Pan}, \&
  {Bizyaev}}]{Shen+15}
{Shen}, Y., {Greene}, J.~E., {Ho}, L.~C., {et~al.} 2015, \apj, 805, 96,
  \dodoi{10.1088/0004-637X/805/2/96}

\bibitem[{{Shen} {et~al.}(2019){Shen}, {Wu}, {Jiang}, {Ba{\~n}ados}, {Fan},
  {Ho}, {Riechers}, {Strauss}, {Venemans}, {Vestergaard}, {Walter}, {Wang},
  {Willott}, {Wu}, \& {Yang}}]{Shen+19}
{Shen}, Y., {Wu}, J., {Jiang}, L., {et~al.} 2019, \apj, 873, 35,
  \dodoi{10.3847/1538-4357/ab03d9}

\bibitem[{{Shen} {et~al.}(2024){Shen}, {Grier}, {Horne}, {Stone}, {Li}, {Yang},
  {Homayouni}, {Trump}, {Anderson}, {Brandt}, {Hall}, {Ho}, {Jiang},
  {Petitjean}, {Schneider}, {Tao}, {Donnan}, {AlSayyad}, {Bershady}, {Blanton},
  {Bizyaev}, {Bundy}, {Chen}, {Davis}, {Dawson}, {Fan}, {Greene},
  {Gr{\"o}ller}, {Guo}, {Ibarra-Medel}, {Jiang}, {Keenan}, {Kollmeier},
  {Lejoly}, {Li}, {de la Macorra}, {Moe}, {Nie}, {Rossi}, {Smith}, {Tee},
  {Weijmans}, {Xu}, {Yue}, {Zhou}, {Zhou}, \& {Zou}}]{Shen+24}
{Shen}, Y., {Grier}, C.~J., {Horne}, K., {et~al.} 2024, \apjs, 272, 26,
  \dodoi{10.3847/1538-4365/ad3936}

\bibitem[{{Silverman} {et~al.}(2025){Silverman}, {Li}, {Ding}, {Onoue},
  {Strauss}, {Matsuoka}, {Izumi}, {Jahnke}, {Treu}, {Volonteri}, {Phillips},
  {Andika}, {Aoki}, {Arita}, {Baba}, {Bosman}, {Eilers}, {Fan}, {Fujimoto},
  {Habouzit}, {Haiman}, {Imanishi}, {Inayoshi}, {Iwasawa}, {Kashikawa},
  {Kawaguchi}, {Lee}, {Lupi}, {Nagao}, {Schindler}, {Schramm}, {Shimasaku},
  {Toba}, {Trakhtenbrot}, {Umehata}, {Vestergaard}, {Walter}, {Wang}, \&
  {Yang}}]{Silverman+2025}
{Silverman}, J., {Li}, J., {Ding}, X., {et~al.} 2025, arXiv e-prints,
  arXiv:2507.23066, \dodoi{10.48550/arXiv.2507.23066}

\bibitem[{{Singh} {et~al.}(2023){Singh}, {Monaco}, \& {Tan}}]{Singh+23}
{Singh}, J., {Monaco}, P., \& {Tan}, J.~C. 2023, \mnras, 525, 969,
  \dodoi{10.1093/mnras/stad2346}

\bibitem[{{Smith} {et~al.}(2017){Smith}, {Becerra}, {Bromm}, \&
  {Hernquist}}]{Smith_DCBH2017}
{Smith}, A., {Becerra}, F., {Bromm}, V., \& {Hernquist}, L. 2017, \mnras, 472,
  205, \dodoi{10.1093/mnras/stx1993}

\bibitem[{{Smith} \& {Bromm}(2019)}]{Smith+19}
{Smith}, A., \& {Bromm}, V. 2019, Contemporary Physics, 60, 111,
  \dodoi{10.1080/00107514.2019.1615715}

\bibitem[{{Smith} {et~al.}(2018){Smith}, {Regan}, {Downes}, {Norman}, {O'Shea},
  \& {Wise}}]{Smith+18}
{Smith}, B.~D., {Regan}, J.~A., {Downes}, T.~P., {et~al.} 2018, \mnras, 480,
  3762, \dodoi{10.1093/mnras/sty2103}

\bibitem[{{Song} {et~al.}(2016){Song}, {Finkelstein}, {Ashby}, {Grazian}, {Lu},
  {Papovich}, {Salmon}, {Somerville}, {Dickinson}, {Duncan}, {Faber}, {Fazio},
  {Ferguson}, {Fontana}, {Guo}, {Hathi}, {Lee}, {Merlin}, \&
  {Willner}}]{Song+16}
{Song}, M., {Finkelstein}, S.~L., {Ashby}, M. L.~N., {et~al.} 2016, \apj, 825,
  5, \dodoi{10.3847/0004-637X/825/1/5}

\bibitem[{{Spinoso} {et~al.}(2023){Spinoso}, {Bonoli}, {Valiante}, {Schneider},
  \& {Izquierdo-Villalba}}]{Spinoso+23}
{Spinoso}, D., {Bonoli}, S., {Valiante}, R., {Schneider}, R., \&
  {Izquierdo-Villalba}, D. 2023, \mnras, 518, 4672,
  \dodoi{10.1093/mnras/stac3169}

\bibitem[{{Stalevski} {et~al.}(2016){Stalevski}, {Ricci}, {Ueda}, {Lira},
  {Fritz}, \& {Baes}}]{Stalevski+16}
{Stalevski}, M., {Ricci}, C., {Ueda}, Y., {et~al.} 2016, \mnras, 458, 2288,
  \dodoi{10.1093/mnras/stw444}

\bibitem[{{Stefanon} {et~al.}(2021){Stefanon}, {Bouwens}, {Labb{\'e}},
  {Illingworth}, {Gonzalez}, \& {Oesch}}]{Stefanon+21}
{Stefanon}, M., {Bouwens}, R.~J., {Labb{\'e}}, I., {et~al.} 2021, \apj, 922,
  29, \dodoi{10.3847/1538-4357/ac1bb6}

\bibitem[{{Stone} {et~al.}(2024){Stone}, {Lyu}, {Rieke}, {Alberts}, \&
  {Hainline}}]{Stone+24}
{Stone}, M.~A., {Lyu}, J., {Rieke}, G.~H., {Alberts}, S., \& {Hainline}, K.~N.
  2024, \apj, 964, 90, \dodoi{10.3847/1538-4357/ad2a57}

\bibitem[{{Storey} \& {Zeippen}(2000)}]{Storey+00}
{Storey}, P.~J., \& {Zeippen}, C.~J. 2000, \mnras, 312, 813,
  \dodoi{10.1046/j.1365-8711.2000.03184.x}

\bibitem[{{Tanaka} \& {Haiman}(2009)}]{Tanaka+09}
{Tanaka}, T., \& {Haiman}, Z. 2009, \apj, 696, 1798,
  \dodoi{10.1088/0004-637X/696/2/1798}

\bibitem[{{Tanaka} {et~al.}(2025){Tanaka}, {Silverman}, {Ding}, {Jahnke},
  {Trakhtenbrot}, {Lambrides}, {Onoue}, {Andika}, {Bongiorno}, {Faisst},
  {Gillman}, {Hayward}, {Hirschmann}, {Koekemoer}, {Kokorev}, {Liu}, {Magdis},
  {Renzini}, {Casey}, {Drakos}, {Franco}, {Gozaliasl}, {Kartaltepe}, {Liu},
  {McCracken}, {Rhodes}, {Robertson}, \& {Toft}}]{Tanaka+25}
{Tanaka}, T.~S., {Silverman}, J.~D., {Ding}, X., {et~al.} 2025, \apj, 979, 215,
  \dodoi{10.3847/1538-4357/ad9d0a}

\bibitem[{{Taylor} {et~al.}(2025){Taylor}, {Finkelstein}, {Kocevski}, {Jeon},
  {Bromm}, {Amor{\'\i}n}, {Arrabal Haro}, {Backhaus}, {Bagley}, {Banados},
  {Bhatawdekar}, {Brooks}, {Calabr{\`o}}, {Ch{\'a}vez Ortiz}, {Cheng}, {Cleri},
  {Cole}, {Davis}, {Dickinson}, {Donnan}, {Dunlop}, {Ellis}, {Fern{\'a}ndez},
  {Fontana}, {Fujimoto}, {Giavalisco}, {Grazian}, {Guo}, {Hathi}, {Holwerda},
  {Hirschmann}, {Inayoshi}, {Kartaltepe}, {Khusanova}, {Koekemoer}, {Kokorev},
  {Larson}, {Leung}, {Lucas}, {McLeod}, {Napolitano}, {Onoue}, {Pacucci},
  {Papovich}, {P{\'e}rez-Gonz{\'a}lez}, {Pirzkal}, {Somerville}, {Trump},
  {Wilkins}, {Yung}, \& {Zhang}}]{Taylor+25}
{Taylor}, A.~J., {Finkelstein}, S.~L., {Kocevski}, D.~D., {et~al.} 2025, \apj,
  986, 165, \dodoi{10.3847/1538-4357/add15b}

\bibitem[{{Trakhtenbrot} {et~al.}(2011){Trakhtenbrot}, {Netzer}, {Lira}, \&
  {Shemmer}}]{Trakhtenbrot+11}
{Trakhtenbrot}, B., {Netzer}, H., {Lira}, P., \& {Shemmer}, O. 2011, \apj, 730,
  7, \dodoi{10.1088/0004-637X/730/1/7}

\bibitem[{{Trenti} \& {Stiavelli}(2008)}]{Trenti+08}
{Trenti}, M., \& {Stiavelli}, M. 2008, \apj, 676, 767, \dodoi{10.1086/528674}

\bibitem[{{{\"U}bler} {et~al.}(2023){{\"U}bler}, {Maiolino}, {Curtis-Lake},
  {P{\'e}rez-Gonz{\'a}lez}, {Curti}, {Perna}, {Arribas}, {Charlot}, {Marshall},
  {D'Eugenio}, {Scholtz}, {Bunker}, {Carniani}, {Ferruit}, {Jakobsen}, {Rix},
  {Rodr{\'\i}guez Del Pino}, {Willott}, {Boeker}, {Cresci}, {Jones}, {Kumari},
  \& {Rawle}}]{Ubler+23}
{{\"U}bler}, H., {Maiolino}, R., {Curtis-Lake}, E., {et~al.} 2023, \aap, 677,
  A145, \dodoi{10.1051/0004-6361/202346137}

\bibitem[{{Valentino} {et~al.}(2025){Valentino}, {Heintz}, {Brammer}, {Ito},
  {Kokorev}, {Whitaker}, {Gallazzi}, {de Graaff}, {Weibel}, {Frye},
  {Kamieneski}, {Jin}, {Ceverino}, {Faisst}, {Farcy}, {Fujimoto}, {Gillman},
  {Gottumukkala}, {Hamadouche}, {Harrington}, {Hirschmann}, {Jespersen},
  {Kakimoto}, {Kubo}, {Lagos}, {Lee}, {Magdis}, {Man}, {Onodera}, {Rizzo},
  {Shimakawa}, {Setton}, {Tanaka}, {Toft}, {Wu}, \& {Zhu}}]{Valentino+25}
{Valentino}, F., {Heintz}, K.~E., {Brammer}, G., {et~al.} 2025, \aap, 699,
  A358, \dodoi{10.1051/0004-6361/202553908}

\bibitem[{{Volonteri} \& {Begelman}(2010)}]{Volonteri+10}
{Volonteri}, M., \& {Begelman}, M.~C. 2010, \mnras, 409, 1022,
  \dodoi{10.1111/j.1365-2966.2010.17359.x}

\bibitem[{{Wang} {et~al.}(2016){Wang}, {Wu}, {Fan}, {Yang}, {Yi}, {Bian},
  {McGreer}, {Yang}, {Ai}, {Dong}, {Zuo}, {Jiang}, {Green}, {Wang}, {Cai},
  {Wang}, \& {Yue}}]{Wang+16}
{Wang}, F., {Wu}, X.-B., {Fan}, X., {et~al.} 2016, \apj, 819, 24,
  \dodoi{10.3847/0004-637X/819/1/24}

\bibitem[{{Wang} {et~al.}(2019){Wang}, {Yang}, {Fan}, {Wu}, {Yue}, {Li},
  {Bian}, {Jiang}, {Ba{\~n}ados}, {Schindler}, {Findlay}, {Davies}, {Decarli},
  {Farina}, {Green}, {Hennawi}, {Huang}, {Mazzuccheli}, {McGreer}, {Venemans},
  {Walter}, {Dye}, {Lyke}, {Myers}, \& {Nunez}}]{Wang+19}
{Wang}, F., {Yang}, J., {Fan}, X., {et~al.} 2019, \apj, 884, 30,
  \dodoi{10.3847/1538-4357/ab2be5}

\bibitem[{{Wang} {et~al.}(2021){Wang}, {Yang}, {Fan}, {Hennawi}, {Barth},
  {Banados}, {Bian}, {Boutsia}, {Connor}, {Davies}, {Decarli}, {Eilers},
  {Farina}, {Green}, {Jiang}, {Li}, {Mazzucchelli}, {Nanni}, {Schindler},
  {Venemans}, {Walter}, {Wu}, \& {Yue}}]{Wang+21}
---. 2021, \apjl, 907, L1, \dodoi{10.3847/2041-8213/abd8c6}

\bibitem[{{Xu} {et~al.}(2022){Xu}, {Heckman}, {Henry}, {Berg}, {Chisholm},
  {James}, {Martin}, {Stark}, {Aloisi}, {Amor{\'\i}n}, {Arellano-C{\'o}rdova},
  {Bordoloi}, {Charlot}, {Chen}, {Hayes}, {Mingozzi}, {Sugahara}, {Kewley},
  {Ouchi}, {Scarlata}, \& {Steidel}}]{Xu+22}
{Xu}, X., {Heckman}, T., {Henry}, A., {et~al.} 2022, \apj, 933, 222,
  \dodoi{10.3847/1538-4357/ac6d56}

\bibitem[{{Yang} {et~al.}(2022){Yang}, {Boquien}, {Brandt}, {Buat},
  {Burgarella}, {Ciesla}, {Lehmer}, {Ma{\l}ek}, {Mountrichas}, {Papovich},
  {Pons}, {Stalevski}, {Theul{\'e}}, \& {Zhu}}]{Yang+22}
{Yang}, G., {Boquien}, M., {Brandt}, W.~N., {et~al.} 2022, \apj, 927, 192,
  \dodoi{10.3847/1538-4357/ac4971}

\bibitem[{{Yang} {et~al.}(2019){Yang}, {Wang}, {Fan}, {Yue}, {Wu}, {Li},
  {Bian}, {Jiang}, {Ba{\~n}ados}, \& {Beletsky}}]{Yang+19}
{Yang}, J., {Wang}, F., {Fan}, X., {et~al.} 2019, \aj, 157, 236,
  \dodoi{10.3847/1538-3881/ab1be1}

\bibitem[{{Yang} {et~al.}(2020){Yang}, {Wang}, {Fan}, {Hennawi}, {Davies},
  {Yue}, {Banados}, {Wu}, {Venemans}, {Barth}, {Bian}, {Boutsia}, {Decarli},
  {Farina}, {Green}, {Jiang}, {Li}, {Mazzucchelli}, \& {Walter}}]{Yang+20}
---. 2020, \apjl, 897, L14, \dodoi{10.3847/2041-8213/ab9c26}

\bibitem[{{Young} {et~al.}(2025){Young}, {Hayes}, {Saldana-Lopez}, {Runnholm},
  {Cammelli}, {Tan}, {Ellis}, {Keller}, {Melinder}, \& {Singh}}]{Young+25}
{Young}, A.~R., {Hayes}, M.~J., {Saldana-Lopez}, A., {et~al.} 2025, arXiv
  e-prints, arXiv:2508.15905, \dodoi{10.48550/arXiv.2508.15905}

\bibitem[{{Yue} {et~al.}(2024{\natexlab{a}}){Yue}, {Eilers}, {Ananna},
  {Panagiotou}, {Kara}, \& {Miyaji}}]{Yue+2024b}
{Yue}, M., {Eilers}, A.-C., {Ananna}, T.~T., {et~al.} 2024{\natexlab{a}},
  \apjl, 974, L26, \dodoi{10.3847/2041-8213/ad7eba}

\bibitem[{{Yue} {et~al.}(2024{\natexlab{b}}){Yue}, {Eilers}, {Simcoe},
  {Mackenzie}, {Matthee}, {Kashino}, {Bordoloi}, {Lilly}, \& {Naidu}}]{Yue+24}
{Yue}, M., {Eilers}, A.-C., {Simcoe}, R.~A., {et~al.} 2024{\natexlab{b}}, \apj,
  966, 176, \dodoi{10.3847/1538-4357/ad3914}

\bibitem[{{Zhang} {et~al.}(2023){Zhang}, {Behroozi}, {Volonteri}, {Silk},
  {Fan}, {Aird}, {Yang}, \& {Hopkins}}]{Zhang+23}
{Zhang}, H., {Behroozi}, P., {Volonteri}, M., {et~al.} 2023, \mnras, 523, L69,
  \dodoi{10.1093/mnrasl/slad060}

\bibitem[{{Zhang} {et~al.}(2025){Zhang}, {Liu}, {Bromm}, {Jeon},
  {Boylan-Kolchin}, \& {K{\"u}hnel}}]{Zhang_PBH2025}
{Zhang}, S., {Liu}, B., {Bromm}, V., {et~al.} 2025, \apj, 987, 185,
  \dodoi{10.3847/1538-4357/adddb4}

\bibitem[{{Zhuang} \& {Ho}(2023)}]{Zhuang+23}
{Zhuang}, M.-Y., \& {Ho}, L.~C. 2023, Nature Astronomy, 7, 1376,
  \dodoi{10.1038/s41550-023-02051-4}

\end{thebibliography}
\bibliographystyle{aasjournal}

\appendix 

\counterwithin{figure}{section}

\section{Completness correction}
\label{app1: incompleteness}

\begin{figure*}
    \centering
    \includegraphics[width=\linewidth]{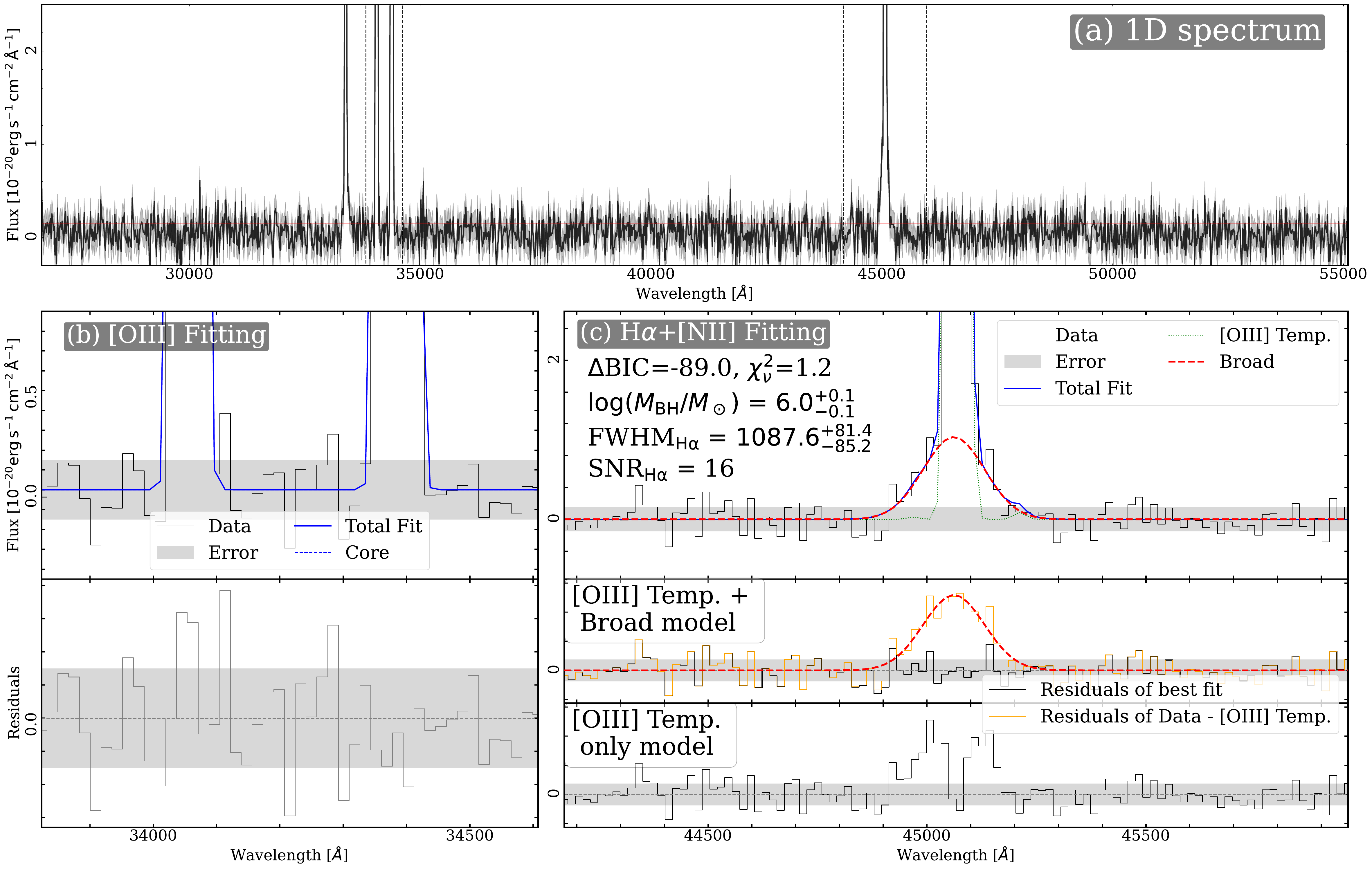}
    \caption{
    \textbf{Illustration of our mock spectrum.}
    Panel (a) displays the mock spectrum of the. In this case the continuum is hard to be detected. Panel (b) presents the fit to the \oiii\ emission line doublet. The mock spectrum and the uncertainty are shown as black histogram and a gray shaded region. The best-fit model is shown as a blue line. Panel (c) shows the fit to the \ha+\nii\ emission complex. Similar to Figure~\ref{fig1: fitting}, the top sub-panel displays the data (black histogram), the total best-fit model (blue line), the broad \ha\ component (red dashed line), and the narrow-line template derived from the \oiii\ profile (green dotted line). The lower two sub-panels show the fit residuals. The middle sub-panel confirms the quality of the total fit, while the bottom sub-panel demonstrates that a model lacking the broad \ha\ component provides a significantly poorer fit to the data. This mock observation was generated for a BH with a mass of $\log M_{\rm BH}/M_\odot=5.99$ and a magnification of $\mu=3.0$, located at a redshift of $z=5.8656$.
    }
\end{figure*}

Resembling the method described in \cite{Taylor+25}, we calculate the line-detection incompleteness through generating the mock spectra. 
In contrast to \cite{Taylor+25}, who synthesized mock spectra by sampling the observational parameter space, our analysis constructs mock spectra by directly sampling the primary parameters of interest: black hole mass ($M_{\rm BH}$) and gravitational lensing magnification ($\mu$). 

\subsection{Mock spectra}
\label{secA1: mock}

Here we introduce the procedure for generating mock spectra from our model parameters. First, the bolometric luminosity ($L_{\rm bol}$) is calculated from the black hole mass, assuming a characteristic Eddington ratio for the source's redshift. The AGNs continuum luminosity at 5100\AA ($L_{\lambda 5100}$), is then derived from $L_{\rm bol}$ via a standard bolometric correction. Subsequently, the luminosity ($L_{\rm H\alpha}$) and Full Width at Half Maximum (FWHM) of the broad \ha\ emission line are estimated using the empirical scaling relations from the local universe presented by \cite{Greene+05}.

For the narrow-line components, the luminosity of the \oiii$\lambda 5007$ line is drawn from a uniform distribution spanning $10^{41}-10^{43}\,\rm erg\,s^{-1}$, consistent with typical values observed at these redshifts \citep[e.g.,][]{Matthee+23}. The luminosities of the narrow \hb\ and \ha\ lines are then derived from the [OIII] luminosity. This is achieved by assuming a typical \oiii/\hb\ ratio and a theoretical Balmer decrement for Case B recombination ($f_{\rm H\alpha}/f_{\rm H\beta}=2.86$), neglecting dust extinction. The FWHMs of all narrow lines are coupled and assigned a single value drawn from a range of 50 to 200\,$\rm km\,s^{-1}$, following the methodology of \cite{Meyer+24}. For simplicity, we do not include a blue-shifted outflow component in the narrow-line profiles.

Finally, the underlying stellar continuum from the host galaxy is modeled as a power law with a spectral slope $\beta$ varying between -1 and -3, centered on the typical value of -2 found in high-redshift \citep{Dunlop+12, Bouwens+14}. The final intrinsic spectrum for each object is constructed by the linear combination of these components: the AGN continuum, the broad and narrow emission lines, and the host galaxy continuum.

To model realistic observational conditions, we introduced instrumental broadening and observational noise to the intrinsic spectra. The JWST Exposure Time Calculator (ETC) engine, \texttt{Pandeia}, is not utilized for this purpose due to potential discrepancies between its implemented Line Spread Function (LSF) and the on-orbit performance, as noted by \cite{deGraaff+2024}.
Instead, we constructed a custom LSF following the methodology outlined in \cite{deGraaff+2024}. For simplicity in this LSF generation, all sources are assumed to be perfectly centered within their respective slitlets. The intrinsic mock spectra are then convolved with this LSF to simulate the effect of instrumental broadening.
Finally, Gaussian noise is injected into the broadened spectra. The standard deviation of this noise is empirically determined from the spectral noise in our own observational data, specifically measured in the spectral regions adjacent to the Balmer emission lines ($\sim 0.10-0.30\times 10^{-20}\rm \,erg\,s^{-1}$). This ensures that the simulated noise levels are representative of those in the actual observations.

Following the method introduced above, we generate a sample of 30,000 black holes with masses uniformly distributed over the range of $10^4-10^7\,M_\odot$, and the gravitational lensing effect uniformly distributed with $0<\log \mu<1.7$. We admitted that we didn't consider the real black mass function and its distribution on the sky. But these conditions don't affect our completeness estimation. 

\subsection{Completeness}
\label{secA2: comp}

\begin{figure}
    \centering
    \includegraphics[width=\linewidth]{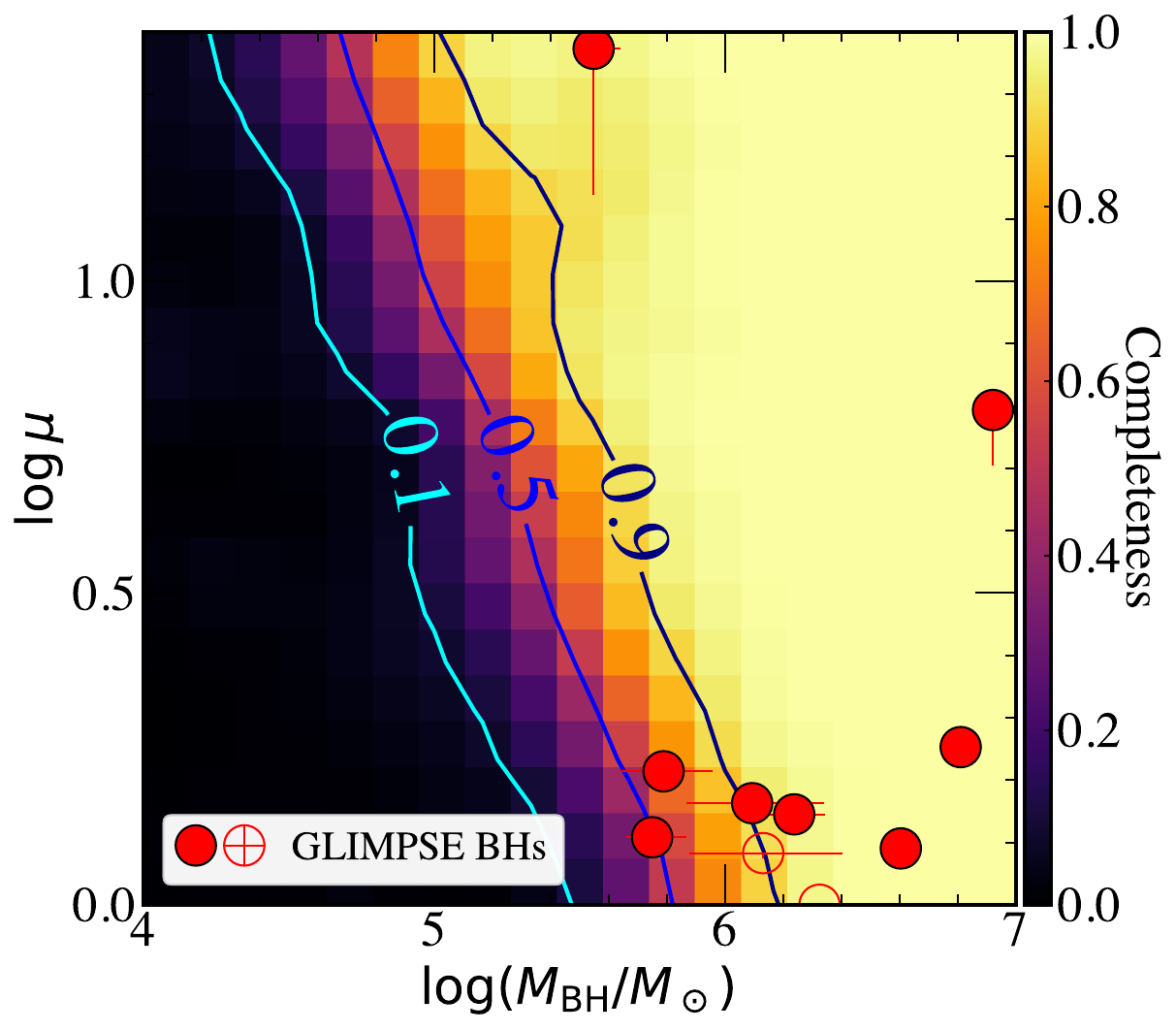}
    \caption{
    \textbf{Completeness of broad \ha\ emission line detection. }
    The colored contours denote 0.1, 0.5, and 0.9. It is noticed that at high-mass regime ($M_{\rm BH}\geq 10^6\,M_\odot$), the broad \ha\ emission line can be detected with high completeness (near unity). In the intermediate mass regime ($10^5<M_{\rm BH}<10^6\,M_\odot$), the broad \ha\ emission lines can be detected with the assistance of gravitational lensing effect. A high magnification ($\mu$) is necessary to detect BHs with lower masses ($M_{\rm BH}<10^5\,M_\odot$). Our sources (red circles) are settled within high-completeness regime.}
    \label{fig5: mock}
\end{figure}

By applying the fitting algorithm and selection criteria detailed in Sections \ref{sec3.1: line} and \ref{sec3.2: selection} to our mock spectral catalog, we evaluate both the detection completeness ($f_{\rm comp}$) and the accuracy of our BH mass measurements. 

We define the detection completeness ($f_{\rm comp}$) within each parameter bin as the fraction of recovered sources:
\begin{eqnarray}
    f_{\rm comp}\equiv \frac{n_{\rm broad}}{n_{\rm tot}}
\end{eqnarray}
where $n_{\rm broad}$ is the number of sources successfully identified as broad-line emitters using our selection method, and n $n_{\rm tot}$ is the total number of input sources within that same bin.

These results are presented as a function of BH mass and magnification in Figure \ref{fig5: mock}. 
The analysis of the mock observations reveals that the deetection completeness is a strong function of both $M_{\rm BH}$ and $\mu$.
For high-mass BHs with $M_{\rm BH}\geq 10^6\,M_\odot$, the instrumental sensitivity of our survey is sufficient to achieve a detection completeness approaching unity. This high rate of recovery is largely independent of the magnification factor, indicating that nearly all active BHs in this mass range are successfully identified.
The incompleteness becomes significant at lower masses. The $f_{\rm comp}$ first declines to approximately 0.90 at $M_{\rm BH}=10^6\,M_\odot$ for unmagnified sources ($\mu=1$). It further decreases to 0.5 at $M_{\rm BH}=10^{5.6}\,M_\odot$, which implies that our observational strategy can identify only half of the AGN population with this BH mass without lensing assistance. For the lowest masses probed ($M_{\rm BH}\leq 10^{5.3}\,M_\odot$), the detection of a broad \ha\ line is contingent upon significant gravitational lensing, requiring a magnification of $\mu>2$ for a source to meet our selection criteria. 

This mock observational underscores the significant challenge of detecting BHs with $M_{\rm BH}\lesssim 10^5\,M_\odot$. Due to the inherent faintness of the broad \ha\ line, these sources can be detected only with strong lensing effect ($\mu > 10$). Moreover, the FWHM of their broad \ha\ lines is expected to be relatively narrow ($\rm FWHM \lesssim 500\,km\,s^{-1}$). This creates a potential degeneracy with the galaxy's spectra (or potential outflow). This degeneracy makes it difficult to unambiguously distinguish the AGN signature. Consequently, these factors highlight the considerable challenges and risks associated with identifying the lowest-mass BHs, even with the advanced capabilities of the JWST.

By overlaying our detected sources onto the completeness map (red circles, Figure \ref{fig5: mock}), we demonstrate that our sample is situated within the high-completeness regime. This positioning provides strong evidence for the reliability of our identifications and indicates a low probability of misidentification within our final sample.

\subsection{Accuracy}
\label{secA3: accuracy}

\begin{figure}
    \centering
    \includegraphics[width=\linewidth]{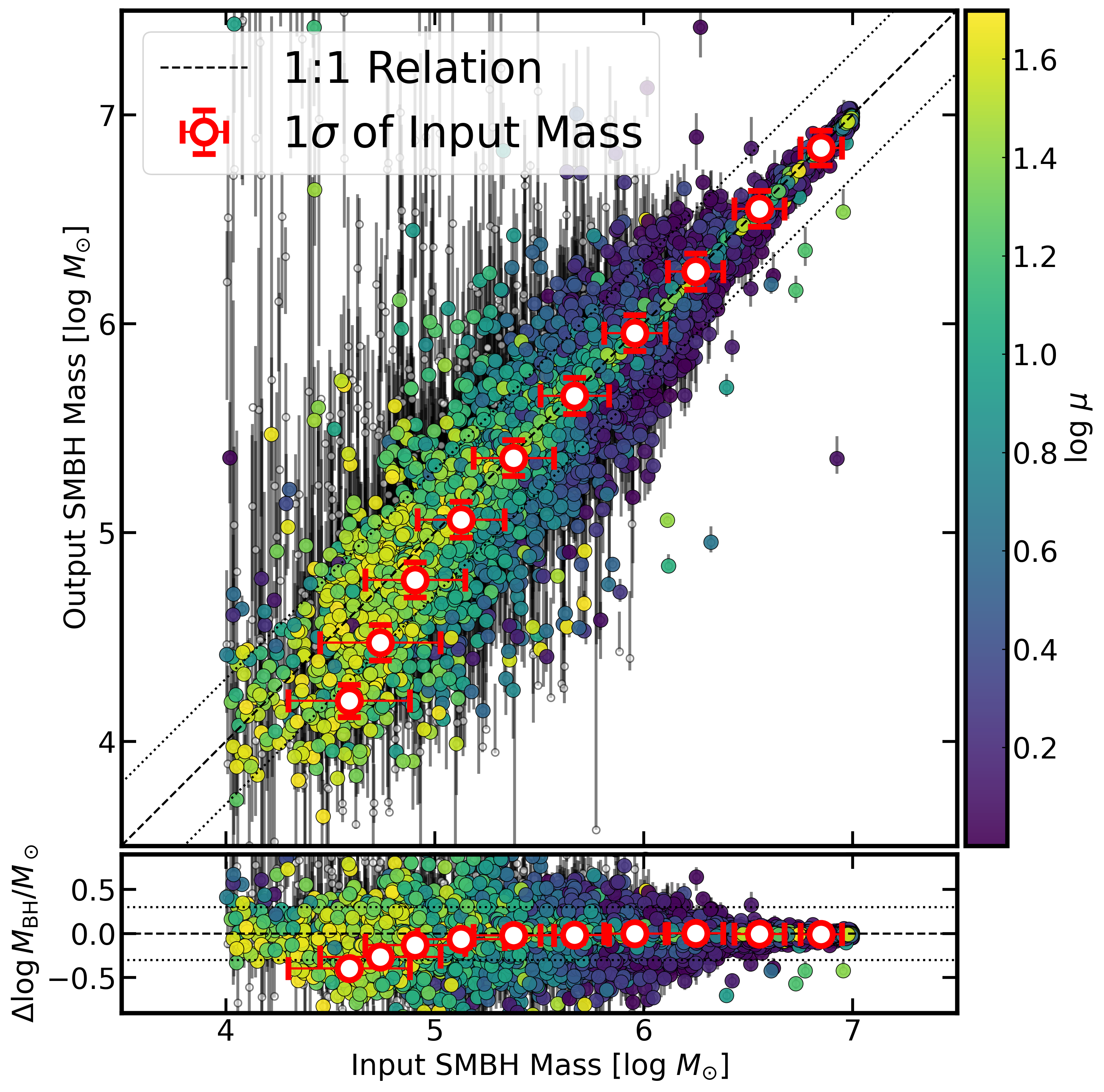}
    \caption{
    \textbf{Accuracy of our BH mass measurement.}
    \textbf{Top:} output BH masses compared to the input BH masses. 
    \textbf{Bottom:} the difference between output and input BH masses ($\Delta \log M_{\rm BH}=\log M_{\rm BH,output}-\log M_{\rm BH, input}$) compared to the input BH masses.
    Colored data points denote sources with $\rm SNR_{\rm H\alpha,broad}>5$ while gray circles presenting those with lower SNR. 
    The color of each data point denotes the input magnification for our sources. The circles present the mean values of data points in each 0.3 dex bin of the output $M_{\rm BH}$. The black dashed line denotes the 1:1 line and the black dotted lines present 0.3 dex.  
    }
    \label{fig6: comp}
\end{figure}

In the final step, we evaluate the accuracy of our method by comparing the recovered BH masses ($M_{\rm BH,output}$) with the true input mass ($M_{\rm BH,input}$). This comparison is shown in Figure~\ref{fig6: comp}. For high-mass BHs ($M_{\rm BH}\geq 10^6\,M_\odot$), the recovered masses exhibit excellent agreement with their input values. The data points cluster tightly around the one-to-one relation (the 1:1 line), with a small scatter of less than 0.1 dex, as determined from the binned data (red circles). This indicates a high degree of accuracy and precision for our measurements in this mass range. In the intermediate-mass range ($10^5\leq M_{\rm BH}<10^6\,M_\odot$), the results remain robust. While the scatter increases to approximately 0.3 dex, the measurements are still centered on the 1:1 relation. The absence of a significant deviation from this line suggests that our method introduces no or negligible systematic offset when measuring these BH masses. This comparison reveals a systemic issue of BH mass measurement for the low-mass regime ($M_{\rm BH}<10^5\,M_\odot$). Firstly, detections in this regime are limited to sources experiencing extreme gravitational lensing ($\mu >20$; Figure~\ref{fig5: mock}). Secondly, the mass measurements for these objects are subject to considerable uncertainty, characterized by a large scatter of 0.5 dex and a notable systematic offset from the identity line. Furthermore, we identify a population of catastrophic outliers: sources with low input masses ($M_{\rm BH}<10^5\,M_\odot$) that are misclassified as high-mass BHs (visible as data points in the upper-left corner of the figure). This misidentification is attributed to strong noise fluctuations in the mock spectra, which are similar to the condition of real targets like ID\,38548.

Through this mock observation and analysis, we confirm the feasibility of 

\section{All fitting results}
\label{app2: fitting}

\begin{figure*}
    \centering
    \includegraphics[width=\linewidth]{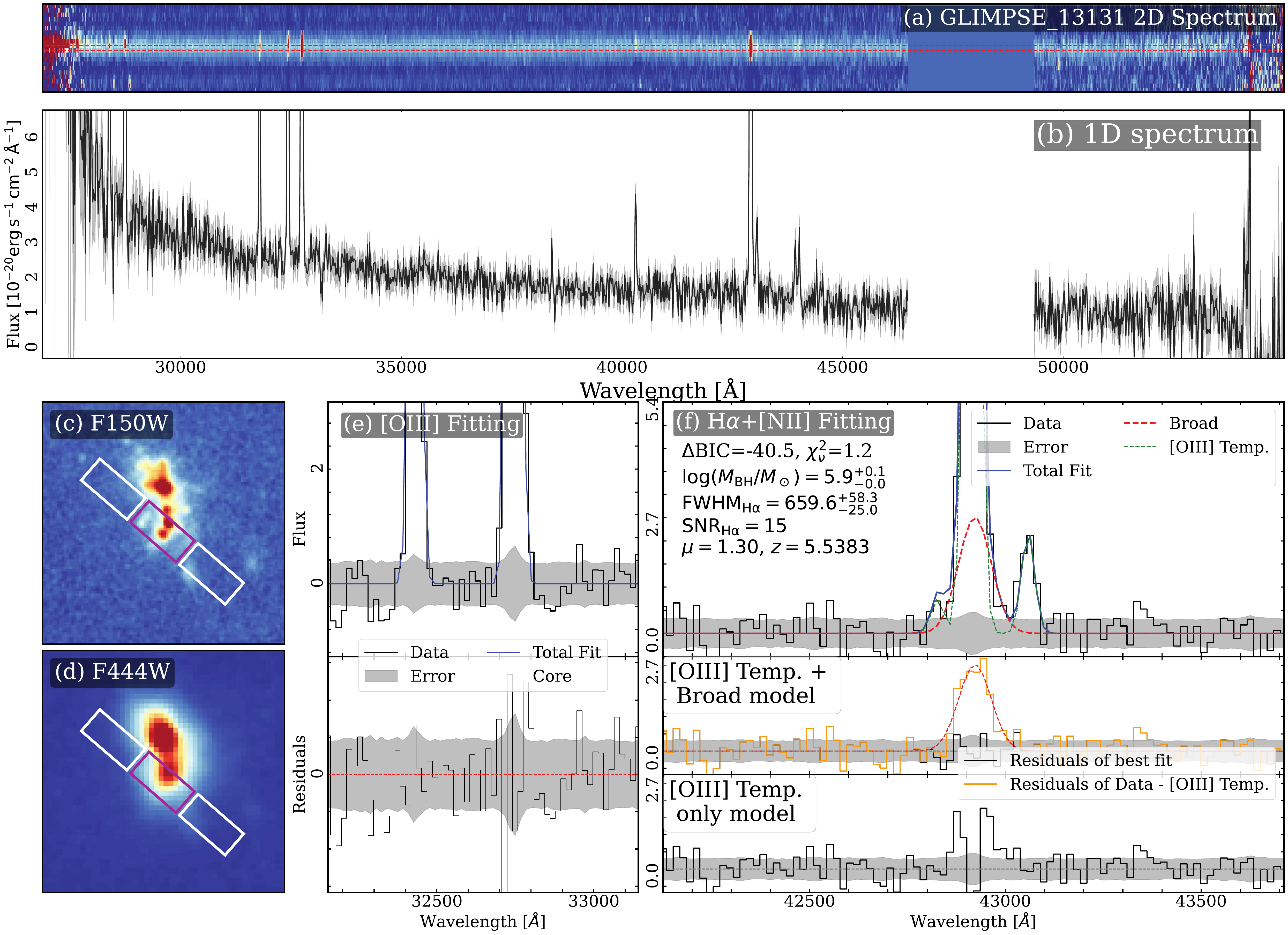}
    \caption{
    Panel (a) displays the 2D NIRSpec spectrum, while panel (b) shows the corresponding 1D spectrum, extracted from the spatial region indicated by the red lines in the upper panel. Panels (c) and (d) present JWST/NIRCam image cutouts of the source in the F150W and F444W filters, respectively. The spectroscopic observation was conducted using a three-shutter slit with a three-point nodding pattern; the central shutter targeting the source is highlighted in magenta, and the adjacent shutters used for background subtraction are outlined in white. Panel (e) presents a multi-component fit to the \oiii\ emission line doublet. The observed data and its associated uncertainty are shown as a black histogram and a gray shaded region. The best-fit model (blue line) is composed of a narrow 'core' component (blue dotted line) and a broader, outflow component (green dotted line; the outflow component is not necessary for this source). Panel (f) shows the fit to the \ha+\nii\ emission complex. The top sub-panel displays the data (black histogram), the total best-fit model (blue dashed line), the broad \ha\ component, and the narrow-line template based on the \oiii\ profile. The lower two sub-panels show the fit residuals (data minus model). The middle sub-panel confirms the quality of the total fit, while the bottom sub-panel demonstrates that a model lacking the broad \ha\ component provides a significantly poorer fit to the data, thereby confirming the AGN detection.}
    \label{fig1: fitting}
\end{figure*}

\begin{figure*}
    \centering
    \includegraphics[width=0.9\linewidth]{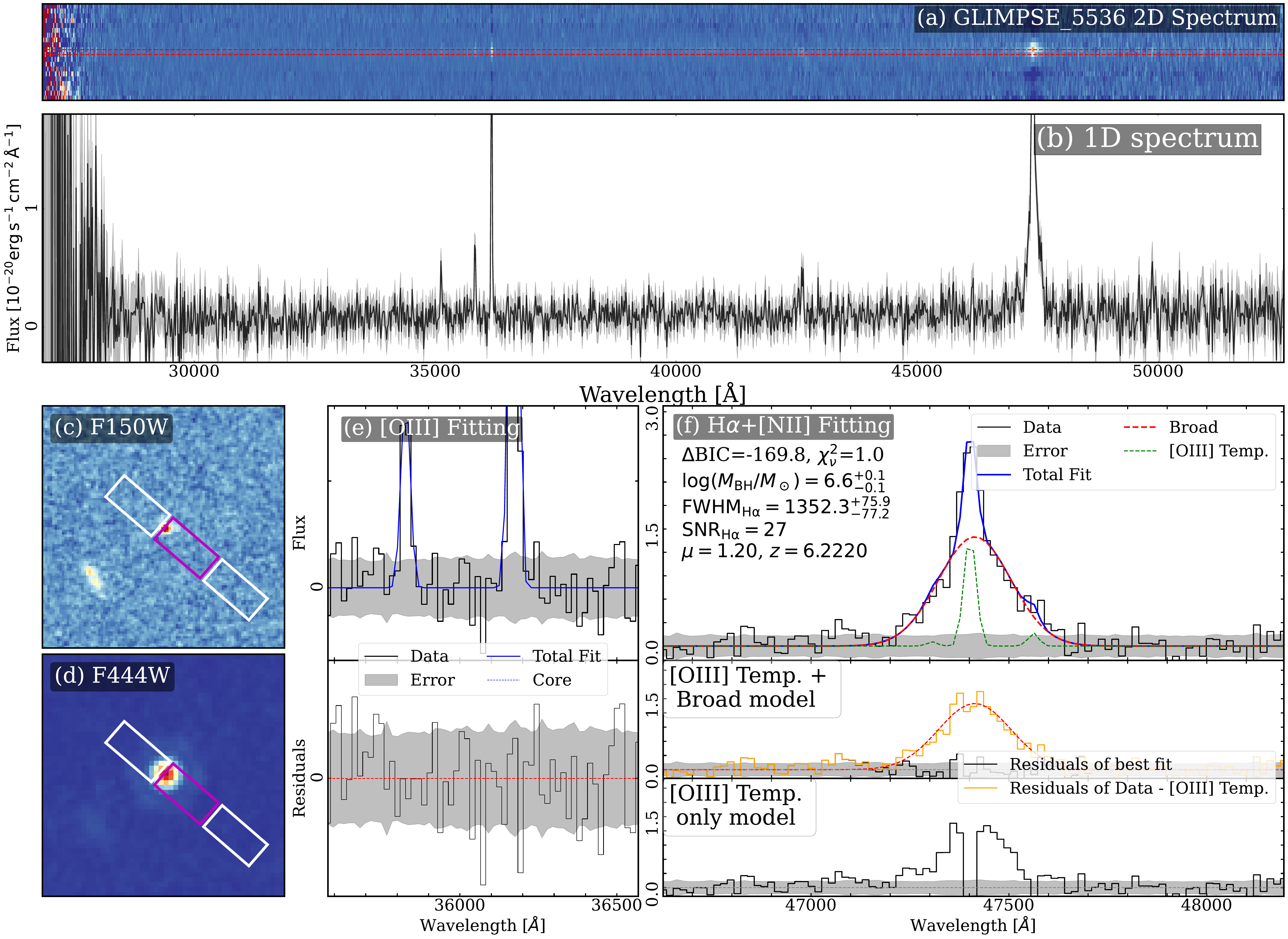}
    \includegraphics[width=0.9\linewidth]{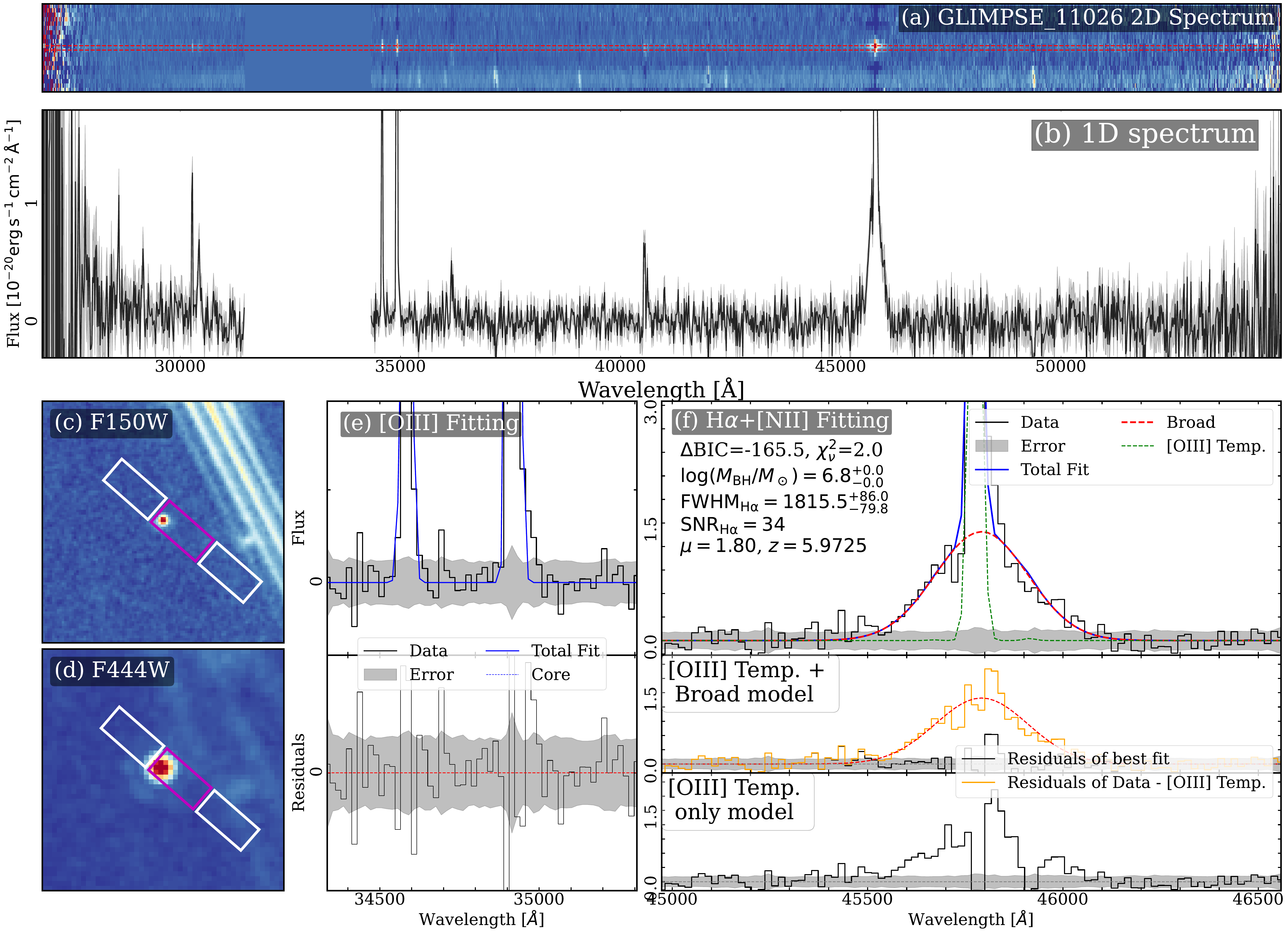}
    \caption{Fitting results. Similar to that shown in Figure~\ref{fig1: fitting}.}
\end{figure*}

\begin{figure*}
    \centering
    \includegraphics[width=0.9\linewidth]{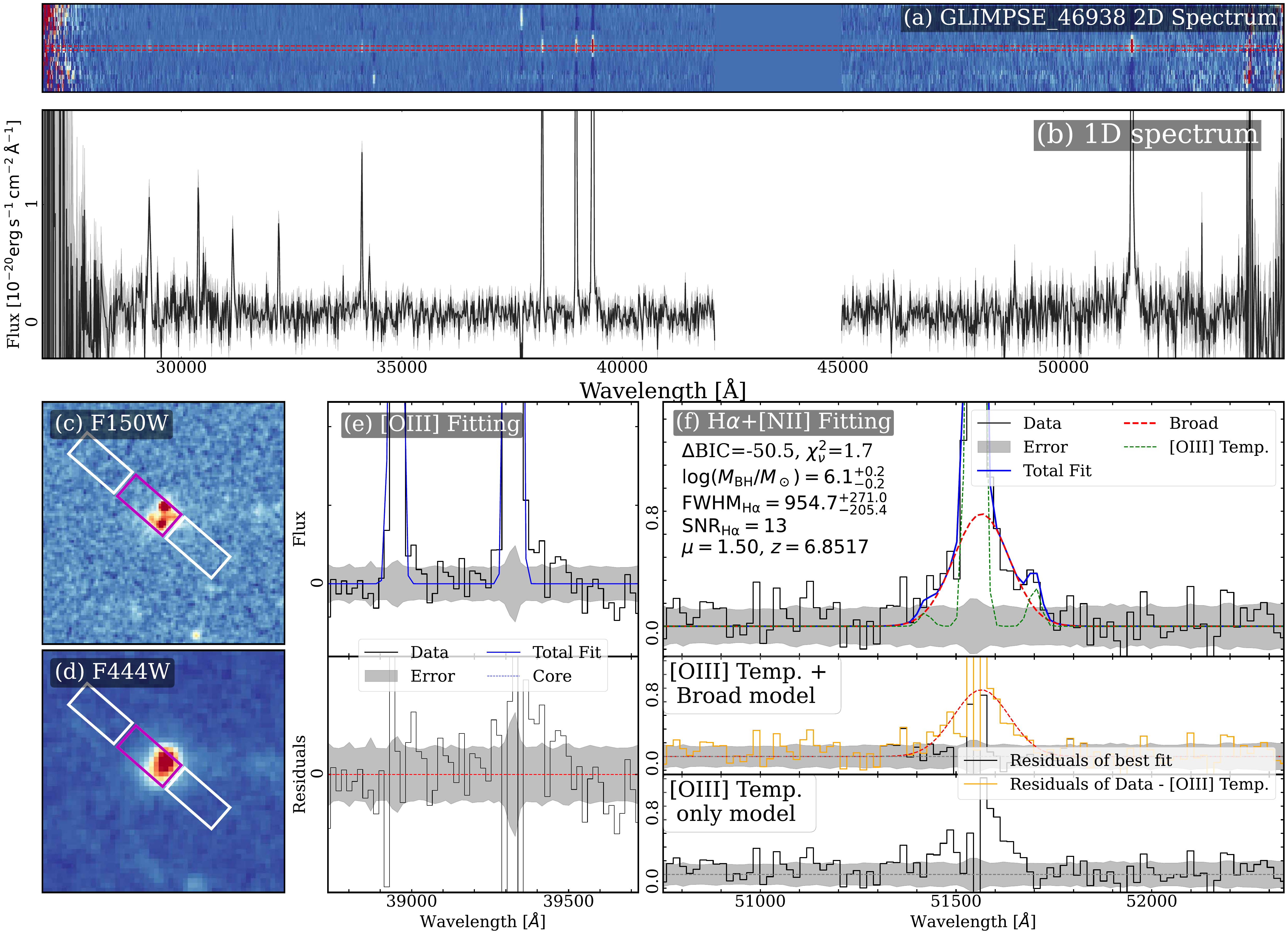}
    \includegraphics[width=0.9\linewidth]{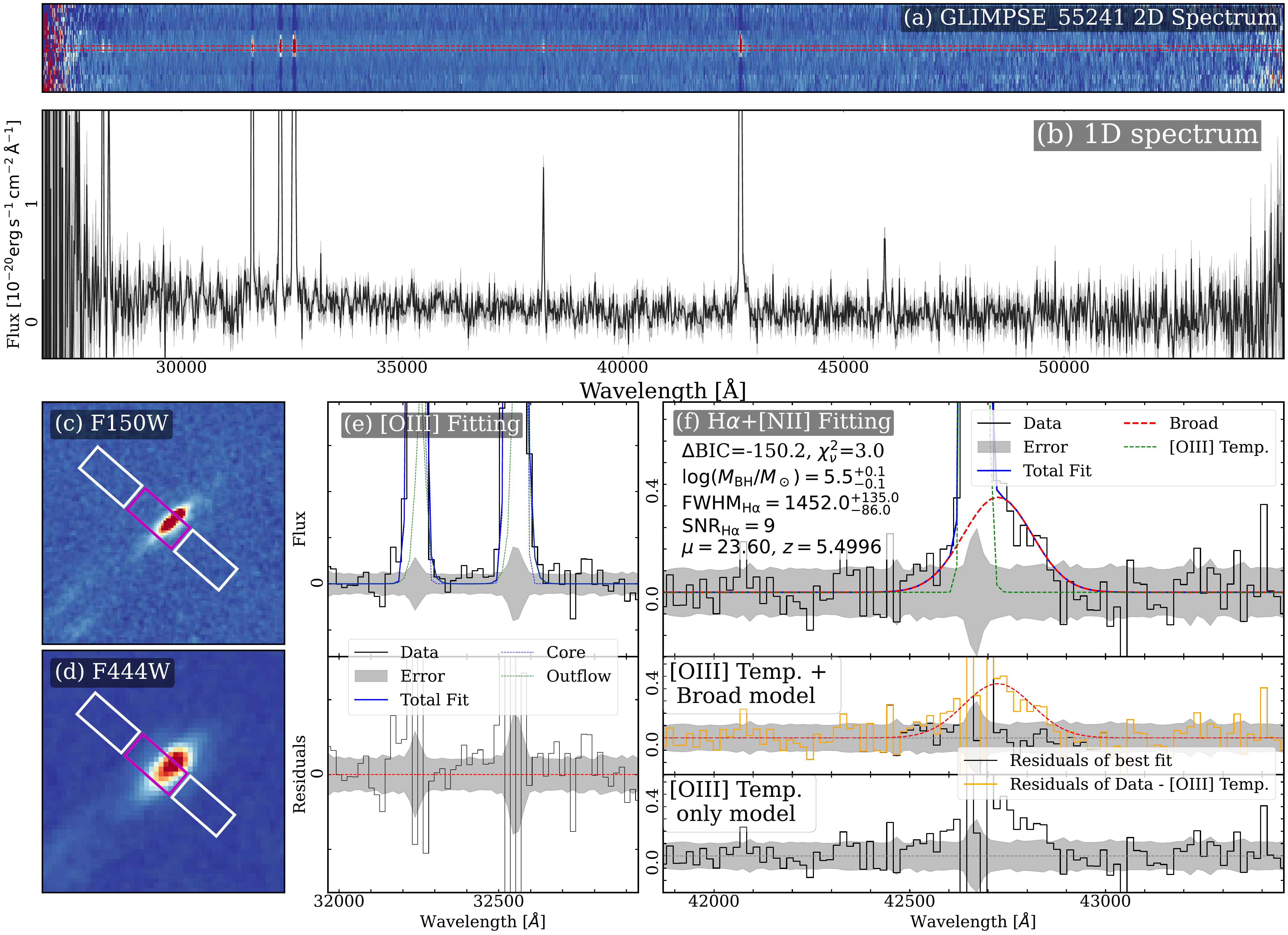}
    \caption{Fitting results. Similar to that shown in Figure~\ref{fig1: fitting}.}
\end{figure*}

\begin{figure*}
    \centering
    \includegraphics[width=0.9\linewidth]{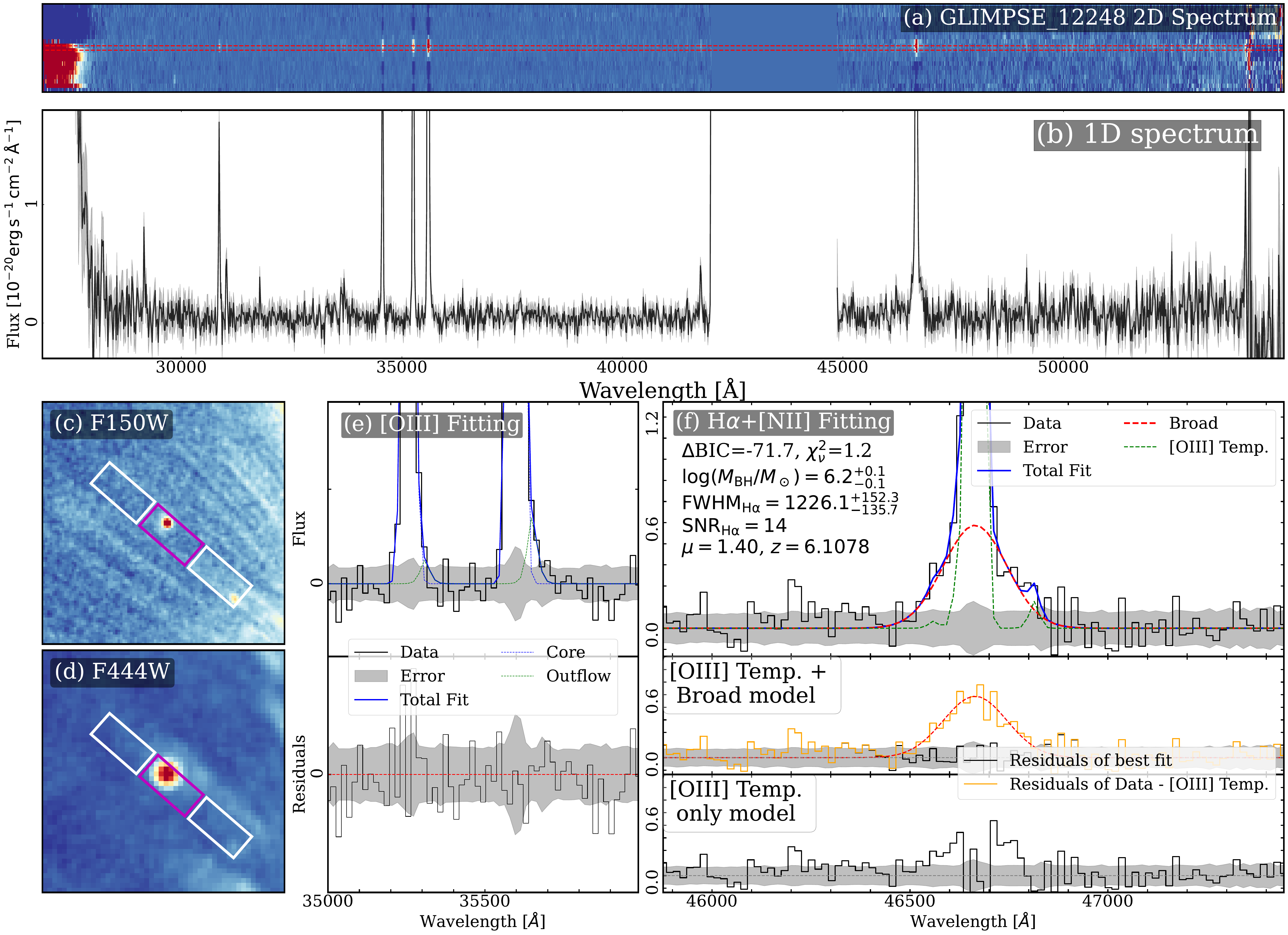}
    \includegraphics[width=0.9\linewidth]{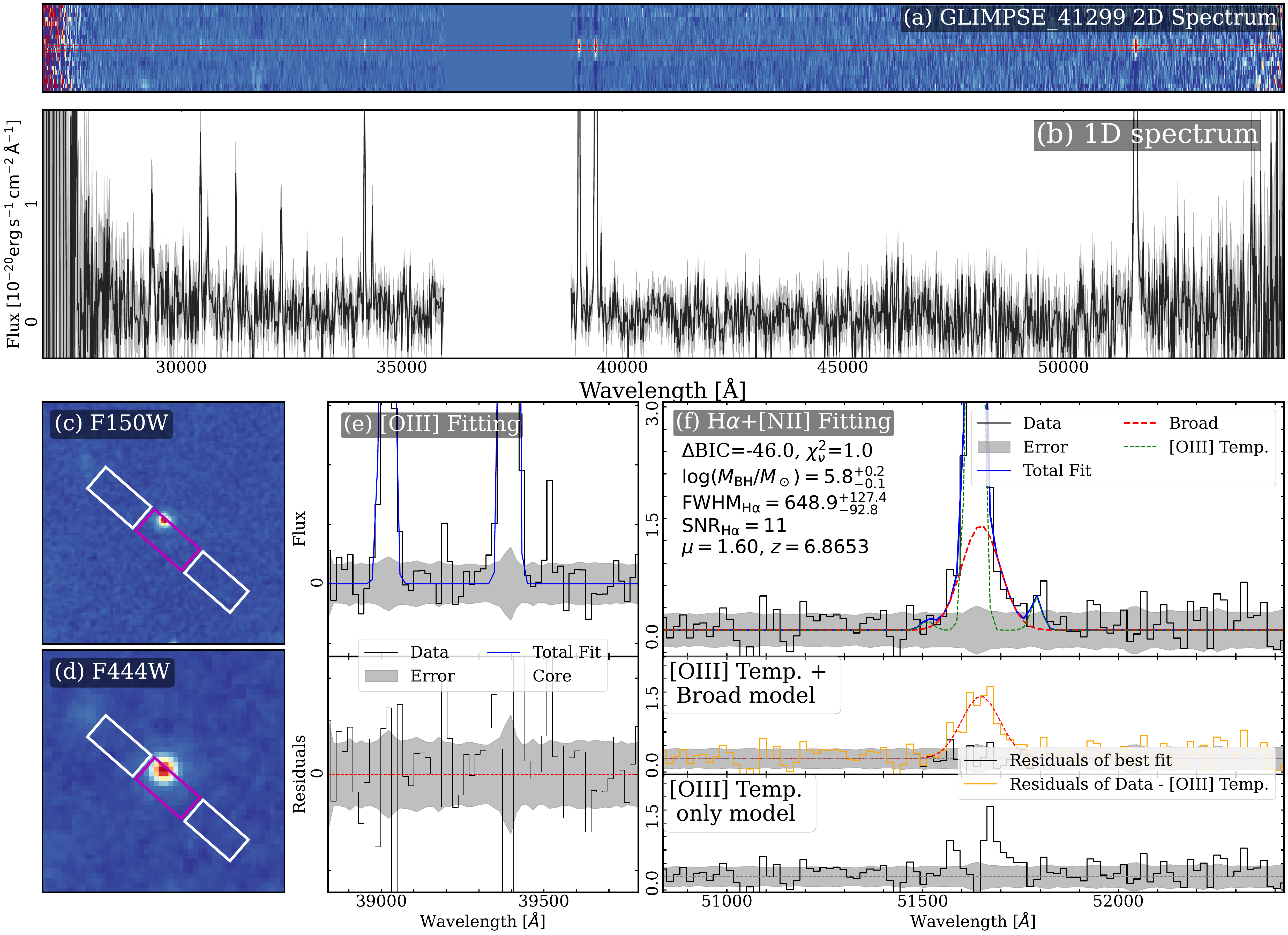}
    \caption{Fitting results. Similar to that shown in Figure~\ref{fig1: fitting}.}
\end{figure*}

\begin{figure*}
    \centering
    \includegraphics[width=\linewidth]{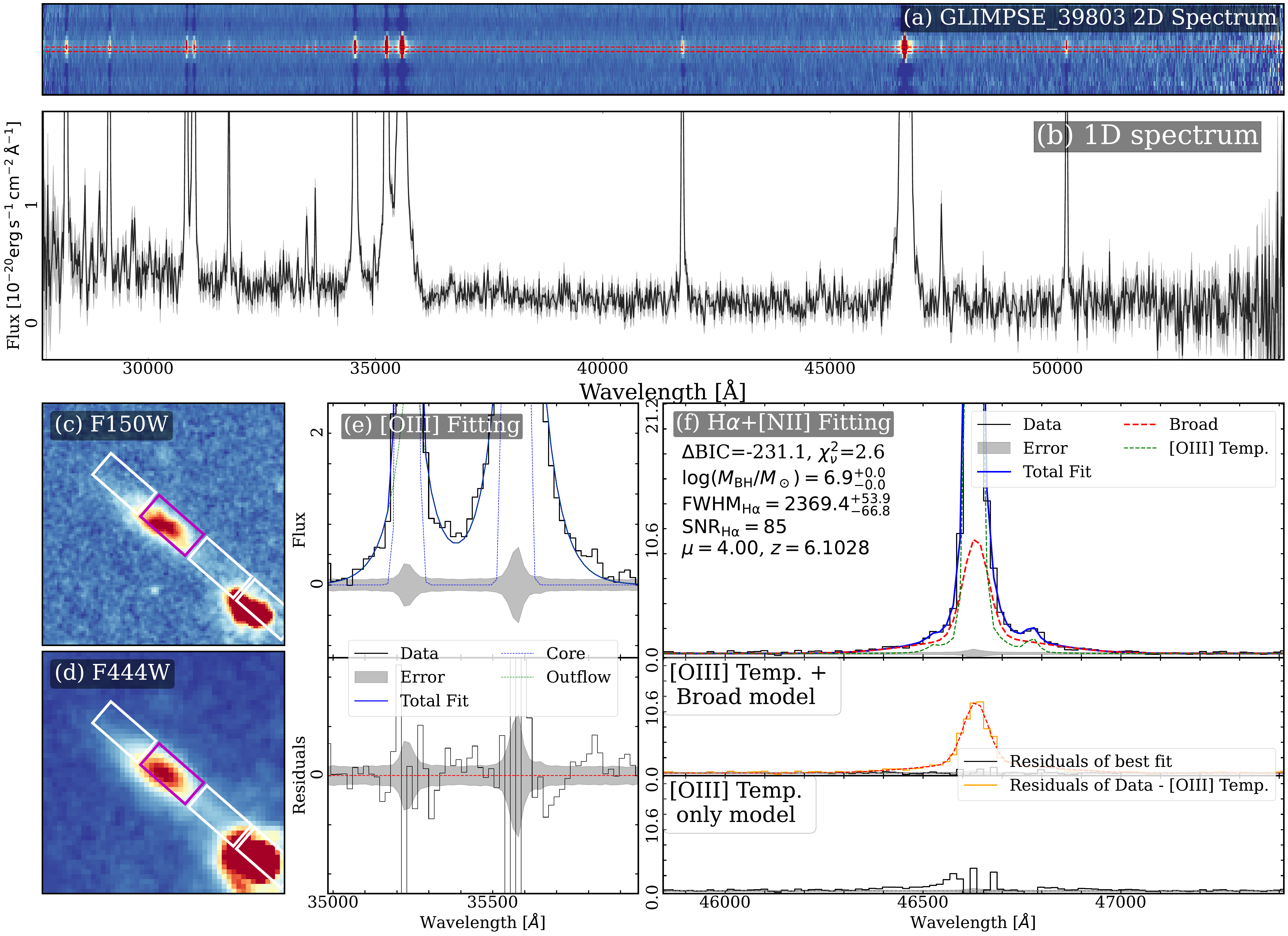}
    \caption{Fitting results. Similar to that shown in Figure~\ref{fig1: fitting}.}
\end{figure*}

\begin{figure*}
    \centering
    \includegraphics[width=0.9\linewidth]{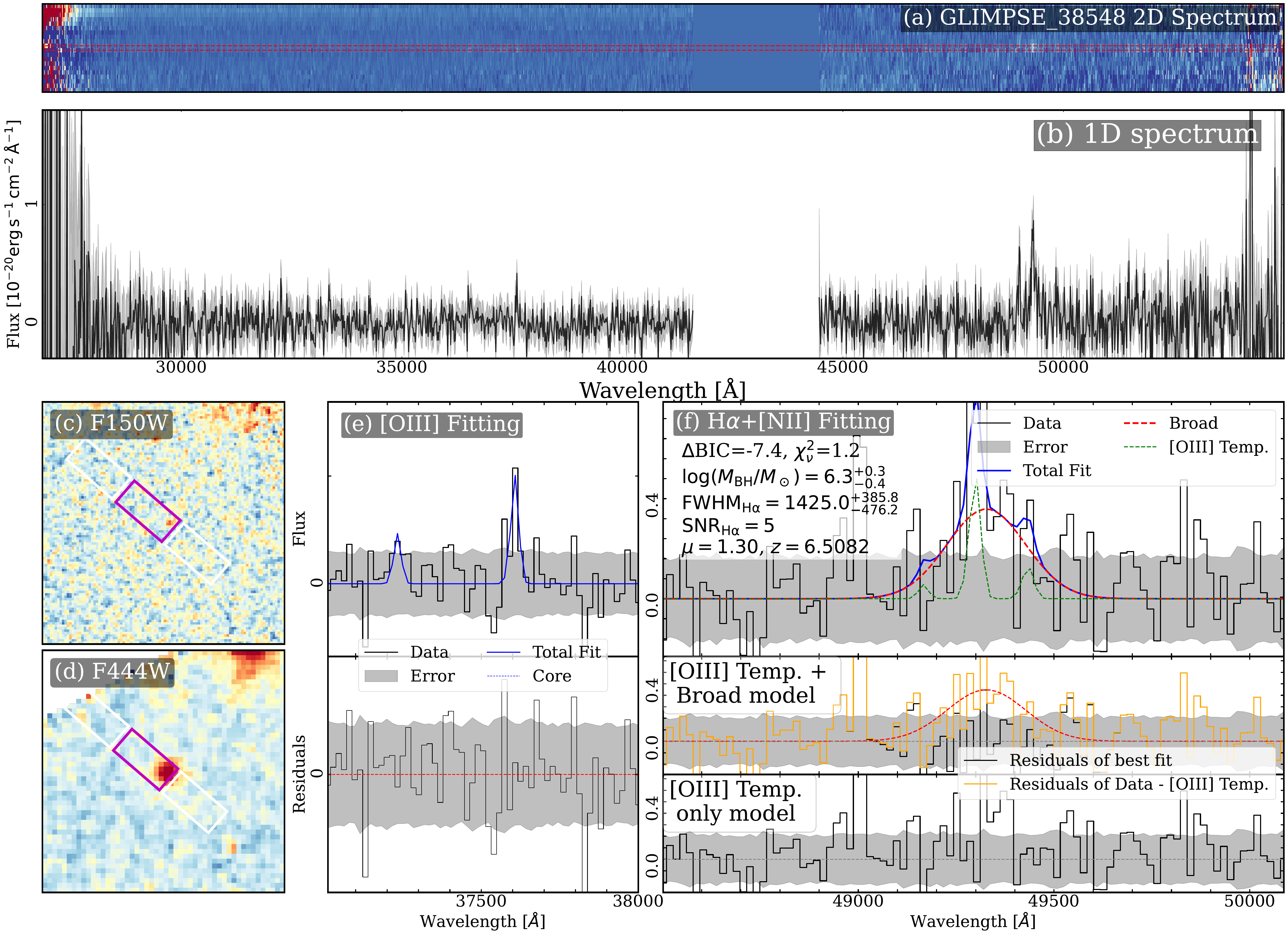}
    \includegraphics[width=0.9\linewidth]{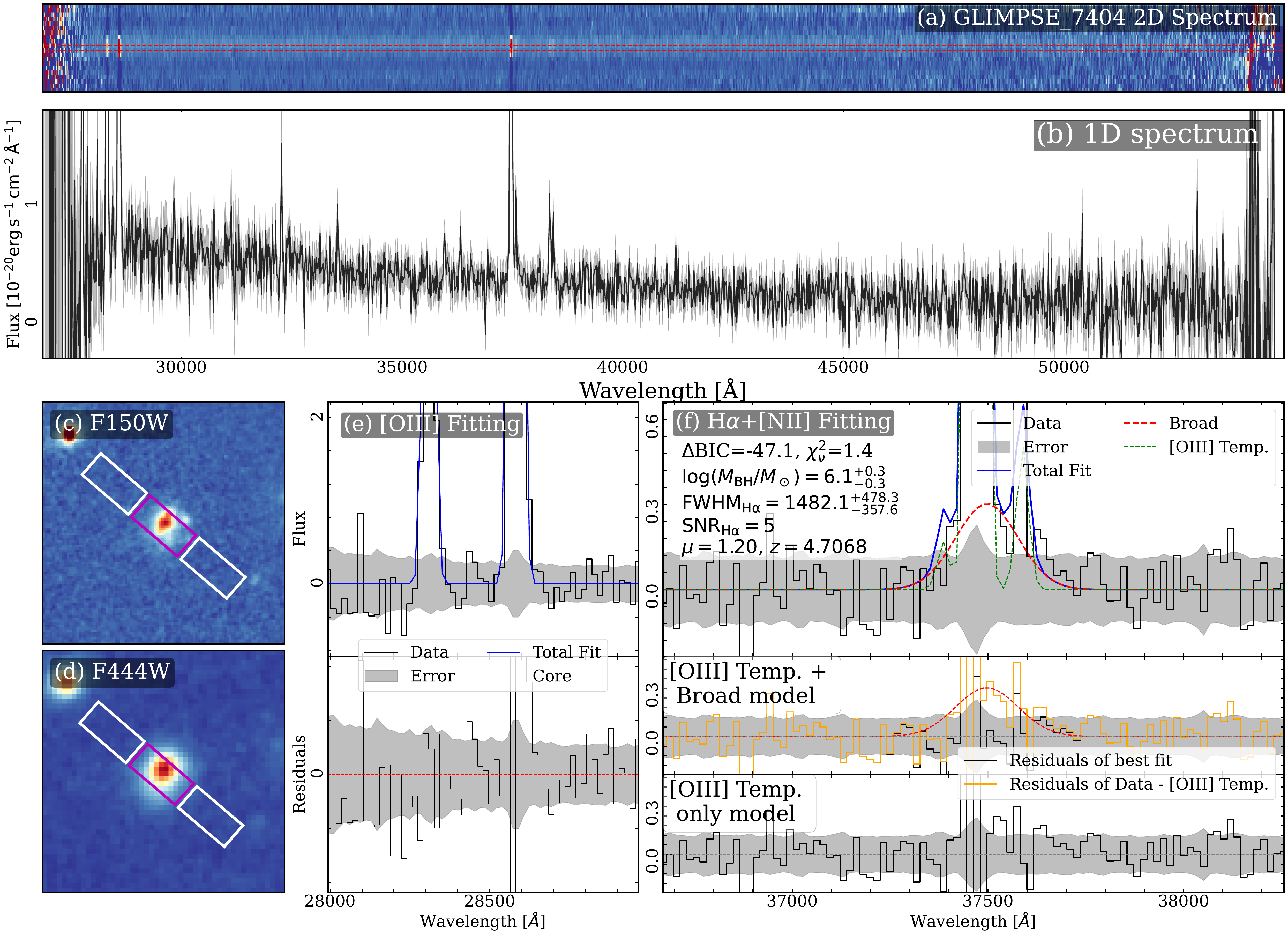}
    \caption{Fitting results, but for two tentative sources. Similar to that shown in Figure~\ref{fig1: fitting}.}
\end{figure*}

\section{All SED fitting}
\label{app3: sed}

\begin{figure*}
    \centering
    \includegraphics[width=0.24\linewidth]{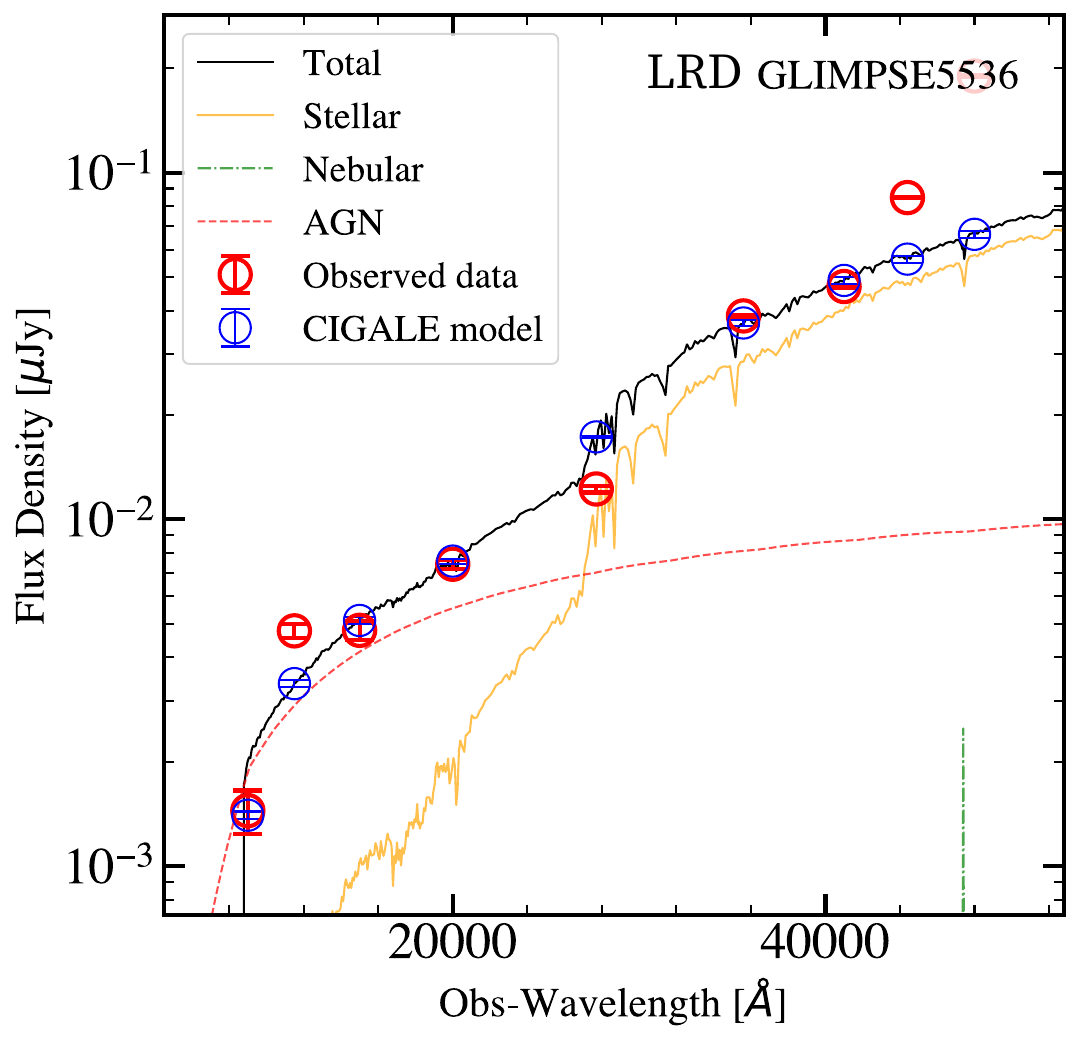}
    \includegraphics[width=0.24\linewidth]{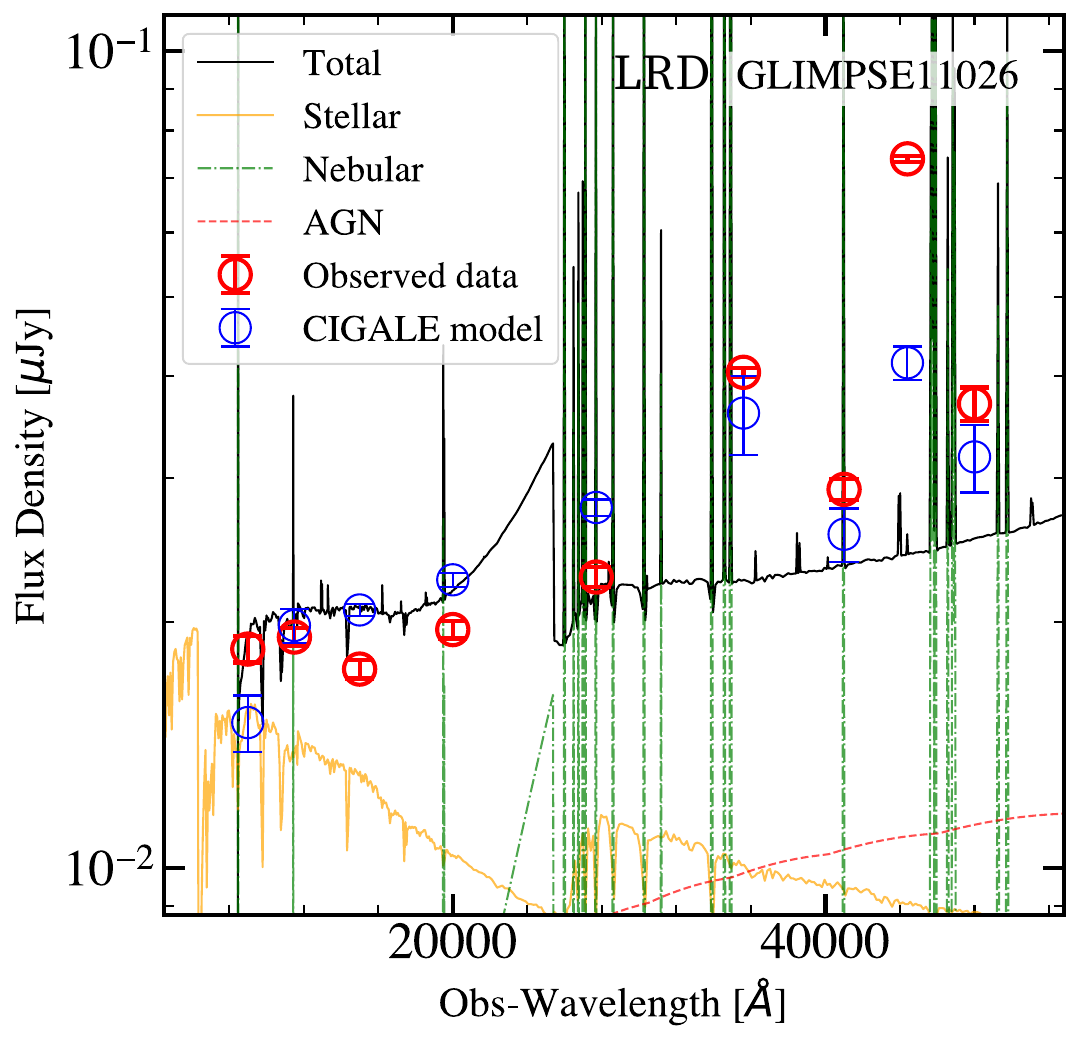}
    \includegraphics[width=0.24\linewidth]{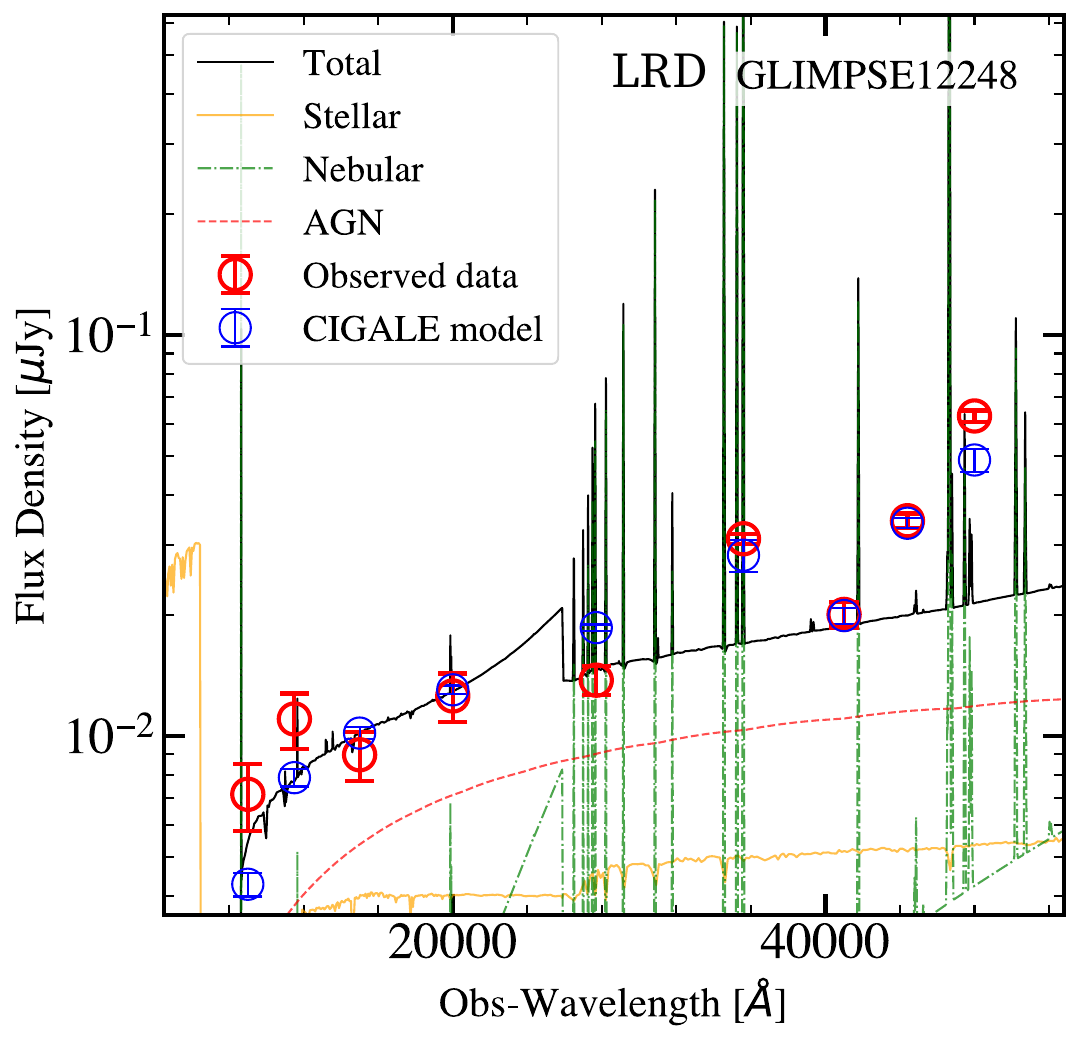}
    \includegraphics[width=0.24\linewidth]{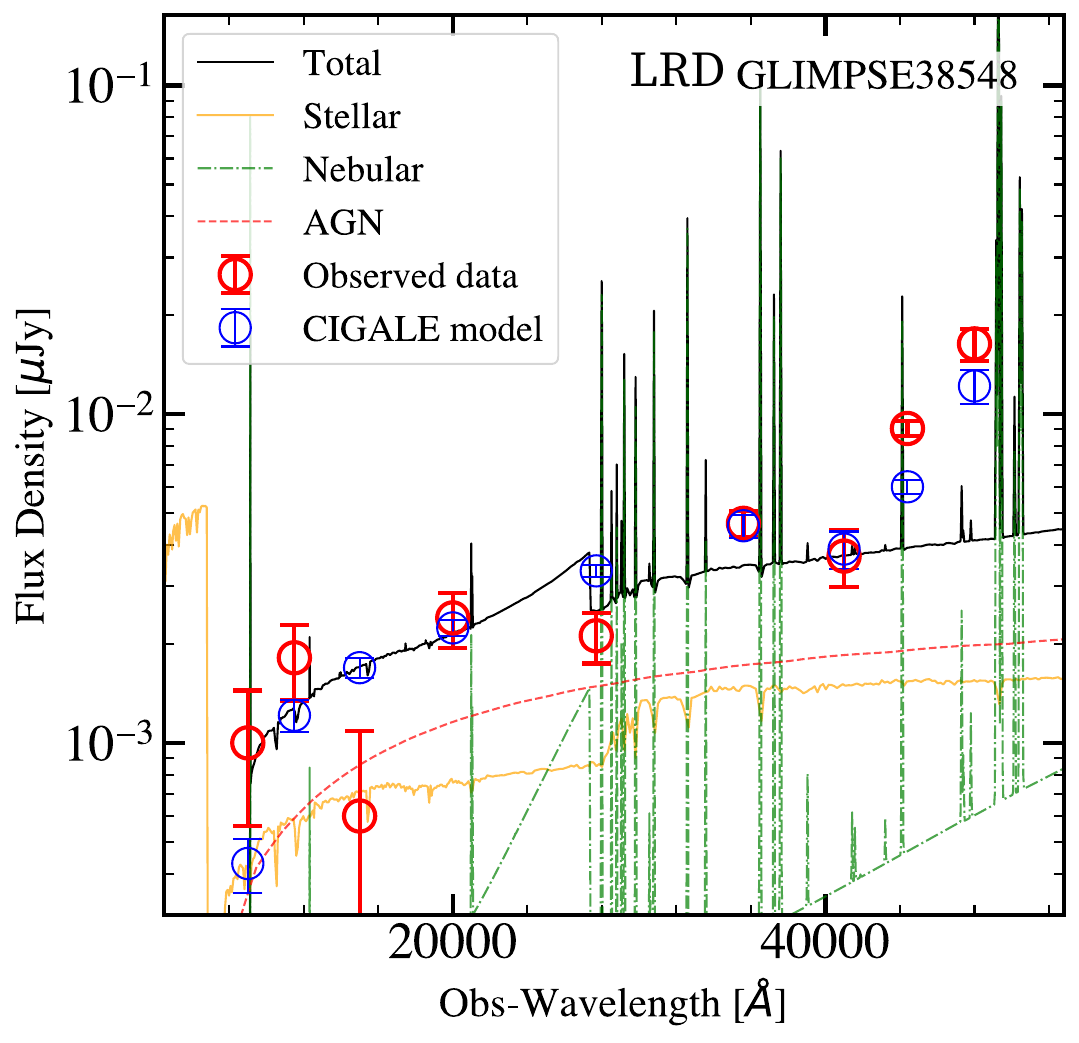}
    \includegraphics[width=0.24\linewidth]{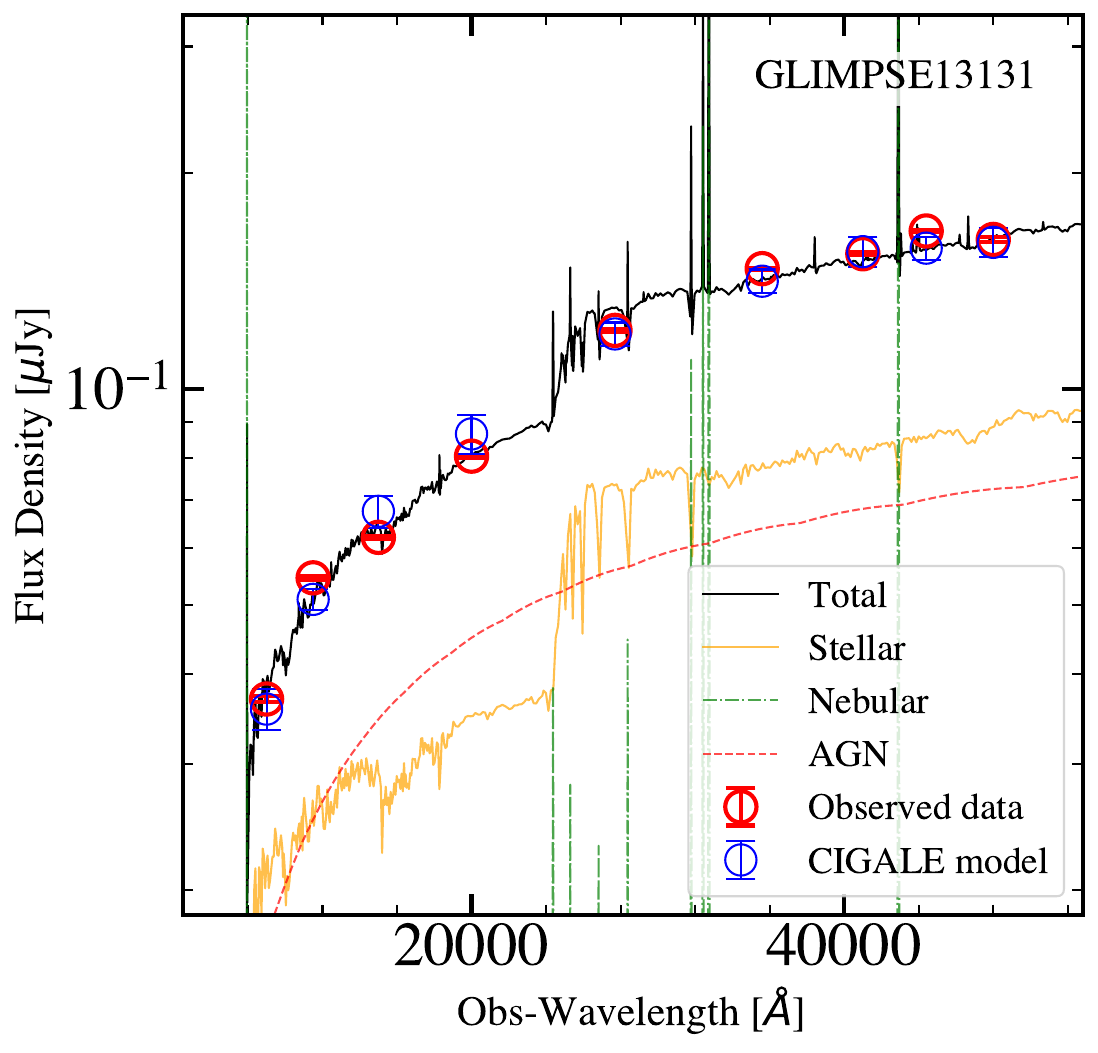}
    \includegraphics[width=0.24\linewidth]{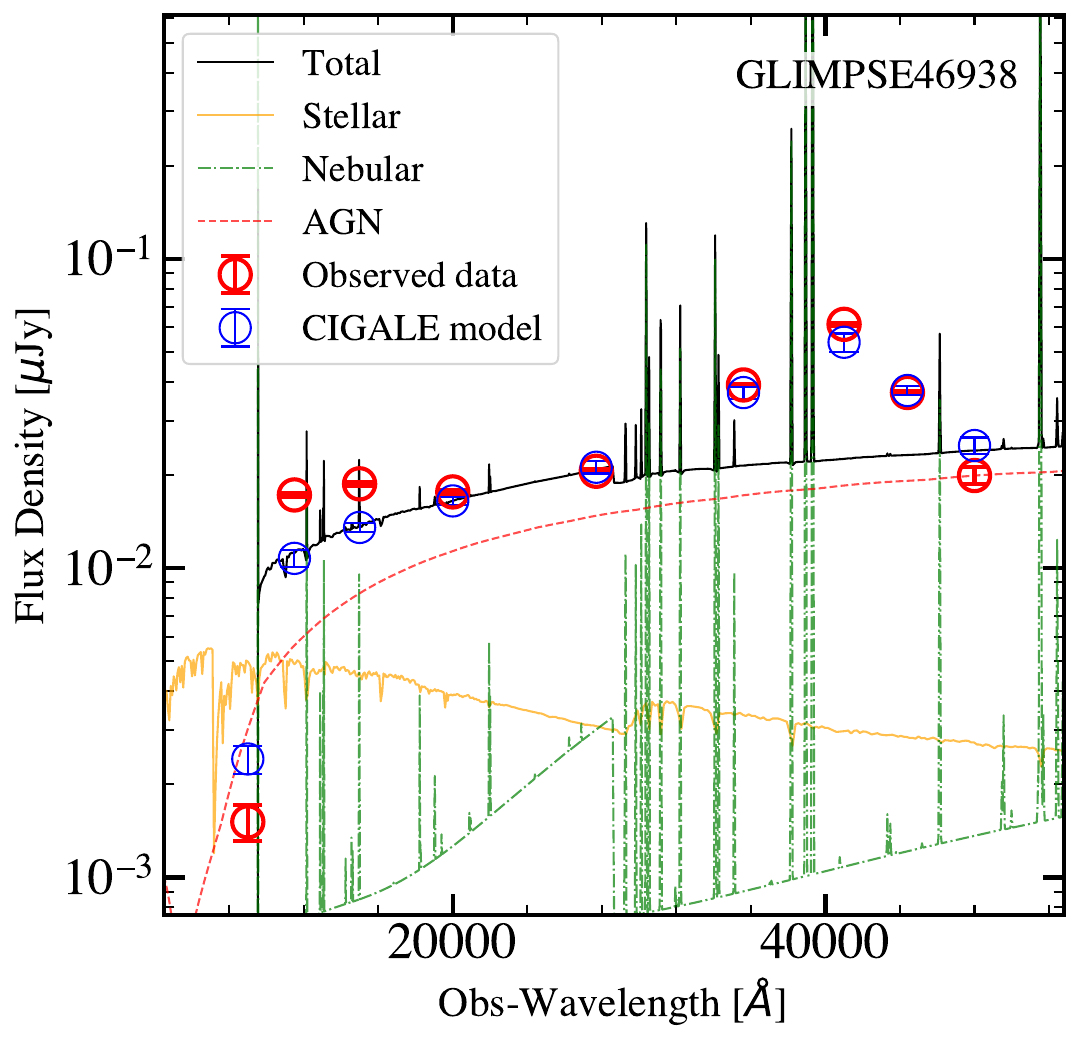}
    \includegraphics[width=0.24\linewidth]{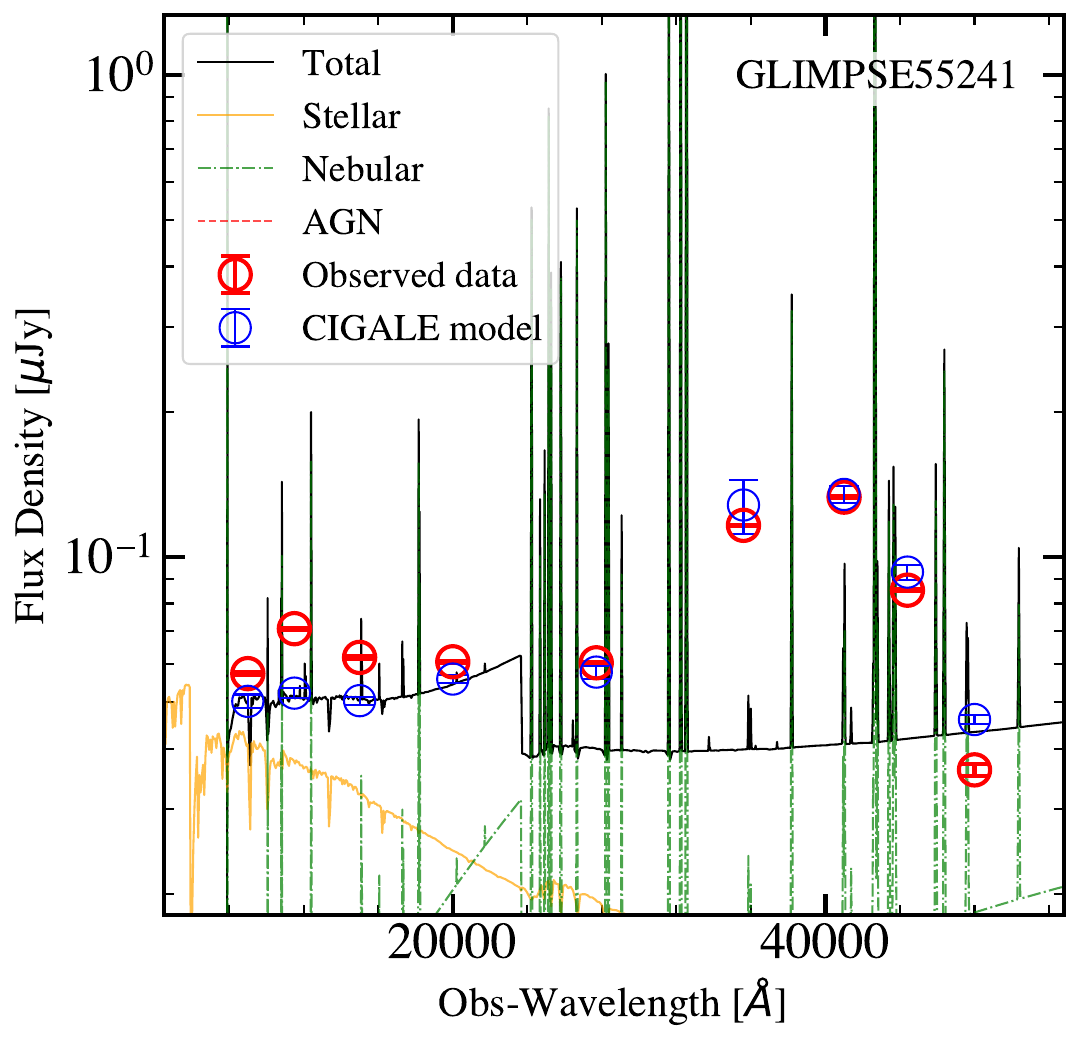}
    \includegraphics[width=0.24\linewidth]{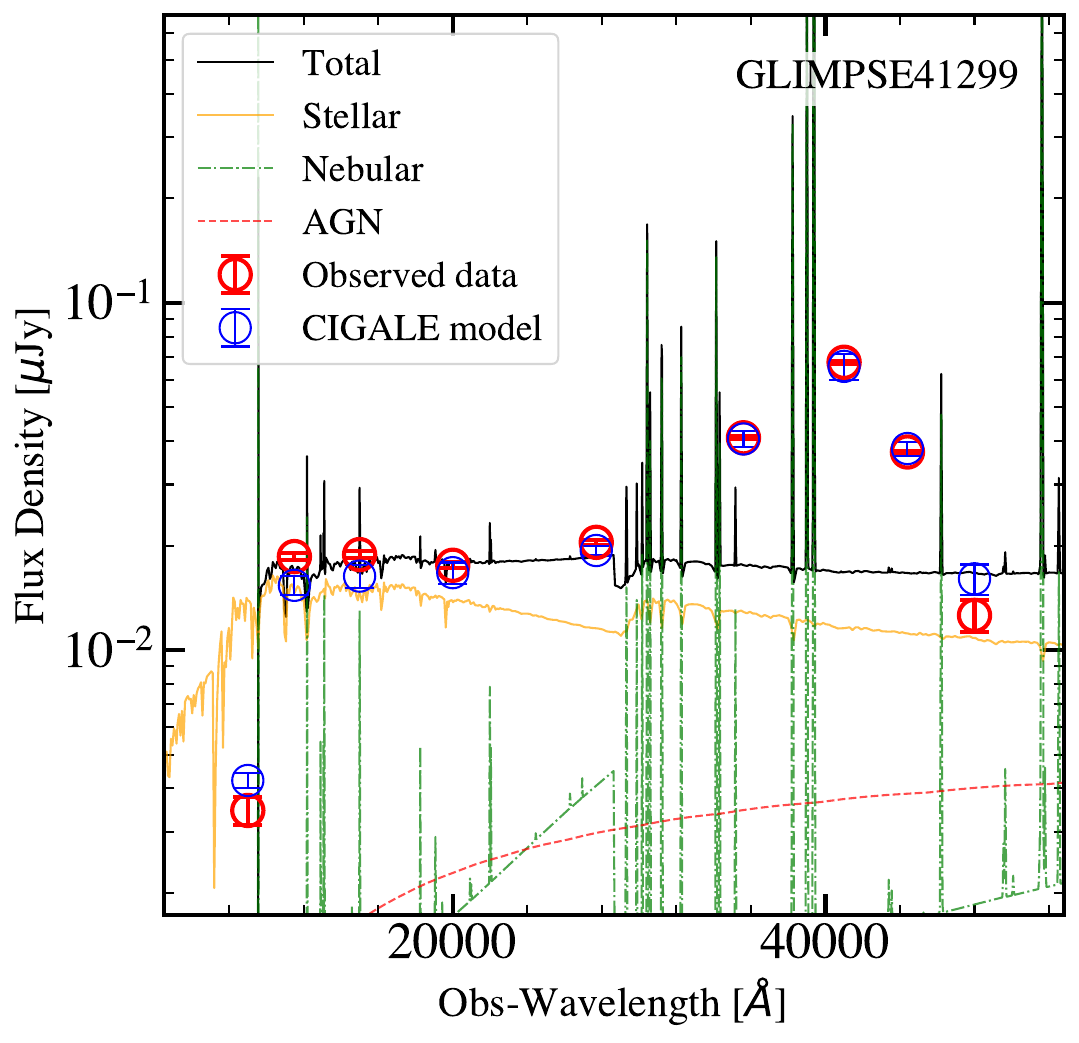}
    \includegraphics[width=0.24\linewidth]{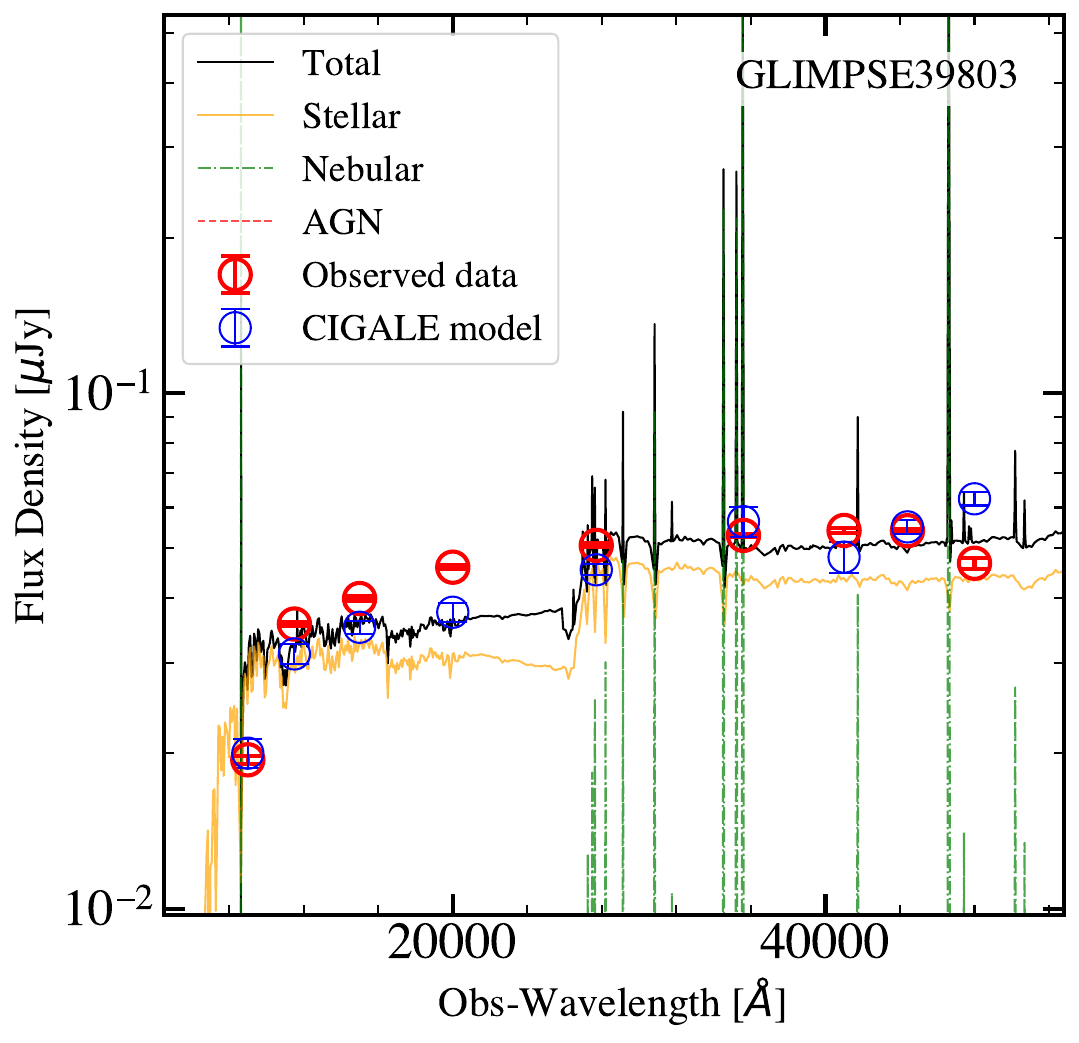}
    \includegraphics[width=0.24\linewidth]{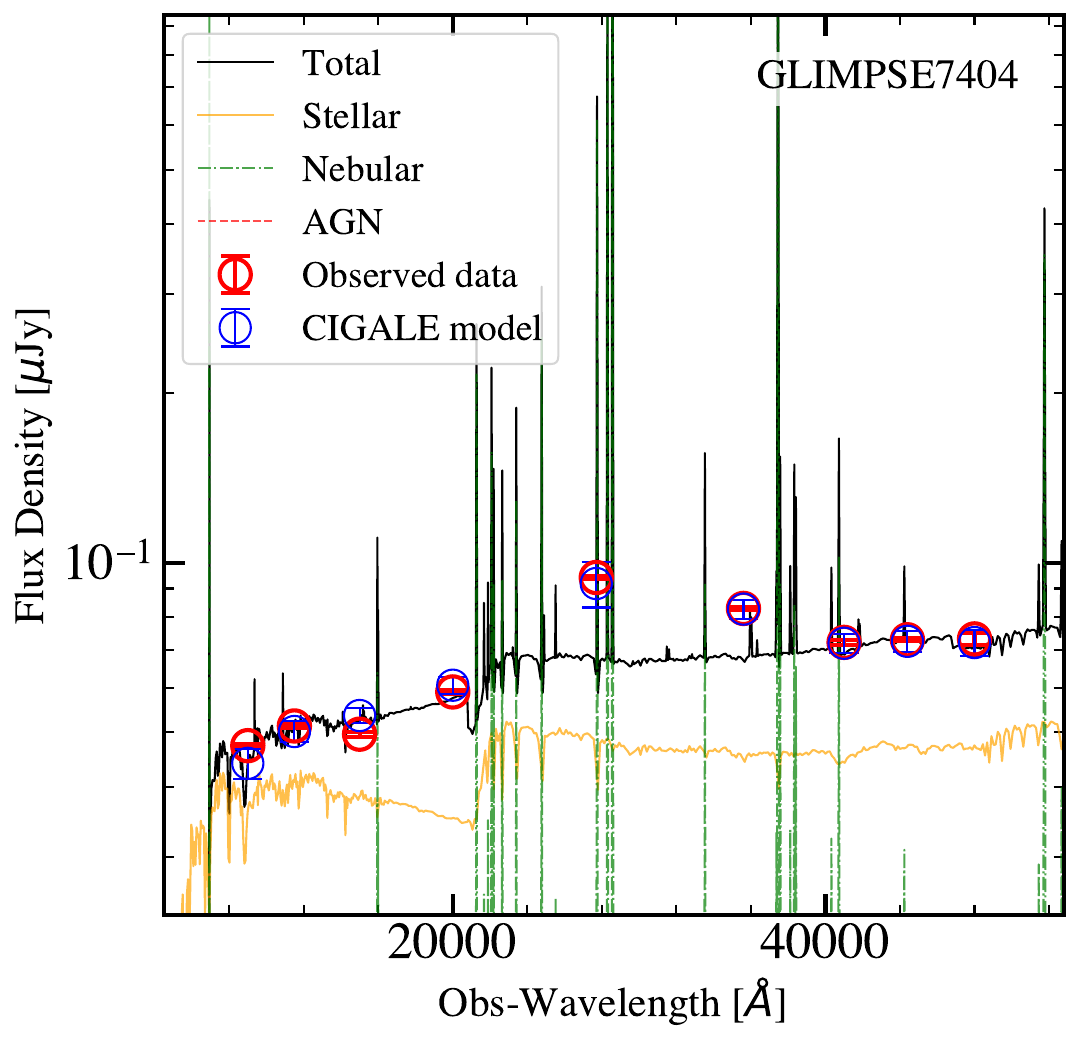}
    \caption{\texttt{CIGALE} fitting result for all our sources. Labels in this plot are as same as those in Figure~\ref{fig2: sed}. Four LRDs are explicitly labeled and listed in the first row. }
    \label{figapp3: seds}
\end{figure*}

\begin{figure}
    \centering
    \includegraphics[width=\linewidth]{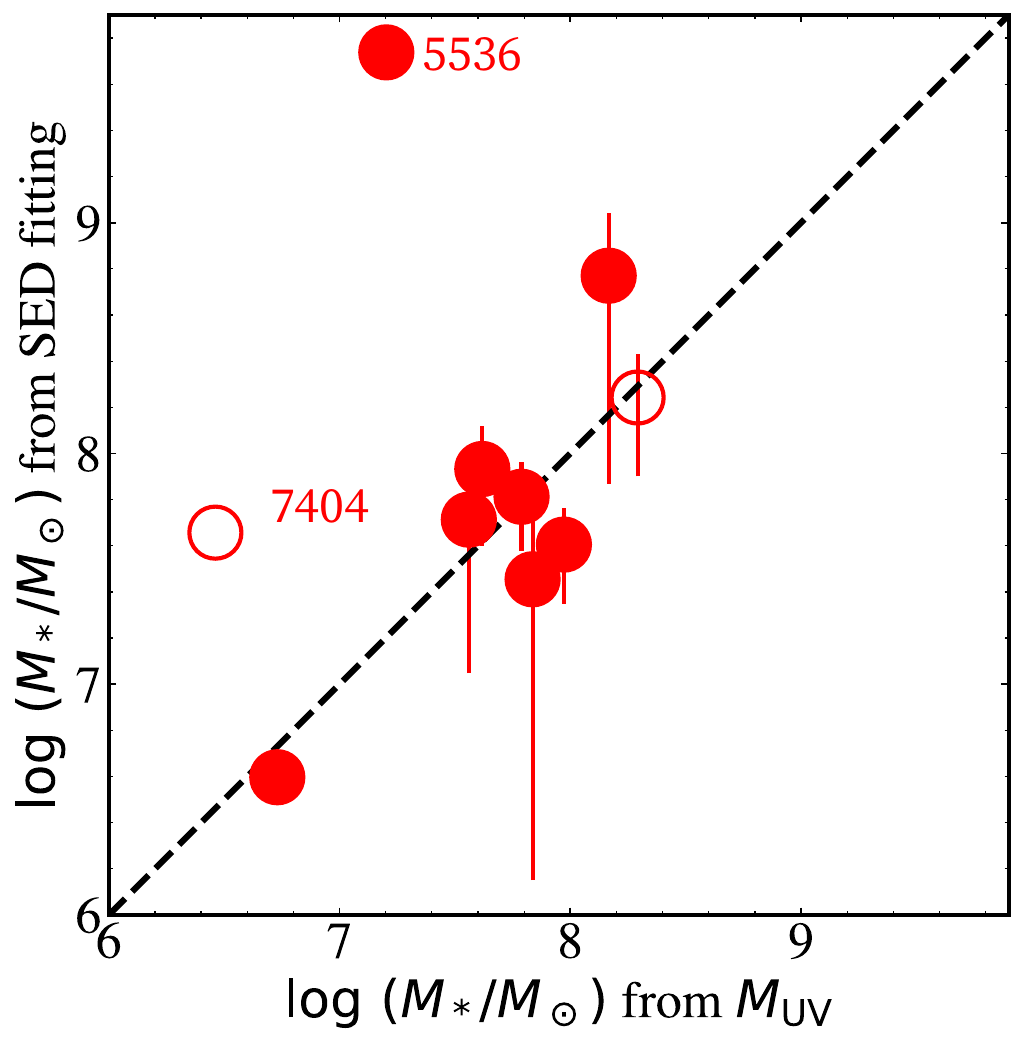}
    \caption{Comparison between $M_*$ derived from two different methods. Two outliers, ID~5536 and ID~7404 are explicitly labeled. The dashed black line represents the 1:1 relation. }
    \label{figapp3:mm-comparison}
\end{figure}



\end{document}